\documentclass[a4paper,10pt,headings=normal,bibliography=totoc]{scrartcl}

\makeatletter
\DeclareOldFontCommand{\rm}{\normalfont\rmfamily}{\mathrm}
\makeatother

\usepackage{scrhack}  % Fix a LaTeX warning

\usepackage{authblk}

\usepackage[utf8]{inputenc}
\usepackage{fancyhdr}
\usepackage{amsmath}
\usepackage{amssymb}
\usepackage{amsthm}
\usepackage{bm}
\usepackage{colonequals}
\usepackage{overpic}
\usepackage{url}
\usepackage{xspace}
\usepackage[square,numbers,sort]{natbib}

\usepackage[colorinlistoftodos]{todonotes}
\usepackage{environ}
\usepackage{enumitem}
\usepackage{listings}
\usepackage{float}
\usepackage{minibox}

\usepackage{tikz}
\usetikzlibrary{arrows}
\tikzset{
    treenode/.style = {
        align=center,
        inner sep=0pt,
        text centered,
        font=\sffamily,
        rectangle,
        rounded corners=3mm,
        draw=black,
        minimum width=2em,
        minimum height=2em,
        inner sep=1mm,
        outer sep=0mm
    },
    smalltreenode/.style = {
        treenode,
        font=\footnotesize
    },
    basisnode/.style = {
        treenode,
        minimum width=9mm,
    },
    smallbasisnode/.style = {
        basisnode,
        font=\footnotesize
    }
}

\pgfmathsetseed{\number\pdfrandomseed}
\usepackage{hyperref}

\hypersetup{pdftitle={Function space bases in the dune-functions module},
            pdfauthor={Christian Engwer, Carsten Gräser, Steffen Müthing, and Oliver Sander} }

\usepackage{attachfile2}
\usepackage{fancyvrb}

\definecolor{interfacecolor}{rgb}{0.95,0.95,1}
\lstdefinestyle{Example}{}
\lstdefinestyle{Interface}{backgroundcolor=\color{interfacecolor},frame=single}

\newcommand{\cpp}[1]{\lstinline[basicstyle=\ttfamily]!#1!}
\newcommand{\cppbreak}[1]{\lstinline[basicstyle=\ttfamily,breaklines]!#1!}

\newtheorem{definition}{Definition}

\newcommand{\R}{\mathbb{R}}
\newcommand{\N}{\mathbb{N}}
\newcommand{\abs}[1]{{\lvert#1\rvert}}
\newcommand{\norm}[1]{\lVert#1\rVert}
\newcommand{\op}[1]{\operatorname{#1}}
\newcommand{\st}{\; : \;}
\renewcommand{\div}{\operatorname{div}}

\newcommand{\dune}{\textsc{Dune}\xspace}
\newcommand{\program}[1]{\textsc{#1}\xspace}

\IfFileExists{stokes-taylorhood.cc}{\newcommand{\PATH}{.}}{\newcommand{\PATH}{../../examples}}

\newcommand{\dunemodule}[1]{\texttt{#1}}
\newcommand{\file}[1]{\texttt{#1}}

\definecolor{lightblue}{HTML}{55AAFF}

\graphicspath{{gfx/}}

\title{Function space bases in the dune-functions module}
\author[1]{Christian Engwer}
\author[2]{Carsten Gräser}
\author[3]{Steffen Müthing}
\author[4]{Oliver Sander}

\affil[1]{Universität Münster, Institute for Computational und Applied Mathematics, christian.engwer@uni-muenster.de}
\affil[2]{Freie Universität Berlin, Institut für Mathematik, graeser@mi.fu-berlin.de}
\affil[3]{Universität Heidelberg, Institut für Wissenschaftliches Rechnen, steffen.muething@iwr.uni-heidelberg.de}
\affil[4]{TU Dresden, Institute for Numerical Mathematics, oliver.sander@tu-dresden.de}
\begin{document}

\maketitle

\begin{center}
\minibox[frame,c]{
  This work is licensed under a\\
  \emph{Creative Commons Attribution-NoDerivatives 4.0 International License}.\\
  \includegraphics[width=3cm]{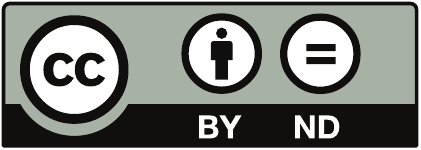}\\
  The full license text is available here:\\
  \url{https://creativecommons.org/licenses/by-nd/4.0/legalcode}
  }
\end{center}

\begin{abstract}
 The \dunemodule{dune-functions} \dune module provides interfaces for functions and function space bases.
 It forms one abstraction level above grids, shape functions, and linear algebra, and provides infrastructure
 for full discretization frameworks like \dunemodule{dune-pdelab} and \dunemodule{dune-fem}.
 This document describes the function space bases provided by \dunemodule{dune-functions}.  These are
 based on an abstract description of bases for product spaces as trees of simpler bases.
 From this description, many different numberings of degrees of freedom by multi-indices can be
 derived in a natural way. We describe the abstract concepts, document the programmer interface,
 and give a complete example program that solves the stationary Stokes equation using Taylor--Hood elements.
\end{abstract}

\section*{Introduction}

The core modules of the \dune software system focus on low-level infrastructure for
implementations of simulation algorithms for partial differential equations.  Modules like
\dunemodule{dune-grid} and \dunemodule{dune-istl} provide programmer interfaces (APIs) to finite element grids
and sparse linear algebra, respectively, but little more. Actual finite element functions only
appear in the \dunemodule{dune-localfunctions} module, which deals with discrete function spaces
on single grid elements exclusively.

On top of these core modules, various other modules in the \dune ecosystem implement finite element and finite volume assemblers
and solvers, and the corresponding discrete function spaces. The most prominent ones are
\dunemodule{dune-pdelab}%
\footnote{\url{https://dune-project.org/modules/dune-pdelab}}
and \dunemodule{dune-fem},%
\footnote{\url{https://dune-project.org/modules/dune-fem}}
but smaller ones like \dunemodule{dune-fufem}%
\footnote{\url{https://dune-project.org/modules/dune-fufem}}
exist as well.  The functionality of these modules overlaps to a considerable extent, even though
each such module has a different focus.

The \dunemodule{dune-functions}  module was written to partially overcome this fragmentation,
and to unify parts of the competing implementations.
It picks a well-defined aspect of finite element assembly---finite element spaces and functions---and,
in the \dune spirit, provides abstract interfaces that try to be both extremely flexibly
and efficient.  The hope is that other implementations of the same functionality
eventually replace their implementations by a dependence on \dunemodule{dune-functions}.
Indeed, at the time of writing at least \dunemodule{dune-pdelab} and \dunemodule{dune-fufem} are in the process
of migrating, and have stated their clear intention to complete this migration eventually.

Of the two parts of \dunemodule{dune-functions} functionality, the APIs for discrete and
closed-form functions have already been described in a separate paper~\cite{engwer_graeser_muething_sander:2015}.
The present document focuses on spaces of discrete functions.  However,
the central concept is not the function space itself, but rather the {\em basis} of the function space.
This is because even though finite element spaces play a central role in theoretical considerations of
the finite element method,
actual computations use coefficient vectors, which are defined with respect to particular bases.  Also,
for various finite element spaces, more than one basis is used in practice.  For example,
the space of second-order Lagrangian finite elements is used both with the nodal (Lagrange) basis~\cite{braess:2013},
and with the
hierarchical basis~\cite{bank:1996}.  Discontinuous Galerkin spaces can be described in terms of Lagrange bases,
monomial bases, Legendre bases, and more~\cite{hesthaven_warburton:2008}.
It is therefore important to be able to distinguish these different
representations of the same space in the application code.
For these reasons, the main \dunemodule{dune-functions} interface represents a basis of a
discrete function space, and not the space itself.

Finite element function space bases frequently exhibit a fair amount of structure.  In particular, vector-valued
and mixed finite element spaces can be written as products of simpler spaces. Even more, such spaces
have a natural structure as a tree, with scalar-valued or otherwise irreducible spaces forming the leaves, and
products forming the inner nodes. The \dunemodule{dune-functions} module
allows to systematically construct new bases by multiplication of existing bases.
The resulting tree structure is reproduced as type information in the code.
This tree construction of finite element spaces has first been systematically worked out in~\cite{muething:2015}.

For the basis functions in such a non-trivial tree structure, there is no single canonical way
to index them.  Keeping all degrees of freedom in a single standard array would require indexing
by a contiguous, zero-starting set of natural numbers. On the other hand, from the tree structure
of the basis follows a natural indexing by multi-indices, which can be used to address nested
vector and matrix data types, like the ones provided by \dunemodule{dune-istl}. Closer inspection
reveals that these two possibilities are just two extreme cases of a wider scale of indexing rules.
The \dunemodule{dune-functions} module therefore provides a systematic way to construct such
rules.  While some of them are somewhat contrived, many others really are useful
in applications.

This document describes Version~2.6 of the \dunemodule{dune-functions} module.
The module is hosted on the \dune project homepage \url{www.dune-project.org}.
Installation instructions and an up-to-date class documentation can be found there.

% Force a new page here to avoid spurious automatic newpage right after 'Contents'
\newpage

\setcounter{tocdepth}{2}  % Show only sections and subsections, but no more
\tableofcontents

\section{Function space bases}
\label{sec:finite_element_trees}

Before we can explain the programmer interface for bases of discrete function spaces in Chapter~\ref{sec:function_space_bases_implementation},
we need to say a few words about how these bases can be endowed with an abstract tree structure.
Readers who are only interested in finite element spaces of scalar-valued functions may try to proceed directly to
Chapter~\ref{sec:function_space_bases_implementation}.  They should only know that whenever a
local finite element tree
is mentioned there, this tree consists of a single node only, which is the local finite element basis.
Similarly, for a scalar finite element space the tree of multi-indices used to index the
basis functions simply represents a contiguous, zero-starting set of natural numbers.

\subsection{Trees of function spaces}

Throughout this paper we assume that we have a single fixed domain $\Omega$, and all function spaces
that we consider are defined on this domain.  The focus is on spaces of functions that are
piecewise polynomial with respect to a grid, but that is not actually required yet.

For a set $R$ we denote by $R^\Omega \colonequals \{f:\Omega \to R\}$
the set of all functions mapping from $\Omega$ to $R$. For domains $\Omega\subset \R^d$
we write $P_k(\Omega) \subset \R^\Omega$ for the space of all
scalar-valued continuous piecewise polynomials of degree at most $k$ on $\Omega$
with respect to some given triangulation.
We will omit the domain if it can be inferred from the context.

Considering the different finite element spaces that appear in the literature, there are some that we will
call \emph{irreducible}.  By this term we mean all bases of scalar-valued functions,
but also others like the Raviart--Thomas basis that cannot easily be written as a combination
of simpler bases.
Many other finite element spaces arise naturally as a combination of simpler ones.
There are primarily two ways how two vector spaces $V$ and $W$ can be combined
to form a new one: sums and products.%
\footnote{While these are also called
internal and external sums, respectively, we stick to the terminology
\emph{sum} and \emph{product} in the following.
}

For sums, both spaces need to
have the same range space $R$, and thus both be subspaces of $R^\Omega$.
Then the vector space sum
\begin{equation*}
  V + W
  \colonequals
  \{ v + w \; : \; v \in V, \; w \in W \}
\end{equation*}
in $R^\Omega$ will have that same range space.
For example, a $P_2$-space
can be viewed as a $P_1$-space plus a hierarchical extension spanned by bubble functions~\cite{bank:1996}.
XFEM spaces~\cite{moes_dolbow_belytschko:1999} are constructed by adding particular weighted Heaviside
functions to a basic space to capture free discontinuities.
The \dunemodule{dune-functions} module does not currently support constructing sums
of finite element bases, but this may be added in later versions.

The second way to construct finite element spaces from simpler ones uses Cartesian products.
Let $V \subset (\R^{r_1})^\Omega$ and $W \subset (\R^{r_2})^\Omega$ be two function spaces.
Then we define the product of $V$ and $W$ as
\begin{align*}
  V \times W
    \colonequals \big\{ (v,w) \st v \in V, \; w \in W \big\}.
\end{align*}
Functions from this space take values in $\R^{r_1} \times \R^{r_2} = \R^{r_1 + r_2}$.
It should be noted that the Cartesian product of
vector spaces must not be confused with the tensor product of these spaces.
Rather, the $k$-th power
of a single space can be viewed as the tensor product of that space with $\R^k$, i.e,
\begin{align*}
    (V)^k
    = \underbrace{V \times \dots \times V}_{k-\text{times}}
    = \R^k \otimes V.
\end{align*}

The product operation allows to build vector-valued and mixed finite element spaces of arbitrary complexity.
For example, the space of
first-order Lagrangian finite elements with values in $\R^3$ can be seen as the product $P_1 \times P_1 \times P_1$.
The lowest-order Taylor--Hood element is the product $P_2 \times P_2 \times P_2 \times P_1$
of $P_2 \times P_2 \times P_2$ for the velocities with $P_1$ for the pressure.
More factor bases can be included easily, if necessary.  We call such products of
spaces \emph{composite spaces}.

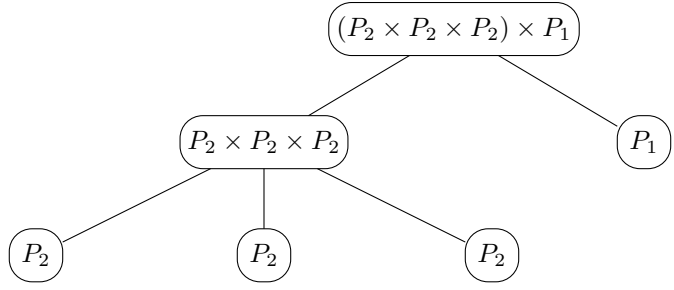
\begin{figure}
    \begin{center}
        \begin{tikzpicture}[
                level/.style={
                    sibling distance = (3-#1)*2cm + 1cm,
                    level distance = 1.5cm
                }
            ]
            \node [treenode] {$(P_2\times P_2 \times P_2) \times P_1$}
                child{ node [treenode] {$P_2 \times P_2 \times P_2$}
                    child{ node [treenode] {$P_2$} }
                    child{ node [treenode] {$P_2$} }
                    child{ node [treenode] {$P_2$} }
                }
                child{ node [treenode] {$P_1$} };
        \end{tikzpicture}
    \end{center}
    \caption{Function space tree of the Taylor--Hood space $(P_2 \times P_2 \times P_2)\times P_1$}
    \label{fig:taylor_hood_space_tree}
\end{figure}

In the Taylor--Hood space, the triple
$P_2 \times P_2 \times P_2$ forms a semantic unit---it contains the components of a velocity field.
The associativity of the product allows to write the Taylor--Hood space
as $(P_2 \times P_2 \times P_2) \times P_1$, which makes the semantic relationship clearer.
Grouped expressions of this type are conveniently visualized as tree structures.  This
suggests to interpret composite
finite element spaces as tree structures.  In these structures, leaf nodes represent scalar or otherwise irreducible spaces,
and inner nodes represent products of their children.  Subtrees then represent composite
finite element spaces.  Figure~\ref{fig:taylor_hood_space_tree} shows the Tayor--Hood finite element
space in such a tree representation. Note that in this document all trees are \emph{rooted} and \emph{ordered},
i.e., they have a dedicated root note, and the children of each node have a fixed given ordering.
Based on this child ordering we associate to each child the corresponding zero-based index.

While the inner tree nodes may initially appear like useless artifacts of the tree representation, they are often extremely useful
because we can treat the subtrees rooted in those nodes as individual trees in their own right.
This often allows to
reuse existing algorithms that expect to operate on those subtrees in more complex settings.

\subsection{Trees of function space bases}
\label{sec:basistree}

The multiplication of finite-dimensional spaces naturally induces a corresponding operation on bases
of such spaces.  We introduce a generalized tensor product notation:
Consider linear ranges $R_0,\dots,R_{m-1}$ of function spaces $R_0^\Omega,\dots,R_{m-1}^\Omega$,
and the $i$-th canonical basis vector $\mathbf{e}_i$ in $\R^m$.
Then
\begin{align*}
  \mathbf{e}_i \otimes f
  \colonequals (0,\dots,0,\underbrace{f}_{\text{$i$-th entry}},0,\dots,0)
  \in \prod_{j=0}^{m-1} \Bigl(R_j^\Omega\Bigr) = \Bigl(\prod_{j=0}^{m-1} R_j\Bigr)^\Omega,
\end{align*}
where $0$ in the $j$-th position denotes the zero-function in $R_j^\Omega$.
Let $\Lambda_i$ be a function space basis of the space $V_i = \operatorname{span} \Lambda_i$
for $i=0,\dots,m-1$. Then a natural basis $\Lambda$ of the product space
\begin{align*}
  V_0 \times \dots \times V_{m-1}
  = \prod_{i=0}^{m-1} V_i
  = \prod_{i=0}^{m-1} \operatorname{span}\Lambda_i
\end{align*}
is given by
\begin{align}
  \label{eq:basis_product}
  \Lambda =
    \Lambda_0 \sqcup \dots \sqcup \Lambda_{m-1}
    = \bigsqcup_{i=0}^{m-1} \Lambda_i
    \colonequals \bigcup_{i=0}^{m-1} \mathbf{e}_i \otimes \Lambda_i.
\end{align}
The product $\mathbf{e}_i \otimes \Lambda_i$ is to be understood element-wise,
and the ``disjoint union'' symbol $\sqcup$ is used here
as a simple short-hand notation for \eqref{eq:basis_product}
and not to be understood as an associative binary operation.
Using this new notation we have
\begin{align*}
  \operatorname{span} \Lambda
    = \operatorname{span} \bigl( \Lambda_0 \sqcup \dots \sqcup \Lambda_{m-1} \bigr)
%  = \prod_{i=0}^{m-1} \operatorname{span} \Lambda_i
    = (\operatorname{span} \Lambda_0) \times \dots \times (\operatorname{span} \Lambda_{m-1}).
\end{align*}

\begin{figure}
    \begin{center}
        \begin{tikzpicture}[
                level/.style={
                    sibling distance = (3-#1)*2cm + 1cm,
                    level distance = 1.5cm
                }
            ]
            \node [treenode] {$(\Lambda_{P_2} \sqcup \Lambda_{P_2} \sqcup \Lambda_{P_2}) \sqcup  \Lambda_{P_1}$}
                child{ node [treenode] {$\Lambda_{P_2} \sqcup \Lambda_{P_2} \sqcup \Lambda_{P_2}$}
                    child{ node [treenode] {$\Lambda_{P_2}$} }
                    child{ node [treenode] {$\Lambda_{P_2}$} }
                    child{ node [treenode] {$\Lambda_{P_2}$} }
                }
                child{ node [treenode] {$\Lambda_{P_1}$} };
        \end{tikzpicture}
    \end{center}
    \caption{Function space basis tree of the Taylor--Hood space $(P_2 \times P_2 \times P_2)\times P_1$}
    \label{fig:taylor_hood_basis_tree}
\end{figure}
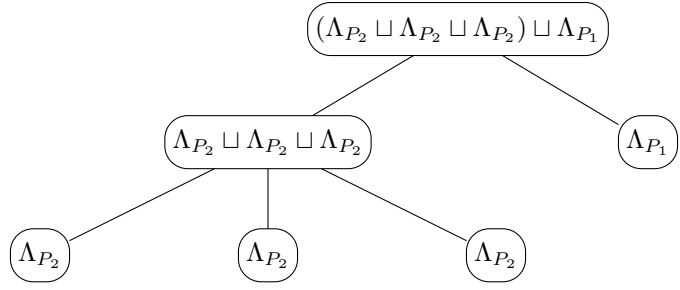

Similarly to the case of function spaces, bases can be interpreted as trees.
If we associate a basis $\Lambda_V$ to each space $V$ in the function space tree,
then the induced natural function space basis tree is obtained by simply replacing
$V$ by $\Lambda_V$ in each node. For the Taylor--Hood basis this leads to the
tree depicted in Figure~\ref{fig:taylor_hood_basis_tree}.

\subsection{Indexing basis functions by multi-indices}
\label{sec:index_trees}

To work with the basis of a finite element space, the basis functions need to be indexed.  Indexing the basis functions
is what allows to address the corresponding vector and matrix coefficients in suitable vector and matrix data structures.
In simple cases, indexing means simply enumerating the basis functions with natural numbers, but for many applications
hierarchically structured matrix and vector data structures are more natural or efficient.  This leads to the idea
of hierarchically structured multi-indices.
\begin{definition}[Multi-indices]
 A tuple $I \in \N_0^k$ for some $k \in \N_0$ is called a multi-index of length $k$,
 and we write $|I| \colonequals k$.
 The set of all multi-indices is denoted by
 $\mathcal{N} = \bigcup_{k \in \N_0} \N_0^k$.
\end{definition}
To establish some structure in a set of multi-indices it is convenient to consider prefixes.
\begin{definition}[Multi-index prefixes]\mbox{}  % Force a line break, otherwise the enumeration looks funny
    \begin{enumerate}
        \item
            If $I \in \mathcal{N}$ takes the form $I = (I^0,I^1)$ for $I^0,I^1 \in \mathcal{N}$,
            then we call $I^0$ a prefix of $I$.
            If additionally $|I^1|>0$, then we call $I^0$ a strict prefix of $I$.
        \item
            For $I,I^0 \in \mathcal{N}$ and a set $\mathcal{M} \subset \mathcal{N}$:
            \begin{enumerate}
              \item
                We write $I=(I^0,\dots)$, if $I^0$ is a prefix of $I$,
              \item
                we write $I=(I^0,\bullet,\dots)$, if $I^0$ is a strict prefix of $I$,
              \item
                we write $(I^0,\dots) \in \mathcal{M}$, if $I^0$ is a prefix of some
                $I \in \mathcal{M}$,
              \item
                we write $(I^0,\bullet,\dots) \in \mathcal{M}$, if $I^0$ is a strict prefix of some
                $I \in \mathcal{M}$.
            \end{enumerate}
    \end{enumerate}
\end{definition}

\begin{figure}
    \begin{center}
        \begin{tikzpicture}[
                level/.style={
                    sibling distance = (3-#1)*2cm + 1cm,
                    level distance = 1.5cm
                }
            ]
            \node [treenode] {$(\Lambda_{P_2} \sqcup \Lambda_{P_2} \sqcup \Lambda_{P_2}) \sqcup  \Lambda_{P_1}$}
                child[sibling distance = 7cm]{ node [treenode] {$\Lambda_{P_2} \sqcup \Lambda_{P_2} \sqcup \Lambda_{P_2}$}
                    child [sibling distance = 3.3cm] { node [treenode] {$\Lambda_{P_2}$}
                        child{ node [basisnode] {$\lambda_0^{P_2}$} }
                        child{ node [] {$\dots$} }
                        child{ node [basisnode] {$\lambda_{n_2-1}^{P_2}$} }
                    }
                    child [sibling distance = 3.3cm] { node [treenode] {$\Lambda_{P_2}$}
                        child{ node [basisnode] {$\lambda_0^{P_2}$} }
                        child{ node [] {$\dots$} }
                        child{ node [basisnode] {$\lambda_{n_2-1}^{P_2}$} }
                    }
                    child [sibling distance = 3.3cm] { node [treenode] {$\Lambda_{P_2}$}
                        child{ node [basisnode] {$\lambda_0^{P_2}$} }
                        child{ node [] {$\dots$} }
                        child{ node [basisnode] {$\lambda_{n_2-1}^{P_2}$} }
                    }
                }
                child{ node [treenode] {$\Lambda_{P_1}$}
                    child [sibling distance=1cm] { node [basisnode] {$\lambda_0^{P_1}$} }
                    child [sibling distance=1cm] { node [] {$\dots$} }
                    child [sibling distance=1cm] { node [basisnode] {$\lambda_{n_1-1}^{P_1}$} }
                };
        \end{tikzpicture}
    \end{center}
    \caption{Tree of basis vectors for the Taylor--Hood basis}
    \label{fig:taylor_hood_basis_function_tree}
\end{figure}
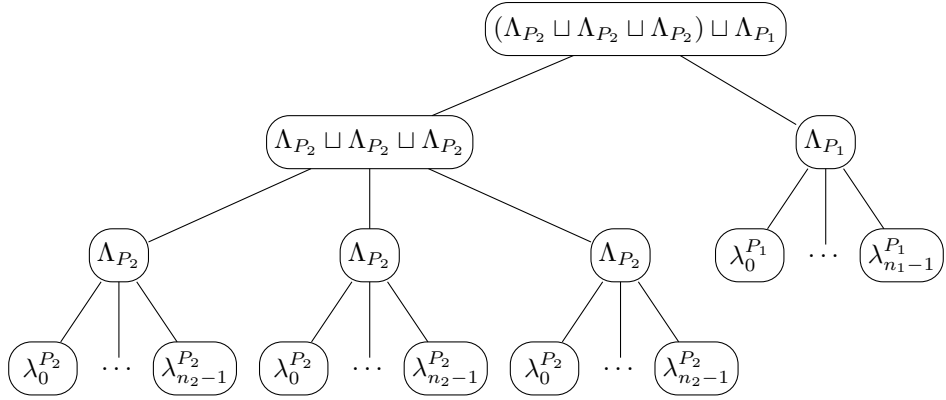

It is important to note that the multi-indices from a given set do not necessarily
all have the same length. For an example,
Figure~\ref{fig:taylor_hood_basis_function_tree} illustrates the set of all basis
functions by extending the basis tree of Figure~\ref{fig:taylor_hood_basis_tree}
by leaf nodes for individual basis functions.
A possible indexing of the basis functions of the Taylor--Hood basis $\Lambda_\text{TH}$ then
uses multi-indices of the form $(0,i,j)$ for velocity components, and $(1,k)$
for pressure components.
For the velocity multi-indices $(0,i,j)$, the $i = 0,\dots,2$ determines the component
of the velocity vector field, and the $j = 0,\dots,n_2-1 \colonequals \abs{\Lambda_{P_2}}-1$ determines the number of the scalar $P_2$ basis
function that determines this component.
For the pressure multi-indices $(0,k)$ the $k= 0,\dots,n_1-1 \colonequals \abs{\Lambda_{P_1}}-1$ determines the number of the $P_1$ basis
function for the scalar $P_1$ function that determines the pressure.

It is evident that the complete set of these multi-indices can again be associated to a rooted tree.
In this tree, the multi-indices correspond to the leaf nodes,
their strict prefixes correspond to interior nodes,
and the multi-index digits labeling the edges are the
indices of the children within the ordered tree.  Prefixes can be
interpreted as paths from the root to a given node.

This latter fact can be seen as the defining property of index trees.  Indeed,
a set of multi-indices (together with all its strict prefixes)
forms a tree as long as it is consistent in the sense that the multi-indices
can be viewed as the paths to the leafs in an ordered tree.
That is, the children of each node are enumerated using consecutive zero-based
indices and paths to the leafs (i.e., the multi-indices) are built by concatenating
those indices starting from the root and ending in a leaf.
Since the full structure of this tree is encoded in the multi-indices associated
to the leafs we will---by a slight abuse of notation---call the set of multi-indices
itself a tree from now on.

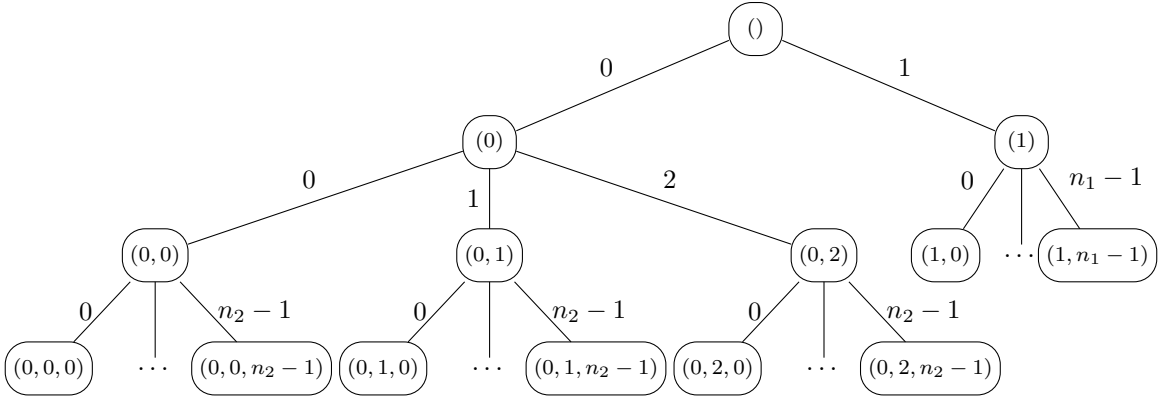
\begin{figure}
  \makebox[\textwidth][c]{
        \begin{tikzpicture}[
                level/.style={
                    sibling distance = (3-#1)*2.5cm + 1cm,
                    level distance = 1.5cm
                }
            ]
            \node [smalltreenode] {$()$}
                child [sibling distance=70mm] { node [smalltreenode] {$( 0 )$}
                    child [sibling distance=44mm] { node [smalltreenode] {$( 0,0 )$}
                        child [sibling distance=14mm] { node [smallbasisnode] {$( 0,0,0 )$} edge from parent node[left] {$0$}}
                        child [sibling distance=14mm] { node [] {$\dots$} }
                        child [sibling distance=14mm] { node [smallbasisnode] {$( 0,0,n_2-1 )$} edge from parent node[right] {$n_2-1$} }
                        edge from parent node[above left] {$0$}
                    }
                    child [sibling distance=44mm] { node [smalltreenode] {$( 0,1 )$}
                        child [sibling distance=14mm] { node [smallbasisnode] {$( 0,1,0 )$} edge from parent node[left] {$0$}}
                        child [sibling distance=14mm] { node [] {$\dots$} }
                        child [sibling distance=14mm] { node [smallbasisnode] {$( 0,1,n_2-1 )$} edge from parent node[right] {$n_2-1$} }
                        edge from parent node[left] {$1$}
                    }
                    child [sibling distance=44mm] { node [smalltreenode] {$( 0,2 )$}
                        child [sibling distance=14mm] { node [smallbasisnode] {$( 0,2,0 )$} edge from parent node[left] {$0$}}
                        child [sibling distance=14mm] { node [] {$\dots$} }
                        child [sibling distance=14mm] { node [smallbasisnode] {$( 0,2,n_2-1 )$} edge from parent node[right] {$n_2-1$} }
                        edge from parent node[above right] {$2$}
                    }
                    edge from parent node[above left] {$0$}
                }
                child [sibling distance=70mm] { node [smalltreenode] {$( 1 )$}
                    child [sibling distance=1cm] { node [smallbasisnode] {$( 1,0 )$} edge from parent node[above left] {$0$} }
                    child [sibling distance=1cm] { node [] {$\dots$} }
                    child [sibling distance=1cm] { node [smallbasisnode] {$( 1,n_1-1 )$} edge from parent node[above right] {$n_1-1$} }
                    edge from parent node[above right] {$1$}
                };
        \end{tikzpicture}
      }
    \caption{Index tree for the Taylor--Hood basis inherited from the basis tree}
    \label{fig:taylor_hood_index_tree}
\end{figure}

\begin{definition}
\label{def:index_tree}
 A set $\mathcal{I} \subset \mathcal{N}$ is called an \emph{index tree}
 if for any $(I,i,\dots) \in \mathcal{I}$ there are also $(I,0,\dots),(I,1,\dots),\dots,(I,i-1,\dots) \in \mathcal{I}$,
 but $I \notin \mathcal{I}$.
\end{definition}
The index tree for the example indexing of the Taylor--Hood basis given above is shown
in Figure~\ref{fig:taylor_hood_index_tree}.

\begin{definition}
Let $(I,\dots) \in \mathcal{I}$, i.e., $I$ is a
prefix of multi-indices in $\mathcal{I}$. Then the size of $\mathcal{I}$ relative
to $I$ is given by
\begin{align}\label{eq:prefix_size}
  \operatorname{deg}^+_{\mathcal{I}}[I] \colonequals  \op{max}\{k \st \exists (I,k,\dots) \in \mathcal{I} \}+1.
\end{align}
\end{definition}
In terms of the ordered tree associated with $\mathcal{I}$ this corresponds
to the out-degree of $I$, i.e., the number of direct children of the node indexed by $I$.

Using the idea of multi-index trees,
an indexing of a function space basis is an injective map from the leaf nodes of a tree of basis functions to the leafs of an
index tree.

\begin{definition}
\label{def:index_map}
  Let $M$ be a finite set and $\iota:M \to \mathcal{N}$ an injective map whose range
  $\iota(M)$ forms an index tree.
  Then $\iota$ is called an \emph{index map} for $M$.
  The index map is called \emph{uniform} if additionally $\iota(M) \subset \mathbb{N}^k_0$ for some $k \in \mathbb{N}$,
  and \emph{flat} if $\iota(M) \subset \mathbb{N}_0$.
\end{definition}

Continuing the Taylor--Hood example, if
all basis functions $\Lambda_\text{TH} = \{\lambda_I \}$ of the whole finite element tree are
indexed by multi-indices of the above given form,
and if $X$ is a coefficient vector that has a compatible hierarchical structure,
then a finite element function $(v_h,p_h)$ with velocity
$v_h$ and pressure $p_h$ defined by the coefficient vector $X$
is given by
\begin{align}
\label{eq:linear_combination}
  (v_h,p_h)
  &= \sum_{i=0}^2\sum_{j=0}^{n_2-1} X_{(0,i,j)}\lambda_{(0,i,j)}
  + \sum_{k=0}^{n_1-1} X_{(1,k)}\lambda_{(1,k)},
\end{align}
with basis functions
\begin{equation*}
  \lambda_{(0,i,j)} = \mathbf{e}_0 \otimes (\mathbf{e}_i \otimes \lambda^{P_2}_j), \qquad i=0,1,2,
    \qquad \text{and} \qquad
    \lambda_{(1,k)} = \mathbf{e}_1 \otimes \lambda^{P_1}_k.
\end{equation*}
Introducing the corresponding index map $\iota : \Lambda_{\text{TH}} \to \mathcal{N}$
with $\iota(\lambda_I)=I$ on the set $\Lambda_{\text{TH}}$ of all basis functions
we can write this in compact form as
\begin{align*}
  (v_h,p_h) &= \sum_{\lambda \in \Lambda_{\text{TH}}} X_{\iota(\lambda)} \lambda
            = \sum_{I \in \iota(\Lambda_{\text{TH}})} X_I \lambda_I.
\end{align*}
Alternatively the individual velocity and pressure fields
$v_h$ and $p_h$ are given by
\begin{align*}
  v_h &= \sum_{i=0}^2 \sum_{j=0}^{n_2-1} X_{(0,i,j)} (\mathbf{e}_i \otimes \lambda^{P_2}_j),
    &
    p_h &= \sum_{k=0}^{n_1-1} X_{(1,k)}\lambda^{P_1}_k.
\end{align*}

\begin{figure}
    \begin{center}
        \begin{tikzpicture}[
                level/.style={
                    sibling distance = (3-#1)*2cm + 1cm,
                    level distance = 1.5cm
                },
            ]
            \node [treenode] {$()$}
                child{ node [treenode] {$( 0 )$}
                        child [sibling distance=2.5cm] { node [smalltreenode] {$( 0,0 )$}
                            child [sibling distance=1.5cm] { node [smallbasisnode] {$( 0,0,0 )$} edge from parent node[above left] {$0$}}
                            child [sibling distance=1.5cm] { node [smallbasisnode] {$( 0,0,1 )$} edge from parent node[left] {$1$}}
                            child [sibling distance=1.5cm] { node [smallbasisnode] {$( 0,0,2 )$} edge from parent node[above right] {$2$}}
                            edge from parent node[above left] {$0$}
                        }
                        child [sibling distance=2.3cm]{ node [] {$\dots$} }
                        child [sibling distance=2.6cm]{ node [smalltreenode] {$( 0,n_2-1 )$}
                            child [sibling distance=2cm] { node [smallbasisnode] {$( 0,n_2 - 1,0 )$} edge from parent node[above left] {$0$}}
                            child [sibling distance=2cm] { node [smallbasisnode] {$( 0,n_2 - 1,1 )$} edge from parent node[left] {$1$}}
                            child [sibling distance=2cm] { node [smallbasisnode] {$( 0,n_2 - 1,2 )$} edge from parent node[above right] {$2$}}
                            edge from parent node[above right] {$n_2-1$}
                        }
                        edge from parent node[above left] {$0$}
                }
                child [sibling distance=6.2cm]{ node [smalltreenode] {$( 1 )$}
                    child [sibling distance=1.3cm] { node [smallbasisnode] {$( 1,0 )$} edge from parent node[above left] {$0$} }
                    child [sibling distance=1.3cm] { node [] {$\dots$} }
                    child [sibling distance=1.3cm] { node [smallbasisnode] {$( 1,n_1-1 )$} edge from parent node[above right] {$n_1-1$} }
                    edge from parent node[above right] {$1$}
                };
        \end{tikzpicture}
    \end{center}
    \caption{Index tree for Taylor--Hood with blocking of velocity components}
    \label{fig:taylor_hood_index_blocked_tree}
\end{figure}
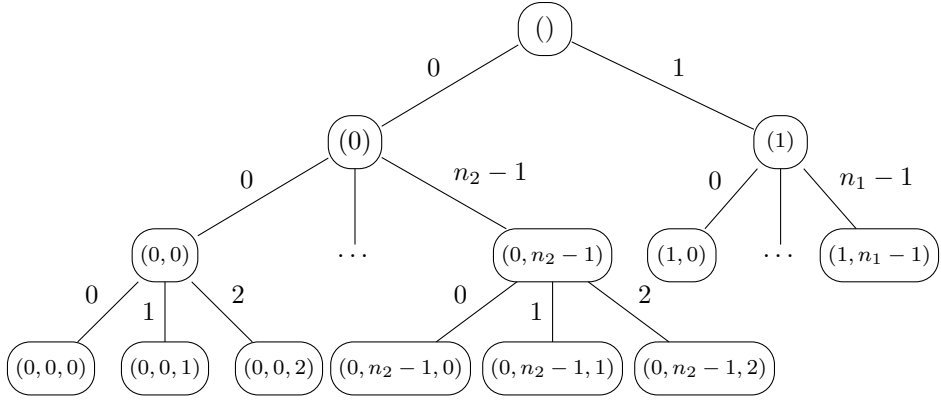

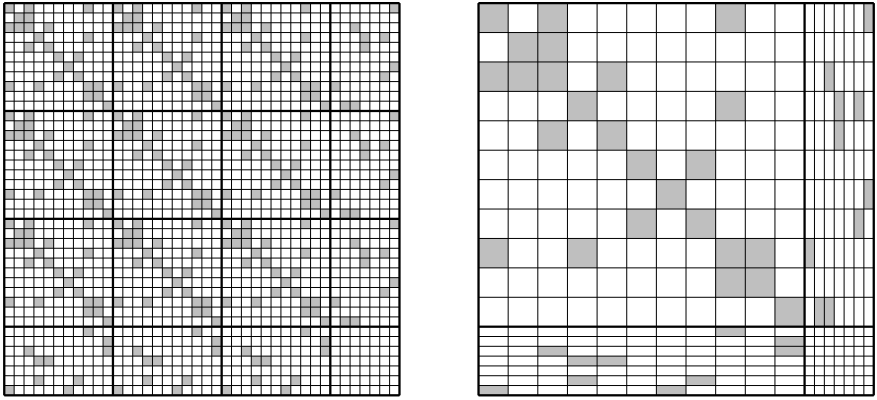
\begin{figure}
 \begin{center}
  \begin{tikzpicture}[scale=.13]

  % Horizontal offset between the two matrices
  \pgfmathsetmacro{\offset}{48}

  %%%%%%%%%%%%%%%%%%%%%%%%%%%%%%%%%%%%%%%%%%%%%%%%%%%%%%%%%%%%%%%%%%%%%%%%%%%%5
  %% Matrix occupation patterns
  %% Both matrices together, because the patterns are to be identical
  %%%%%%%%%%%%%%%%%%%%%%%%%%%%%%%%%%%%%%%%%%%%%%%%%%%%%%%%%%%%%%%%%%%%%%%%%%%%5

  % The occupation pattern
  % Diagonal elements---those have to be there
  \foreach \x in {0,...,10}
  {
    % Left matrix
    \foreach \i in {0,...,2}
      \foreach \j in {0,...,2}
        \fill [lightgray] (11*\i + \x,40- \j*11 - \x) rectangle (11*\i + 1+\x,39 -\j*11 -\x);

    % Right matrix
    \fill [lightgray] (\offset + 0+3*\x,40-3*\x) rectangle (\offset + 3+3*\x,37-3*\x);
  }

  % Off-diagonal
  \foreach \blocknumber in {0,...,9}
  {
    \pgfmathsetmacro{\x}{int(random(0,9))}
    \pgfmathsetmacro{\y}{int(random(\x,9))}

    % Left matrix
    \foreach \i in {0,...,2}
      \foreach \j in {0,...,2}
      {
        \fill [lightgray] (11*\i + \x,40-11*\j - \y) rectangle (11*\i + 1+\x,39-11*\j-\y);  % lower triangular part
        \fill [lightgray] (11*\i + \y,40-11*\j - \x) rectangle (11*\i + 1+\y,39-11*\j-\x);  % upper triangular part
      }

    % Right matrix
    \fill [lightgray] (\offset + 0+3*\x,40-3*\y) rectangle (\offset + 3+3*\x,37-3*\y);  % lower triangular part
    \fill [lightgray] (\offset + 0+3*\y,40-3*\x) rectangle (\offset + 3+3*\y,37-3*\x);  % upper triangular part
  }

  \foreach \blocknumber in {0,...,9}
  {
    \pgfmathsetmacro{\x}{int(random(0,10))}
    \pgfmathsetmacro{\y}{int(random(0,6))}

    % Left matrix
    \foreach \i in {0,...,2}
    {
      \fill [lightgray] (11*\i + \x, \y) rectangle (11*\i + 1+\x,1+\y);  % lower left block
      \fill [lightgray] (40 - \y,40-11*\i - \x) rectangle (39-\y,39-11*\i-\x);  % upper right block
    }

    % Right matrix
    \fill [lightgray] (\offset + 0+3*\x,\y)     rectangle (\offset + 3+3*\x,1+\y);  % lower left block
    \fill [lightgray] (\offset + 40-\y,40-3*\x) rectangle (\offset + 39-\y,37-3*\x);  % upper right block
  }

  %%%%%%%%%%%%%%%%%%%%%%%%%%%%%%%%%%%%%%%%%%%%%%%%%%%%%%%%%%%%%%%%%%%%%%%%%%%%5
  %% Line drawing of the first matrix
  %%%%%%%%%%%%%%%%%%%%%%%%%%%%%%%%%%%%%%%%%%%%%%%%%%%%%%%%%%%%%%%%%%%%%%%%%%%%5

  % The fine grid
  \foreach \x in {0,...,40}
    \draw [line width=0.05mm] (\x,0)--(\x,40);

  \foreach \y in {0,...,40}
    \draw [line width=0.05mm] (0,\y)--(40,\y);

  % The thick-line grid
  \foreach \x in {0,11,22,33,40}
    \draw [line width=0.3mm] (\x,0)--(\x,40);

  \foreach \y in {0,7,18,29,40}
    \draw [line width=0.3mm] (0,\y)--(40,\y);

  %%%%%%%%%%%%%%%%%%%%%%%%%%%%%%%%%%%%%%%%%%%%%%%%%%%%%%%%%%%%%%%%%%%%%%%%%%%%5
  %% Line drawing of the second matrix
  %%%%%%%%%%%%%%%%%%%%%%%%%%%%%%%%%%%%%%%%%%%%%%%%%%%%%%%%%%%%%%%%%%%%%%%%%%%%5

  % The fine grid
  \foreach \x in {0,...,10}
    \draw [line width=0.05mm] (\offset + 3*\x,0)--(\offset + 3*\x,40);

  \foreach \y in {0,...,10}
    \draw [line width=0.05mm] (\offset + 0,40-3*\y)--(\offset + 40,40-3*\y);

  \foreach \x in {0,...,6}
    \draw [line width=0.05mm] (\offset + 40 - 1*\x,0)--(\offset + 40 - 1*\x,40);

  \foreach \y in {0,...,6}
    \draw [line width=0.05mm] (\offset + 0,\y)--(\offset + 40,\y);

  % The thick-line grid
  \foreach \x in {0,33,40}
    \draw [line width=0.3mm] (\offset + \x,0)--(\offset + \x,40);

  \foreach \y in {0,7,40}
    \draw [line width=0.3mm] (\offset + 0,\y)--(\offset + 40,\y);

  \end{tikzpicture}
 \end{center}
 \caption{Two matrix occupation patterns for different indexings of the Taylor--Hood bases.
   Left: Corresponding to the index tree of Figure~\ref{fig:taylor_hood_index_tree}.
   Right: Corresponding to the index tree of Figure~\ref{fig:taylor_hood_index_blocked_tree}.
   }
 \label{fig:matrix_occupation_patterns}
\end{figure}

In the previous example, the index tree was
isomorphic to the basis function tree depicted in Figure~\ref{fig:taylor_hood_basis_function_tree}.
However, one may also be interested in constructing multi-indices
that do not mimic the structure of the basis function tree:
For example, to increase data locality in assembled matrices for the Taylor--Hood basis it may be
preferable to group all velocity degrees of freedom corresponding to a single
$P_2$ basis function together, i.e., to use the index $(0,j,i)$
for the $j$-th $P_2$ basis function for the $i$-th component.
The corresponding alternative index tree is shown in
Figure~\ref{fig:taylor_hood_index_blocked_tree}.
Figure~\ref{fig:matrix_occupation_patterns} shows the corresponding layouts of a hierarchical stiffness matrix.

Alternatively, the case of indexing all basis functions from the Taylor--Hood basis with a single
natural number can be represented by an index tree with $3 n_2 + n_1$ leaf nodes all
directly attached to a single root. Different variations of such a tree differ by how the
degrees of freedom are ordered.

\subsection{Strategy-based construction of multi-indices}
\label{sec:index_strategies}

Let $\Lambda$ be the set of basis functions of a finite element basis tree.
In principle, \dunemodule{dune\-functions} allows any indexing scheme that
is given by an index map, i.e., any map $\iota: \Lambda \to \mathcal{N}$ that
is injective and whose range $\iota(\Lambda)$ is an index tree.  In practice,
out of this large set of maps, \dunemodule{dune-functions} allows to construct the most
important ones systematically using certain transformation rules.

Consider a tree of function space bases in the sense of Section~\ref{sec:basistree}.
We want to construct an indexing for this tree, that is
an index tree $\mathcal{I}$ and a bijection $\iota$ from the set of all basis functions $\Lambda$
to the multi-indices in $\mathcal{I}$. The construction proceeds recursively.
To describe it,
we assume in the following that $\Lambda$ is a node in the function space
basis tree, i.e., it is the set of all basis functions
corresponding to a node $V \colonequals \operatorname{span} \Lambda$
in the function space tree.

To end the recursion, we assume that an index map $\iota : \Lambda \to \mathcal{N}$
is given if $V = \operatorname{span} \Lambda$ is a leaf node of the function space tree.
The most obvious choice would be a flat zero-based index
of the basis functions of $\Lambda$. However, other choices are possible.
For example, in case of a discontinuous finite element space, each
basis function $\lambda \in \Lambda$ could also be associated to a two-digit
multi-index $\iota(\lambda)=(i,k)$, where $i$ is the
index of the grid element that forms the support of $\lambda$, and $k$ is the index of $\lambda$
within this element.

For the actual recursion,
if $\Lambda$ is any non-leaf node in the function space basis tree,
then it takes the form
\begin{align*}
  \Lambda = \Lambda_0 \sqcup \dots \sqcup \Lambda_{m-1}
          = \bigcup_{i=0}^{m-1} \mathbf{e}_i \otimes \Lambda_i,
\end{align*}
where $\Lambda_0, \dots,\Lambda_{m-1}$ are the direct children of $\Lambda$,
i.e., the sets of basis functions of the child
spaces $\{\operatorname{span} \Lambda_i\}_{i=0,\dots,m-1}$ of the product space
\begin{align*}
  \operatorname{span} \Lambda
    = \operatorname{span} \bigl( \Lambda_0 \sqcup \dots \sqcup \Lambda_{m-1} \bigr)
%  = \prod_{i=0}^{m-1} \operatorname{span} \Lambda_i
    = (\operatorname{span} \Lambda_0) \times \dots \times (\operatorname{span} \Lambda_{m-1}).
\end{align*}
For the recursive construction we assume that an index map
$\iota_i : \Lambda_i \to \mathcal{N}$ on $\Lambda_i$ is given for any $i=0,\dots,m-1$.
The task is to construct an index map $\iota: \Lambda \to \mathcal{N}$
from the maps $\iota_i$.
In the following we describe four strategies to achieve this; all
have been implemented in \dunemodule{dune-functions}. When reading about these
strategies, remember that any $\lambda \in \Lambda$ has a unique representation
$\lambda = \mathbf{e}_i \otimes \hat{\lambda}$ for $i \in \{0,\dots,m-1\}$ and some
$\hat{\lambda} \in \Lambda_i$.
It will be necessary to distinguish the special case that all children
$\Lambda_i$ are identical.
\begin{definition}
\label{def:power_node}
  An inner node $\Lambda$ will be called \emph{power node} if all of its children $\Lambda_i$
  are identical and equipped with identical index maps $\iota_i$.
  An inner node that is not a power node is called \emph{composite node}.
\end{definition}
This definition is needed because some of the following strategies can only be applied
to power nodes.
\begin{itemize}
  \item
    \textbf{BlockedLexicographic}: This strategy prepends the child index
    to the multi-index within the child basis. That is, the index map $\iota:\Lambda \to \mathcal{N}$
    is given by
    \begin{align*}
      \iota(\mathbf{e}_i \otimes \hat{\lambda}) = (i,\iota_i(\hat{\lambda})).
    \end{align*}
    It is straightforward to show that $\iota$ is always an index map
    for $\Lambda$.
    To demonstrate the strategy the following table shows the multi-indices at inner nodes,
    when the basis functions of the subtrees $\Lambda_0, \Lambda_1,\dots$ are labeled by
    multi-indices $(I^0), (I^1), \dots$ for $\Lambda_0$, $(K^0), (K^1), \dots$ for $\Lambda_1$,
    and so on.

    \begin{center}
    \begin{tabular}{c|c|c|c}
      indices for $\Lambda_0$ &
      indices for $\Lambda_1$ &
      \hspace{2em}$\dots$\hspace{2em} &
      indices for $\Lambda$ \\
      \hline
      $\iota_0(\hat{\lambda}_{0,0}) = (I^0)$ & & &
        $\iota(\mathbf{e}_0 \otimes \hat{\lambda}_{0,0}) = (0,I^0)$ \\
      $\iota_0(\hat{\lambda}_{0,1}) = (I^1)$ & & &
        $\iota(\mathbf{e}_0 \otimes \hat{\lambda}_{0,1}) = (0,I^1)$ \\
      & $\iota_1(\hat{\lambda}_{1,0}) = (K^0)$ & &
        $\iota(\mathbf{e}_1 \otimes \hat{\lambda}_{1,0}) = (1,K^0)$ \\
      & $\iota_1(\hat{\lambda}_{1,1}) = (K^1)$ & &
        $\iota(\mathbf{e}_1 \otimes \hat{\lambda}_{1,1}) = (1,K^1)$ \\
      & $\iota_1(\hat{\lambda}_{1,2}) = (K^2)$ & &
        $\iota(\mathbf{e}_1 \otimes \hat{\lambda}_{1,2}) = (1,K^2)$ \\
      & & \dots &
        \dots \\
    \end{tabular}
    \end{center}

  \item
    \textbf{BlockedInterleaved}: This strategy is only well-defined for power nodes. It appends the child index
    to the multi-index within the child basis.
    That is, the index map $\iota:\Lambda \to \mathcal{N}$
    is given by
    \begin{align*}
      \iota(\mathbf{e}_i \otimes\hat{\lambda}) = (\iota_i(\hat{\lambda}),i).
    \end{align*}
    An example is given in the following table:

    \begin{center}
    \begin{tabular}{c|c|c|c}
      indices for $\Lambda_0$ &
      indices for $\Lambda_1$ &
      \hspace{2em}$\dots$\hspace{2em} &
      indices for $\Lambda$ \\
      \hline
      $\iota_0(\hat{\lambda}_{0,0}) = (I^0)$ & & &
        $\iota(\mathbf{e}_0 \otimes \hat{\lambda}_{0,0}) = (I^0,0)$ \\
      & $\iota_1(\hat{\lambda}_{1,0}) = (I^0)$ & &
        $\iota(\mathbf{e}_1 \otimes \hat{\lambda}_{1,0}) = (I^0,1)$ \\
      & & \dots &
        \dots \\
      $\iota_0(\hat{\lambda}_{0,1}) = (I^1)$ & & &
        $\iota(\mathbf{e}_0 \otimes \hat{\lambda}_{0,1}) = (I^1,0)$ \\
      & $\iota_1(\hat{\lambda}_{1,1}) = (I^1)$ & &
        $\iota(\mathbf{e}_1 \otimes \hat{\lambda}_{1,1}) = (I^1,1)$ \\
      & & \dots &
        \dots \\
      $\iota_0(\hat{\lambda}_{0,2}) = (I^2)$ & & &
        $\iota(\mathbf{e}_0 \otimes \hat{\lambda}_{0,2}) = (I^2,0)$ \\
      & $\iota_1(\hat{\lambda}_{1,2}) = (I^2)$ & &
        $\iota(\mathbf{e}_1 \otimes \hat{\lambda}_{1,2}) = (I^2,1)$ \\
      & & \dots &
        \dots \\
    \end{tabular}
    \end{center}

    To see that this strategy does not work for general composite nodes,
    consider $\iota_0(\Lambda_0) = \{0\}$ and $\iota_1(\Lambda_1) = \{(0,0)\}$.
    Then $\iota(\Lambda) = \{(0,0), (0,0,1)\}$ which is not an index tree.
\end{itemize}
Unlike the previous two strategies, the following two do not introduce new
multi-index digits. Such strategies are called \emph{flat}.
\begin{itemize}
  \item \textbf{FlatLexicographic}: This strategy merges the roots of all
    index tree $\iota_i(\Lambda_i)$ into a single new one.
    Assume that we split the multi-index
    $\iota_i(\hat{\lambda})$ according to
    \begin{align}\label{eq:multiindex-split}
      \iota_i(\hat{\lambda}) = (i_0,I),
    \end{align}
    where $i_0 \in \mathbb{N}_0$ is the first digit.
    The index map $\iota:\Lambda \to \mathcal{N}$ is then given by
    \begin{align*}
      \iota(\mathbf{e}_i \otimes\hat{\lambda}) = (L_i + i_0, I),
    \end{align*}
    where the offset $L_i$ for the first digit is computed by
    \begin{align*}
      L_i = \sum_{j=0}^{i-1} \operatorname{deg}_{\iota_j(\Lambda_j)}^+[()].
    \end{align*}
    This construction offsets the first digits of
    the multi-indices of all basis functions from $\Lambda_j$ with $j>0$
    such that they form a consecutive sequence.
    This guarantees that $\iota$ is always an index map for $\Lambda$.
    An example is given in the following table:

    \begin{center}
    \begin{tabular}{c|c|c|c}
      indices for $\Lambda_0$ &
      indices for $\Lambda_1$ &
      \hspace{2em}$\dots$\hspace{2em} &
      indices for $\Lambda$ \\
      \hline
      $\iota_0(\hat{\lambda}_{0,0}) = (0,I^0)$ & & &
        $\iota(\mathbf{e}_0 \otimes \hat{\lambda}_{0,0}) = (0,I^0)$ \\
      $\iota_0(\hat{\lambda}_{0,1}) = (1,I^1)$ & & &
        $\iota(\mathbf{e}_0 \otimes \hat{\lambda}_{0,1}) = (1,I^1)$ \\
      & $\iota_1(\hat{\lambda}_{1,0}) = (0,K^0)$ & &
        $\iota(\mathbf{e}_1 \otimes \hat{\lambda}_{1,0}) = (2,K^0)$ \\
      & $\iota_1(\hat{\lambda}_{1,1}) = (0,K^1)$ & &
        $\iota(\mathbf{e}_1 \otimes \hat{\lambda}_{1,1}) = (2,K^1)$ \\
      & $\iota_1(\hat{\lambda}_{1,2}) = (1,K^2)$ & &
        $\iota(\mathbf{e}_1 \otimes \hat{\lambda}_{1,2}) = (3,K^2)$ \\
      & & \dots &
        \dots \\
    \end{tabular}
    \end{center}

    The digit zero deliberately appears twice in the column for $\Lambda_1$, to demonstrate
    that a consecutive first digit is not required.

  \item \textbf{FlatInterleaved}: This strategy again only works for power nodes.
    It also merges
    the roots of all child index trees $\iota_i(\Lambda_i)$
    into a single one, but it interleaves the children.
    Again using the splitting
    $\iota_i(\hat{\lambda}) = (i_0,I)$ introduced in~\eqref{eq:multiindex-split},
    the index map $\iota:\Lambda \to \mathcal{N}$ is given by
    \begin{align*}
      \iota(\mathbf{e}_i \otimes\hat{\lambda}) = (i_0 m + i, I),
    \end{align*}
    where the fixed stride $m$ is given by the number of children of $\Lambda$.
    The following table shows an example:

    \begin{center}
    \begin{tabular}{c|c|c|c}
      indices for $\Lambda_0$ &
      indices for $\Lambda_1$ &
      \hspace{1em}$\dots$\hspace{1em} &
      indices for $\Lambda$ \\
      \hline
      $\iota_0(\hat{\lambda}_{0,0}) = (0,I^0)$ & & &
        $\iota(\mathbf{e}_0 \otimes \hat{\lambda}_{0,0}) = (0,I^0)$ \\
      & $\iota_1(\hat{\lambda}_{1,0}) = (0,I^0)$ & &
        $\iota(\mathbf{e}_1 \otimes \hat{\lambda}_{1,0}) = (1,I^0)$ \\
      & & \dots &
        \dots \\
      $\iota_0(\hat{\lambda}_{0,1}) = (1,I^1)$ & & &
        $\iota(\mathbf{e}_0 \otimes \hat{\lambda}_{0,1}) = (m+0,I^1)$ \\
      & $\iota_1(\hat{\lambda}_{1,1}) = (1,I^1)$ & &
        $\iota(\mathbf{e}_1 \otimes \hat{\lambda}_{1,1}) = (m+1,I^1)$ \\
      & & \dots &
        \dots \\
      $\iota_0(\hat{\lambda}_{0,2}) = (2,I^2)$ & & &
        $\iota(\mathbf{e}_0 \otimes \hat{\lambda}_{0,2}) = (2m+0,I^2)$ \\
      & $\iota_1(\hat{\lambda}_{1,2}) = (2,I^2)$ & &
        $\iota(\mathbf{e}_1 \otimes \hat{\lambda}_{1,2}) = (2m+1,I^2)$ \\
      & & \dots &
        \dots \\
    \end{tabular}
    \end{center}

    Again, for this interleaved strategy, $\iota$ may not be an
    index map for general composite nodes.
\end{itemize}

These four strategies are offered by \dunemodule{dune-functions}, but there are others
that are sometimes useful.  Experimentally, \dunemodule{dune-functions} therefore also
provides a way to use self-implemented custom rules.

\bigskip

\begin{table}
\footnotesize
\makebox[\textwidth][c]{
\begin{tabular}{c|c|c|c|c|c|c|c|c}
    & BL(BL)
    & BL(BI)
    & BL(FL)
    & BL(FI)
    & FL(BL)
    & FL(BI)
    & FL(FL)
    & FL(FI)
    \\
  \hline
  $v_{x_0,0}$
    & $(0,0,0)$
    & $(0,0,0)$
    & $(0,0)$
    & $(0,0+0)$
    & $(0,0)$
    & $(0,0)$
    & $(0)$
    & $(0+0)$
    \\
  $v_{x_0,1}$
    & $(0,0,1)$
    & $(0,1,0)$
    & $(0,1)$
    & $(0,3+0)$
    & $(0,1)$
    & $(1,0)$
    & $(1)$
    & $(3+0)$
    \\
  $v_{x_0,2}$
    & $(0,0,2)$
    & $(0,2,0)$
    & $(0,2)$
    & $(0,6+0)$
    & $(0,2)$
    & $(2,0)$
    & $(2)$
    & $(6+0)$
    \\
  $v_{x_0,3}$
    & $(0,0,3)$
    & $(0,3,0)$
    & $(0,3)$
    & $(0,9+0)$
    & $(0,3)$
    & $(3,0)$
    & $(3)$
    & $(9+0)$
    \\
  $\vdots$ & $\vdots$ & $\vdots$ & $\vdots$ &  $\vdots$ & $\vdots$ & $\vdots$ & $\vdots$ & $\vdots$
  \\
%  $v_{x_0,n_2-1}$
%    & $(0,0,n_2-1)$
%    & $(0,n_2-1,0)$
%    & $(0,n_2-1)$
%    & $(0,3(n_2-1)+0)$
%    & $(0,n_2-1)$
%    & $(n_2-1,0)$
%    & $(n_2-1)$
%    & $(3(n_2-1)+0)$
%    \\
%  \hline
  $v_{x_1,0}$
    & $(0,1,0)$
    & $(0,0,1)$
    & $(0,n_2+0)$
    & $(0,0+1)$
    & $(1,0)$
    & $(0,1)$
    & $(n_2+0)$
    & $(0+1)$
    \\
  $v_{x_1,1}$
    & $(0,1,1)$
    & $(0,1,1)$
    & $(0,n_2+1)$
    & $(0,3+1)$
    & $(1,1)$
    & $(1,1)$
    & $(n_2+1)$
    & $(3+1)$
    \\
  $v_{x_1,2}$
    & $(0,1,2)$
    & $(0,2,1)$
    & $(0,n_2+2)$
    & $(0,6+1)$
    & $(1,2)$
    & $(2,1)$
    & $(n_2+2)$
    & $(6+1)$
    \\
  $v_{x_1,3}$
    & $(0,1,3)$
    & $(0,3,1)$
    & $(0,n_2+3)$
    & $(0,9+1)$
    & $(1,3)$
    & $(3,1)$
    & $(n_2+3)$
    & $(9+1)$
    \\
  $\vdots$ & $\vdots$ & $\vdots$ & $\vdots$ &  $\vdots$ & $\vdots$ & $\vdots$ & $\vdots$ & $\vdots$
  \\
%  \hline
  $v_{x_2,0}$
    & $(0,2,0)$
    & $(0,0,2)$
    & $(0,2n_2+0)$
    & $(0,0+2)$
    & $(2,0)$
    & $(0,2)$
    & $(2n_2+0)$
    & $(0+2)$
    \\
  $v_{x_2,1}$
    & $(0,2,1)$
    & $(0,1,2)$
    & $(0,2n_2+1)$
    & $(0,3+2)$
    & $(2,1)$
    & $(1,2)$
    & $(2n_2+1)$
    & $(3+2)$
    \\
  $v_{x_2,2}$
    & $(0,2,2)$
    & $(0,2,2)$
    & $(0,2n_2+2)$
    & $(0,6+2)$
    & $(2,2)$
    & $(2,2)$
    & $(2n_2+2)$
    & $(6+2)$
    \\
  $v_{x_2,3}$
    & $(0,2,3)$
    & $(0,3,2)$
    & $(0,2n_2+3)$
    & $(0,9+2)$
    & $(2,3)$
    & $(3,2)$
    & $(2n_2+3)$
    & $(9+2)$
    \\
  $\vdots$ & $\vdots$ & $\vdots$ & $\vdots$ &  $\vdots$ & $\vdots$ & $\vdots$ & $\vdots$ & $\vdots$
  \\
%  \hline
%  \hline
  $p_{0}$
    & $(1,0)$
    & $(1,0)$
    & $(1,0)$
    & $(1,0)$
    & $(3+0)$
    & $(n_2+0)$
    & $(3n_2+0)$
    & $(3n_2+0)$
    \\
  $p_{1}$
    & $(1,1)$
    & $(1,1)$
    & $(1,1)$
    & $(1,1)$
    & $(3+1)$
    & $(n_2+1)$
    & $(3n_2+1)$
    & $(3n_2+1)$
    \\
  $p_{2}$
    & $(1,2)$
    & $(1,2)$
    & $(1,2)$
    & $(1,2)$
    & $(3+2)$
    & $(n_2+2)$
    & $(3n_2+2)$
    & $(3n_2+2)$
    \\
  $\vdots$ & $\vdots$ & $\vdots$ & $\vdots$ &  $\vdots$ & $\vdots$ & $\vdots$ & $\vdots$ & $\vdots$
  \\
  \hline
\end{tabular}
}
\caption{Different indexing strategies for the Taylor--Hood basis functions}
\label{tab:th_indexing_variants}
\end{table}

To further illustrate the four index transformation strategies, we return to
the Taylor--Hood example.
While the indexing schemes proposed for this example so far where
introduced in an ad-hoc way, we will now systematically apply
the above given strategies.  Recall that the Taylor--Hood basis is denoted by
\begin{align*}
 \Lambda_\text{TH}
  = (\Lambda_{P_2} \sqcup \Lambda_{P_2} \sqcup \Lambda_{P_2}) \sqcup \Lambda_{P_1}.
\end{align*}
For the bases $\Lambda_{P_1}, \Lambda_{P_2}$
of the elementary spaces $P_1,P_2$ we consider fixed given flat index maps
\begin{align*}
  \iota_{P_1}(\Lambda_{P_1}) &\to \mathbb{N}_0, &
  \iota_{P_2}(\Lambda_{P_2}) &\to \mathbb{N}_0.
\end{align*}
These are typically constructed by enumerating the grid entities
the basis functions are associated to.
Then the interior product space basis
\begin{align*}
  \Lambda_V = \Lambda_{P_2} \sqcup \Lambda_{P_2} \sqcup \Lambda_{P_2}
\end{align*}
together with the index map $\iota_{P_2}$ is a power node in the
sense of Definition~\ref{def:power_node}, while the tree root
\begin{align*}
 \Lambda_\text{TH}
  = \Lambda_V \sqcup \Lambda_{P_1}
\end{align*}
is a composite node.
The basis functions for the $k$-th component of the velocity
are denoted by
\begin{align*}
  v_{x_k,i} = \mathbf{e}_0 \otimes (\mathbf{e}_k \otimes \lambda^{P_2}_i)
\end{align*}
where $i=0,\dots,n_2-1$ for $n_2=|\Lambda_{P_2}|= \operatorname{dim} P_2$
whereas the basis functions for the pressure are denoted by
\begin{align*}
  p_{j} = \mathbf{e}_1 \otimes \lambda^{P_1}_j
\end{align*}
where $j=0,\dots,n_1-1$ for $n_1=|\Lambda_{P_1}|= \operatorname{dim} P_1$.

As two of the above given strategies can be used
for composite nodes, while all four can be applied to power nodes
we obtain eight different index maps for the Taylor--Hood basis
$\Lambda_{\text{TH}}$.
They are listed in Table~\ref{tab:th_indexing_variants}, where the label $X(Y)$
means that strategy $X$ is used for the outer product and strategy $Y$
for the inner product. For $X$ and $Y$ we use the abbreviations BL
(BlockedLexicographic), BI (BlockedInterleaved), FL (FlatLexicographic), and FI (FlatInterleaved).
Notice that the index maps depicted in
Figure~\ref{fig:taylor_hood_index_tree} and Figure~\ref{fig:taylor_hood_index_blocked_tree}
are reproduced for the strategies
BL(BL) and BL(BI), respectively.

\subsection{Localization to single grid elements}
\label{sec:localization}

For the most part, access to finite element bases happens element by element.  It is therefore important
to consider the restrictions of bases to single grid elements.  In contrast to the previous sections
we now require that there is a finite element grid for the domain $\Omega$. For simplicity we will
assume that all bases consist of functions that are defined piecewise with respect to this grid,
but it is actually sufficient to require that the restrictions of all basis functions to elements
of the grid can be constructed cheaply.

Consider the restrictions of all basis functions $\lambda \in \Lambda$ of a given tree to a single fixed grid element $e$.
Of these restricted functions, we discard all those that are constant zero functions on $e$.
All others form the \emph{local basis} on $e$
\begin{equation*}
 \Lambda|_e
 \colonequals
 \{ \lambda|_e \; : \; \lambda \in \Lambda,
         \quad \operatorname{int}(\operatorname{supp} \lambda) \cap e \neq \emptyset \}.
\end{equation*}
The local basis forms a tree that is isomorphic to the original function space basis tree,
with each global function space basis $\Lambda$ replaced by its local counterpart $\Lambda|_e$.

For a given index map $\iota$ of $\Lambda$,
this natural isomorphism from the global to the local tree
naturally induces a localized version of $\iota$ given by
\begin{align*}
  \iota|_e : \Lambda|_e &\to \mathcal{I}, &
  \iota|_e(\lambda_e) &\colonequals \iota(\lambda).
\end{align*}
This is the map that associates shape functions on a given grid element $e$ to
the multi-indices of the corresponding global basis functions.
Note that the map $\iota|_e$ itself is not an index map in the sense of Definition~\ref{def:index_map}
since $\iota|_e(\Lambda|_e)$ is only a subset of the index tree $\iota(\Lambda)$,
and not always an index tree itself.

In order to index the basis functions in $\Lambda|_e$ efficiently we introduce
an additional local index map
\begin{align*}
  \iota^{\text{local}}_{\Lambda|_e}: \Lambda|_e \to \mathcal{N},
\end{align*}
such that $\iota^{\text{local}}_{\Lambda|_e}(\Lambda|_e)$ is an index tree.
The index $\iota^{\text{local}}_{\Lambda|_e}(\lambda|_e)$ is
called the \emph{local index} of $\lambda$ (with respect to $e$).
To distinguish it from the indices generated by $\iota$
we call $\iota(\lambda)$ the \emph{global index} of $\lambda$.
The local index is typically used to address the element stiffness matrix.
In principle, this indexing can use another non-flat index tree,
which does not have to coincide with the index tree for the global basis.
This means that the local index of a shape function can again be a multi-index, but the types,
lengths and orderings can be completely unrelated to the corresponding global indices.
This would allow to use nested types for element stiffness matrices and load vectors.
As explained in Chapter~\ref{sec:function_space_bases_implementation},
the \dunemodule{dune-functions} \emph{implementation} is fairly restrictive here,
and only allows flat local indices, i.e.,
$\iota^{\text{local}}_{\Lambda|_e}(\Lambda|_e) \subset \mathbb{N}_0.$

In addition, we introduce for each leaf local basis $\hat{\Lambda}|_e$
of the full local basis tree another local index map
\begin{align*}
  \iota^{\text{leaf-local}}_{\hat{\Lambda}|_e}: \hat{\Lambda}|_e \to \mathbb{N}_0.
\end{align*}
As there is no hierarchical structure involved, this index is simply a
natural number.
The index $\iota^{\text{leaf-local}}_{\hat{\Lambda}|_e}(\lambda|_e)$ is
called the \emph{leaf-local index} of $\lambda$ (with respect to $e$).

In an actual programming interface one typically accesses
basis functions by indices directly. We will later see that
in \dunemodule{dune-functions} the leaf-local index is the
shape function index of the \dunemodule{dune-localfunctions} module.
Hence the \dunemodule{dune-functions} API needs to implement the map
\begin{align*}
  \iota^{\text{leaf}\to\text{local}}_e \colonequals \iota^{\text{local}}_{\Lambda|_e} \circ (\iota^{\text{leaf-local}}_{\hat{\Lambda}|_e})^{-1}
\end{align*}
mapping leaf-local indices to local indices and
\begin{align*}
  \iota^{\text{local}\to\text{global}}_e \colonequals \iota|_e \circ (\iota^{\text{local}}_{\Lambda|_e})^{-1}
\end{align*}
mapping local indices to global multi-indices.

\section{Programmer interface for function space bases}
\label{sec:function_space_bases_implementation}

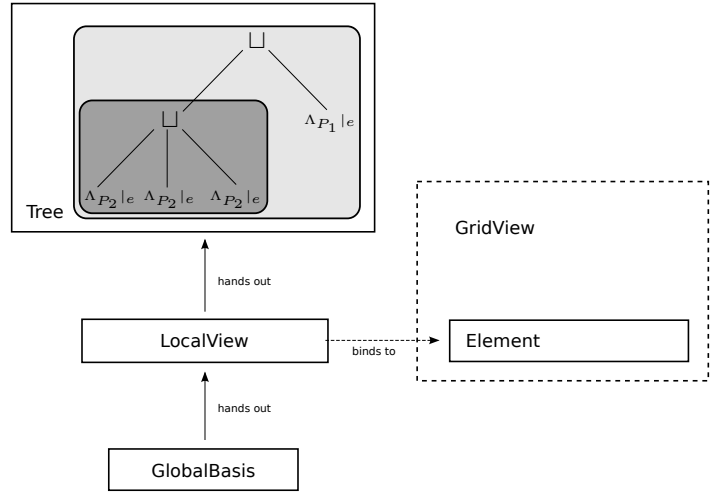
\begin{figure}
 \begin{center}
  \begin{overpic}[width=0.7\textwidth]{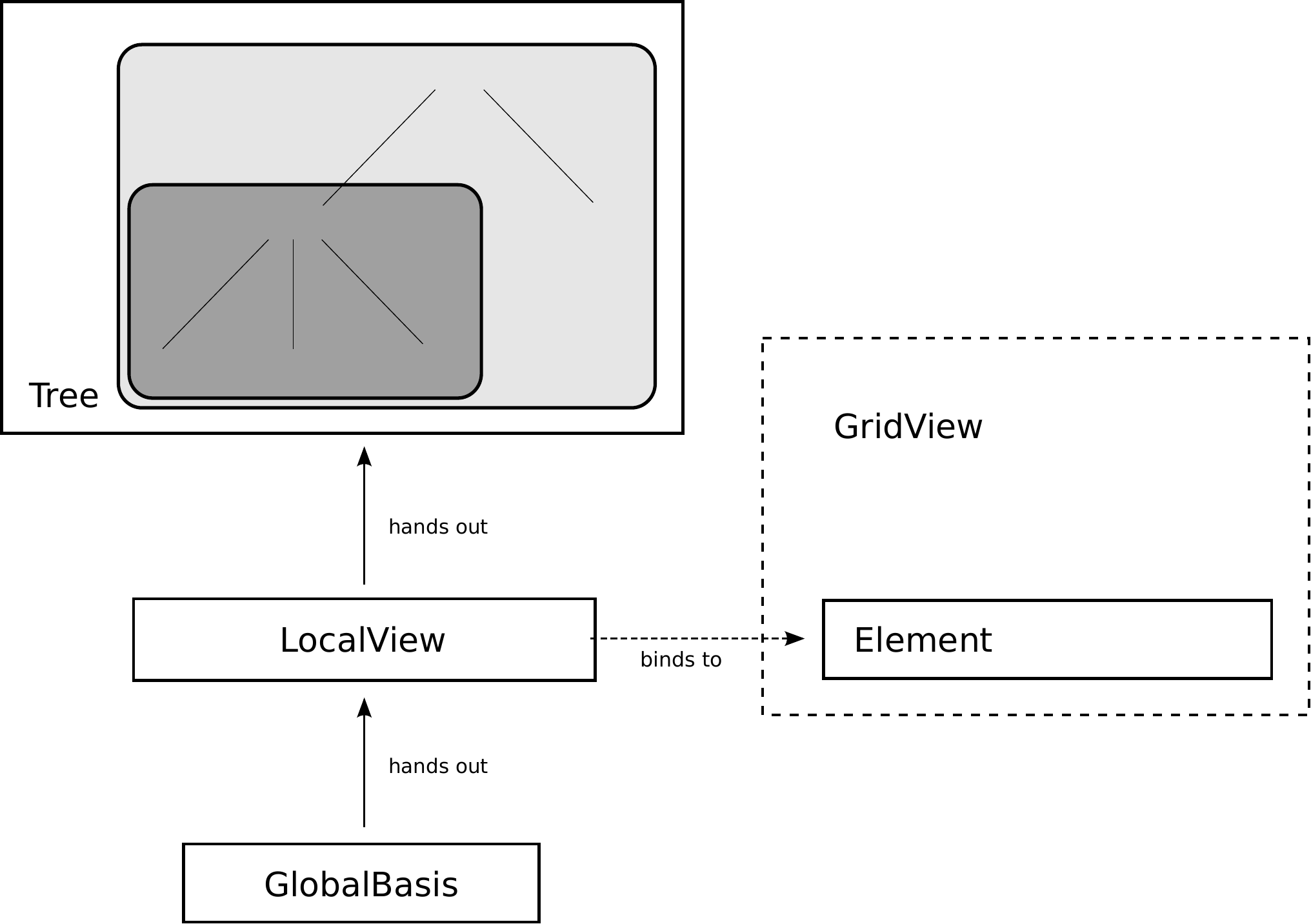}
  \put(34,64){$\sqcup$}
  \put(21.5,52.5){$\sqcup$}
  \put(42,53){\tiny $\Lambda_{P_1}|_e$}
  \put(10.5,42){\tiny $\Lambda_{P_2}|_e$}
  \put(19,  42){\tiny $\Lambda_{P_2}|_e$}
  \put(28.5,42){\tiny $\Lambda_{P_2}|_e$}
  \end{overpic}
 \end{center}
 \caption{Overview of the classes making up the interface to finite element space bases}
 \label{fig:febasis_interface_schematic}
\end{figure}

The design of the \dunemodule{dune-functions} interface for bases of function spaces
follows the ideas of the previous section. The main interface concept are global basis objects
that represent trees of function space bases. These trees can be localized to individual elements
of the grid.  Such a localization provides access to the (tree of) shape functions there,
together with the two shape-function index maps
$\iota^{\text{leaf}\to\text{local}}_e$ and
$\iota^{\text{local}\to\text{global}}_e$.
The structure of the interface is visualized in Figure~\ref{fig:febasis_interface_schematic}.

\subsection*{Notation}

The \dunemodule{dune-functions} module is implemented in the C++ programming languange.
All header include paths
start with \cpp{dune/functions/}, and all code is contained in the namespace \cpp{Dune::Functions}.
Internally, \dunemodule{dune-functions} depends on the \dunemodule{dune-typetree} module,
which implements abstract compile-time tree data structures.
The global basis interface described below is not enforced by deriving
from specific base classes. Instead, \dunemodule{dune-functions} is based on
C++-style duck-typing~\cite{koenig_moo:2005},
i.e., any C++ type providing the required
interface is a valid implementation of that interface.

Throughout this text we will introduce the programmer interfaces
by presenting the interface declaration, explaining
its meaning, and giving examples of its usage.
In order to distinguish interface declarations
from code examples they are formatted differently.
Furthermore, implementation defined types and arguments are highlighted.
The following shows an example of an interface declaration:
\begin{lstlisting}[style=Interface]
// Declaration of type T referring to an implementation-defined type
using T = @@<implementation defined>@@;

// Declaration of method foo
T foo(int);

// Declaration of class Bar with implementation-defined constructor arguments
class Bar {
public:
  Bar(@@<args>@@);
};
\end{lstlisting}
In contrast,
an example for using this interface would be formatted like this:
\begin{lstlisting}[style=Example]
// Call foo() and store result
T t = foo(1);

// Construct an object of type Bar
auto bar = Bar(t, @@<more args>@@);
\end{lstlisting}

\subsection{The interface for a global function space basis}
We start by describing the user interface for global bases.  Since we are discussing duck-typing interfaces,
all class names used below are generic. A tree of global bases is implemented by one class which,
in the following, we will call \cpp{GlobalBasis}, and which
can have an arbitrary number of template parameters.
All types and methods listed in the following
interface declaration shall be public members of
the generic implementation class \cpp{GlobalBasis}.

As each basis implementation may require its own specific data for construction,
we do not enforce a precise set of constructors.
\begin{lstlisting}[style=Interface]
GlobalBasis(@@<implementation defined>@@);
\end{lstlisting}
Each \cpp{GlobalBasis} may have
one or several constructors with implementation-depen\-dent lists of arguments.

The main feature of a \cpp{GlobalBasis} is to give access to basis functions and their indices.
Most of this access happens through the localization of the basis to single grid elements (Section~\ref{sec:localization}).
In the programmer interface, this localization is called \cpp{LocalView}.  Objects of type
\cpp{LocalView} are obtained from \cpp{GlobalBasis} objects through the method
\begin{lstlisting}[style=Interface]
using LocalView = @@<implementation defined>@@;
LocalView localView() const;
\end{lstlisting}
The precise return type of the \cpp{localView} method is implementation-dependent. Objects
created by the method have undefined state, and need to be attached to individual grid
elements in a process called \emph{binding}.
The details are explained in Section~\ref{sec:localview_interface}.

Several methods of a \cpp{GlobalBasis} provide information on the sizes of the bases
contained in the tree.
The total number of basis functions of the global basis is
exported via the method
\begin{lstlisting}[style=Interface]
using size_type = @@<implementation defined>@@;
size_type dimension() const;
\end{lstlisting}
This method can be used to allocate vector containers if flat multi-indices are used.
However, the information provided by the \cpp{dimension} method is generally not
sufficient to allocate hierarchical containers to be
accessed by more general multi-indices.
Therefore, the basis provides more structural
information of those multi-indices via the method
\begin{lstlisting}[style=Interface]
using SizePrefix = ReservedVector@@<implementation defined>@@;
size_type size(const SizePrefix& prefix) const;
\end{lstlisting}
The parameter \cpp{prefix} is a multi-index itself.
If $\mathcal{I}$ is the set of all global multi-indices of the
basis and \cpp{prefix} is a prefix for this set, then
\cpp{size(prefix)} returns the size $\operatorname{deg}^+_{\mathcal{I}}[\text{\cpp{prefix}}]$ of
$\mathcal{I}$ relative to $\text{\cpp{prefix}}$ defined in \eqref{eq:prefix_size},
i.e., the number of direct children of the node \cpp{prefix} in the index tree.
If $\text{\cpp{prefix}}$ is not a prefix for $\mathcal{I}$ the result is undefined.
If $\text{\cpp{prefix}} \in \mathcal{I}$, i.e., the prefix is itself one of the multi-indices
then the result is zero.
The type \cpp{SizePrefix} is always a container of type \cpp{ReservedVector} (from the
\dunemodule{dune-common} module).  More details are given in Section~\ref{sec:multi_indices}.
Like all other types used in the \cpp{GlobalBasis} interface, it is expected to be
exported by the implementation class.
For convenience there is also the method
\begin{lstlisting}[style=Interface]
size_type size() const;
\end{lstlisting}
returning the same value as \cpp{size(\{\})}, i.e., the number of children of the root of
the index tree.
For a scalar basis, this is again the overall number of basis functions.

Finally, each \cpp{GlobalBasis} provides access to the grid view it is defined on
by the method
\begin{lstlisting}[style=Interface]
const GridView& gridView() const;
\end{lstlisting}
The corresponding type
is exported as \cpp{GridView}. If the grid view
was modified (e.g., by local grid refinement), the result of calling any
method of the basis is undefined until the basis has been explicitly updated.
For this, call the method
\begin{lstlisting}[style=Interface]
void update(const GridView & gv);
\end{lstlisting}
which tells the basis to adapt its local state to the new grid view.

\subsection{The user interface for a localized basis}
\label{sec:localview_interface}

The localization of a function space basis to a single grid element is represented
by an interface called \cpp{LocalView}. Objects of type
\cpp{LocalView} are returned by the method \cpp{GlobalBasis::localView()},
and there is no way to construct such objects directly.
All types and methods listed in the following
interface declaration are public members of
the generic class \cpp{LocalView}.

A freshly constructed \cpp{LocalView} object is not completely initialized yet.
To truly have the object represent the basis localization on a particular element,
it must be \emph{bound} to that element.  This is achieved by calling
\begin{lstlisting}[style=Interface]
using GridView = typename GlobalBasis::GridView;
using Element = typename GridView::template Codim<0>::Entity;
void bind(const Element& e);
\end{lstlisting}
Once this method has been called, the \cpp{LocalView} object is fully set up
and can be used.
The call may incorporate expensive computations needed to
precompute the local basis functions and their global indices.
The local view can be
bound to another element at any time by calling the \cpp{bind} method again.
To set the local view back to the unbound state again, call the method
\begin{lstlisting}[style=Interface]
void unbind();
\end{lstlisting}
The local view will store a copy of the
element it is bound to, which is accessible via
\begin{lstlisting}[style=Interface]
const Element& element() const;
\end{lstlisting}

A bound \cpp{LocalView} object provides information
about the size of the local basis at the current element.
The total number of basis functions associated to the
local view at the current element is returned by
\begin{lstlisting}[style=Interface]
using size_type = typename GlobalBasis::size_type;
size_type size() const;
\end{lstlisting}
In the language of Chapter~\ref{sec:finite_element_trees}, this method computes
the number $\abs{\Lambda|_e}$.

To allow preallocation of buffers for local functions, the method
\begin{lstlisting}[style=Interface]
size_type maxSize() const;
\end{lstlisting}
returns the maximum return value of the
\cpp{size} method for all elements in the grid view
associated to the global basis, i.e., it computes $\max_e \abs{\Lambda|_e}$.
As this information does not depend on a particular element,
the method \cpp{maxSize} can even be called in unbound state.

As an example, suppose that \cpp{basis} is an object of type \cpp{Dune::Functions::TaylorHoodBasis},
which implements the Taylor--Hood basis that has been used for examples in the previous section.
The following code loops over all elements of the grid view and prints the numbers
of degrees of freedom per element:
\begin{lstlisting}[style=example]
auto localView = basis.localView();

for (auto&& element : elements(basis.gridView()))
{
  localView.bind(element);
  std::cout << "Element with " << localView.size()
    << " degrees of freedom" << std::endl;
}
\end{lstlisting}

Access to the actual local basis functions is provided
by the method
\begin{lstlisting}[style=Interface]
using Tree = @@<implementation defined>@@;
const Tree& tree() const;
\end{lstlisting}
This encapsulates the set $\Lambda|_e$ of basis functions localized to the
element $e$, organized in the tree of function space bases.
While the tree  itself can be queried in unbound state,
the local view must be bound in order to use most of the
trees methods.
A detailed discussion of the interface of the tree object is
given below.

For any of the local basis functions in the local tree
accessible by \cpp{tree()} the global multi-index
is provided by the method
\begin{lstlisting}[style=Interface]
using MultiIndex = @@<implementation defined>@@;
MultiIndex index(size_type i) const;
\end{lstlisting}
The argument for this method is the local
index of the basis function within the tree as
returned by \cpp{node.localIndex(k)};
here \cpp{node} is a leaf node of the
tree provided by \cpp{tree()}, and \cpp{k}
is the number of the shape function within the corresponding
local finite element (see below).
Hence the \cpp{index} method
implements the map $\iota^{\text{local}\to\text{global}}_e$
introduced in Section~\ref{sec:localization},
which maps local indices to global multi-indices.
Accessing the same global index multiple times
is expected to be cheap, because implementations are supposed to pre-compute
and cache indices during \cpp{bind(Element)}.
The result of calling \cpp{index(size\_type)} in
unbound state is undefined.

Extending the previous example a little, the following loop prints the
global indices for each degree of freedom of each element.
\begin{lstlisting}[style=example]
auto localView = basis.localView();

for (auto&& element : elements(basis.gridView()))
{
  localView.bind(element);
  for (std::size_t i=0; i<localView.size(); i++)
    std::cout << localView.index(i) << std::endl;
}
\end{lstlisting}
When this code is run for a Taylor--Hood basis on a two-dimensional triangle grid,
it will print 15 multi-indices per element, because a Taylor--Hood element has 12
velocity degrees of freedom and 3 pressure degrees of freedom per triangle.

Finally, the global basis of type \cpp{GlobalBasis}
is known by the \cpp{LocalView} object, and exported by the \cpp{globalBasis} method.
\begin{lstlisting}[style=Interface]
using GlobalBasis = @@<implementation defined>@@;
const GlobalBasis& globalBasis() const;
\end{lstlisting}
Therefore, code that is given only a \cpp{LocalView} object can retrieve the
global basis from it, and the grid view from there.

\subsection{The user interface of the tree of local bases}
The local view provides access to local basis functions of an element
by exporting a \cpp{Tree} object, which keeps the local basis functions in its leaves.
The tree structure is encoded in the type of the \cpp{Tree} object,
using the infrastructure of the \dunemodule{dune-typetree} module.

The object returned by the \cpp{LocalView::tree} method is not actually a tree,
but rather (a const reference to) the root node of the tree.  To navigate within
this tree, any non-leaf node allows to access
its children using the two methods
\begin{lstlisting}[style=Interface]
template<class... ChildIndices>
auto child(ChildIndices... childIndices);

template<class ChildTreePath>
auto child(ChildTreePath childTreePath);
\end{lstlisting}
The arguments to these methods are the paths from the current node to the desired
descendants.
Notice that both methods only provide access to strict descendants,
while accessing the node itself using an empty path is not supported.

For the first method, the path is passed
as a sequence of indices. Indices referring
to children of a power node can be passed as integer values, typically of type \cpp{std::size_t}.
Indices referring to children of a composite node
have to be passed statically as objects of type \cpp{Dune::index_constant<i>}.%
\footnote{
\dots which is a shortcut for \cpp{std::integral_constant<std::size_t, i>}.
}
For convenience global constants \cpp{_0},\cpp{_1}, \dots
of this type are implemented in the \cpp{Dune::Indices}
namespace.
Continuing the example of the previous section, if \cpp{localView} is a
local view of the \cpp{TaylorHoodBasis} localized to a particular grid element,
then the leaf node for the second velocity
component can be obtained by
\begin{lstlisting}[style=Example]
using namespace Indices;  // Import namespace with index constants _0, _1, _2, etc
const auto& node = localView.tree().child(_0, 1);
\end{lstlisting}
Note how the index constant \cpp{Dune::Indices::\_0} is used
to address the velocity node, because the
tree root is a composite node whose child nodes are of different type.
Within the velocity subtree, all three children are identical, and the second one can be
accessed by the run-time integer \cpp{1}.

The second \cpp{child} method allows to pass the tree path in a dedicated container.
Such a container needs to handle sequences of static and run-time values.
The \dunemodule{dune\-functions} module uses \cpp{Dune::TypeTree::HybridTreePath}
for this, which we describe in detail in Section~\ref{sec:multi_indices}.
Using a \cpp{HybridTreePath} object, the example looks as follows:
\begin{lstlisting}[style=Example]
auto treePath = Dune::TypeTree::treePath(_0, 1);
const auto& node = localView.tree().child(treePath);
\end{lstlisting}

At each of its leaf nodes, the localized basis function tree provides the set
of all corresponding shape functions.
The method for this is
\begin{lstlisting}[style=Interface]
using FiniteElement = @@<implementation defined>@@;
const FiniteElement& finiteElement() const;
\end{lstlisting}
The object returned by this method is a \cpp{LocalFiniteElement} as specified
in the \dunemodule{dune-localfunctions} module. As such, it provides access
to shape function values and derivatives, to the evaluation of degrees of freedom
in the sense of~\cite{ciarlet:1978}, and to the assignment of local degrees of
freedom to element faces.
The numbering used by \dunemodule{dune-localfunctions} for the shape functions
coincides with the leaf-local indices defined in Section~\ref{sec:localization}.
For example, if \cpp{node} is a leaf node in the localized Taylor--Hood tree,
the following code prints all shape function values of the leaf shape function set
at the point $(0,0,0)$ in local coordinates of the appropriate reference element:
\begin{lstlisting}[style=Example]
const auto& localBasis = node.finiteElement().localBasis();
std::vector<double> values;
localBasis.evaluate({0,0,0}, values);
for (auto v : values)
  std::cout << v << std::endl;
\end{lstlisting}

To obtain the entries of the element stiffness matrix that corresponds to a
given shape function from a given leaf node, the local index needs to be computed
from the leaf-local index of that shape function.
For each local basis function the method
\begin{lstlisting}[style=Interface]
size_type localIndex(size_type i) const;
\end{lstlisting}
returns the
local index within all local basis functions of the current element associated to the full
local tree.
The argument to this method is the
index of the local basis functions within the leaf.
In other words, the \cpp{localIndex} method
implements the map $\iota^{\text{leaf}\to\text{local}}_e$
introduced in Section~\ref{sec:localization}.
The return value is \emph{not} a multi-index.
While in principle all basis functions of the local subtree could be indexed using
general multi-indices, the \dunemodule{dune-functions} module only supports
flat indices here to keep the implementation simple.

While \cpp{LocalFiniteElement} objects are only available at leaf nodes,
the following methods work at every node in the tree again.
Calling
\begin{lstlisting}[style=Interface]
using size_type = @@<implementation defined>@@;
size_type size() const;
\end{lstlisting}
returns the total number of
local basis functions within the subtree rooted at the
present node.  In particular, calling this method for the tree root
yields the size of the element stiffness matrix.

Finally, all nodes provide access to the
element which they are bound
to via the method
\begin{lstlisting}[style=Interface]
using Element = @@<implementation defined>@@;
const Element& element() const;
\end{lstlisting}

\subsection{Multi-indices}
\label{sec:multi_indices}
Multi-indices appear in several places in \dunemodule{dune-functions}.
They are used as global indices to identify individual
basis functions of a function space basis, and for indexing
inner nodes of basis and index trees as well.
From an implementation point of view, basis and index trees differ
considerably. Only the localized basis tree explicitly appears
in the programmer interface, whereas index trees appear only implicitly
in the form of sets of indices with the appropriate structure (Definition~\ref{def:index_tree}).
These differences require separate multi-index
implementations for the different types of trees.  We discuss implementations
for both types of trees in turn.

\subsubsection{Multi-index implementations for basis trees}

The tree of localized basis functions is the only tree that explicitly appears
in the \dunemodule{dune-functions} programmer interface. The tree structure
is encoded as C++ type information using the tools from the \dunemodule{dune-typetree}
module. Navigation in this tree requires to manipulate paths from the root to
particular nodes.  In principle, such a path is a sequence of integers.

To understand the implementation,
remember that non-leaf tree nodes can be of two types, \emph{power} and
\emph{composite} (Section~\ref{sec:index_strategies}).
Since composite nodes have children of different
types, it is not possible to access those children using a dynamic run-time
index. Instead the child index in a composite node has to be encoded in a static
way.  For such situations, \dunemodule{dune-common} offers the type
\begin{lstlisting}[style=Interface]
template <std::size_t i> Dune::index_constant<i>;
\end{lstlisting}
which turns the number \cpp{i} into a type, and by this makes it accessible
to compile-time expressions.
On the other hand, all children of a power node have
the same C++ type, and can be accessed using a dynamic index of type \cpp{std::size_t}.

In a typical tree, composite and power nodes appear together.  It is therefore
necessary to have a container that can store both compile-time and run-time integers.
This is achieved by the class
\begin{lstlisting}[style=Interface]
template <class... I>
class HybridTreePath;
\end{lstlisting}
from the \dunemodule{dune-typetree} module.

Conceptually, a \cpp{HybridTreePath} is a fixed-size container, where each entry
can be of different type. The types of the individual entries
are passed as template parameters. If the type used for an entry is
\cpp{std::size_t}, then this entry will have a dynamic value.
If, on the other hand, the type is \cpp{Dune::index_constant<i>}, then its value is static,
and can be used for compile-time decisions.

An object of type \cpp{HybridTreePath} can be used to access the nodes of a
localized basis tree if dynamic tree path entries only appear as child indices
for power nodes in the tree while all other entries are static.
For example, to access the leaf nodes corresponding
to the velocity components in the Taylor--Hood ansatz tree depicted
in Figure~\ref{fig:taylor_hood_basis_tree} one would use multi-indices
of the type
\begin{lstlisting}[style=Example]
HybridTreePath<Dune::index_constant<0>, std::size_t>;
\end{lstlisting}
whereas the multi-index for the pressure leaf node would use the type
\begin{lstlisting}[style=Example]
HybridTreePath<Dune::index_constant<1>>;
\end{lstlisting}
To construct objects of these types, call
\begin{lstlisting}[style=Example]
using namespace Dune::Indices;
HybridTreePath<Dune::index_constant<0>, std::size_t> i00(_0,0);
HybridTreePath<Dune::index_constant<0>, std::size_t> i01(_0,1);
HybridTreePath<Dune::index_constant<0>, std::size_t> i02(_0,2);
\end{lstlisting}
for the velocity leaf nodes, and
\begin{lstlisting}[style=Example]
HybridTreePath<Dune::index_constant<1>> i1(_1);
\end{lstlisting}
for the pressure node.  The constants \cpp{_0}, \cpp{_1}, and \cpp{_2}
are predefined in the namespace \cpp{Dune::Indices}.

This way of construction is overly verbose, because static indices have
to be provided both as template and as constructor arguments.
To simplify the construction of such objects, the \dunemodule{dune-typetree}
module provides the helper function
\begin{lstlisting}[style=Interface]
template <class... I>
auto TypeTree::treePath(I... i);
\end{lstlisting}
which creates a \cpp{HybridTreePath} with the given entries.
As this is a free method rather than a constructor, the entries have to be
given only once, and their types are inferred.
The multi-indices of the previous example can be constructed using
\begin{lstlisting}[style=Example]
auto i00 = TypeTree::treePath(_0, 0);
auto i01 = TypeTree::treePath(_0, 1);
auto i02 = TypeTree::treePath(_0, 2);
auto i1  = TypeTree::treePath(_1);
\end{lstlisting}
which is much shorter.
To access entries of a \cpp{HybridTreePath} object, the object has
the method
\begin{lstlisting}[style=Interface]
template<std::size_t i>
constexpr decltype(auto) operator[](const Dune::index_constant<i>&) const;
\end{lstlisting}
Depending on the template parameter \cpp{i}, the return type is either \cpp{std::size_t}
or \cpp{Dune::index_constant<i>}.
Note that this construction is necessary although the function is already
\cpp{constexpr}:
Since \cpp{HybridTreePath} objects are intended to select
children of different type in run-time contexts, they have to encode
compile time index values into the run-time index objects.
The latter is only possible by making the types of those index objects
dependent on their compile time values.

However, as objects of type \cpp{Dune::index_constant} can be implicitly converted
to \cpp{std::size_t}, there is also
\begin{lstlisting}[style=Interface]
auto operator[](std::size_t) const;
\end{lstlisting}
Hence, to get the first digit of a tree path it is possible to write
\begin{lstlisting}[style=Example]
std::size_t a = myHybridTreePath[0];
\end{lstlisting}
For a tree path in the Taylor--Hood tree this will return \cpp{0} or \cpp{1}
as expected.  However, this return value is not usable in compile-time situations
anymore.

These are just the more important methods of the \cpp{HybridTreePath} class.
For a complete description see the online
documentation of the \dunemodule{dune-typetree} module.

\subsubsection{Multi-index implementations for index trees}
\label{sec:multi_indicies_for_index_trees}

Index trees are formed by the multi-indices that are used to label basis functions.
Conceptually, there are two such trees in the \dunemodule{dune-functions} interface:
the tree of global indices, and the tree of local indices.  To keep the implementation simple,
\dunemodule{dune-functions} only allows flat (i.e., single-digit) multi-indices for the local index tree.
Therefore, only data types for global indices need to be discussed.

Unlike the tree paths of the previous section,
global indices are run-time constructs.  A single C++ type represents all such
indices for a given basis, even if that basis has a non-trivial tree structure.
The exact type is selected by the basis implementation, and can differ from basis to basis.
It mainly depends on whether the index is uniform, i.e., whether all indices from the
set have the same number of digits.
Having purely dynamic multi-indices can be inconvenient when accessing
containers such as \cpp{std::tuple}
or \cpp{MultiTypeBlockVector} (from the \dunemodule{dune-istl}
module). However, it has the advantage that standard run-time loops can be used
to iterate over the indices.

Dynamic multi-indices are random-access containers holding entries of a fixed integer type.
All implement a common interface, consisting of two member functions
\begin{lstlisting}[style=Interface]
std::size_t size() const;
auto operator[](std::size_t) const;
\end{lstlisting}
The \cpp{size} method returns the number of digits of the multi-index,
and \cppbreak{operator[]} allows to access
each entry by its position. Since multi-indices are
typically not changed by user code, both methods are \cpp{const}.
The type used to represent
the individual digits of multi-indices can be selected when instantiating \cpp{GlobalBasis}
objects.  The default type is \cpp{std::size_t}.

In the following we will
give an overview of the types used to represent multi-indices
in \dunemodule{dune-functions}.
In the most general case, not all multi-indices for a given basis have
the same number of digits.  As examples, consider columns 1, 2, 5, and 6
of Table~\ref{tab:th_indexing_variants}, which give such numberings for the Taylor--Hood basis.
In these cases, multi-indices are typically represented
by the class
\begin{lstlisting}[style=Interface]
template <class T, int k>
class ReservedVector;
\end{lstlisting}
from the \dunemodule{dune-common} module, which is parameterized
by the entry type \cpp{T} and a capacity \cpp{k}.
It implements an STL-compatible random-access container with a dynamic size,
which may not exceed \cpp{k} entries.
In contrast to a fully dynamic vector implementation
like \cpp{std::vector<T>}, the class \cpp{ReservedVector} stores its entries
on the stack.  This avoids dynamic memory management, and makes the
implementation much more efficient. The global multi-indices typically have
a small number of digits only with a known upper bound.
Hence the overhead of always using a buffer of size \cpp{k} even
for indices with less than \cpp{k} digits will typically be small.

However, many bases can be indexed by uniform index trees, i.e., sets of indices where
all indices have the same number of digits.  In that case, the capacity of a
\cpp{ReservedVector} can be set to the correct length, and no buffer space is wasted.
However, in addition to the buffer, each \cpp{ReservedVector} object has to store
the container length, which is not needed when the index set is known to be uniform.
\cpp{GlobalBasis} objects that implement uniform index sets can therefore
opt to use a fixed-size container type like \cpp{std::array} instead of
\cpp{ReservedVector}.

Finally, if the basis is indexed with a flat index, i.e., a multi-index with only a single digit,
then using an array can be a bit cumbersome.  Morally, flat multi-indices
are simply natural numbers.  However, if \cpp{i} is a \cpp{std::array} of length~1,
using it to access the corresponding entry of a \cpp{std::vector} called \cpp{vec} has to be
written as
\begin{lstlisting}
auto value = vec[i[0]];
\end{lstlisting}
To allow the more intuitive syntax
\begin{lstlisting}
auto value = vec[i];
\end{lstlisting}
\dunemodule{dune-functions} implements the \cpp{FlatMultiIndex} class for the
case that the index
of a basis tree is flat.  Objects of type \cpp{FlatMultiIndex}
behave like objects of type \cpp{std::array<T,1>}, but additionally, they allow to cast
their content to \cpp{T&}.  Therefore, objects of type \cpp{FlatMultiIndex} can be directly used
like number types, and like multi-index types as well.

\section{Constructing trees of function space bases}

There are various ways to construct finite element bases in \dunemodule{dune-functions}.
A set of standard bases is provided directly.  These can then be combined to form trees.
Conversely, subtrees can be extracted, and they act like complete bases in their own right.

\subsection{Basis implementations provided by \texorpdfstring{\dunemodule{dune-functions}}{dune-functions}}
\label{subsec:available_bases}

The \dunemodule{dune-functions} module contains a collection of standard finite element bases.
These can be directly used in finite element simulation codes. At the time of writing there are:

\begin{itemize}
 \item \cpp{LagrangeBasis}: Lagrange basis of order $k$, where $k$ is a compile-time parameter.
   This implementation works on all kinds of conforming grids, including grids with more
   than one element type.  At the time of writing, higher-order spaces are implemented only partially.
   Check the online class documentation for the current status.

 \item \cpp{LagrangeDGBasis}: Implements a $k$-th order Discontinuous-Galerkin (DG) basis with Lagrange shape functions.
   As a DG basis, it also
   works well on non-conforming grids.  The polynomial order $k$ is again a compile-time parameter.

 \item \cpp{RannacherTurekBasis}: An $H^1$-nonconforming scalar basis, which adapts the idea
   of the Crouzeix--Raviart basis to cube grids~\cite{rannacher_turek:1992}.

 \item \cpp{BSplineBasis}:  Implements a B-Spline basis on a structured, axis-aligned grid as described,
   e.g., in~\cite{cottrell_hughes_bazilevs:2009}.  Arbitrary orders, dimensions, and knot vectors are supported,
   allowing, e.g., to work with $C^1$ elements for fourth-order differential equations.

   Each \cpp{BSplineBasis} object implements a basis on a single patch, and the grid must correspond to this
   patch. For this to work, several restrictions apply for the grid.  It must be structured and axis-aligned,
   and consist of (hyper-)cube elements only.  Further, the element indices must be lexicographic and
   increase from the lower left to the upper right domain corner.  The element spacing must match the knot spans.
   Unfortunately, not all these requirements can be checked for by the basis, so users have to be a bit
   careful.  Using \cpp{YaspGrid} objects works well.

   Unlike in standard finite element bases, in a B-spline basis the basis functions cannot be associated
   to grid entities such as vertices, edges, or elements.  The \dunemodule{dune-localfunctions}
   programmer interface of a B-spline basis nevertheless mandates that a
   \cpp{LocalCoefficient} object must be available on each element, which assigns shape functions
   to faces of the reference element. For the \cpp{BSplineBasis}, the behavior of this
   object is undefined.

 \item \cpp{TaylorHoodBasis}:
   An implementation of a first-order Taylor--Hood basis.  It exists mainly to serve as an example of
   how to directly implement a basis with a non-trivial tree.
   Generally, non-trivial product bases
   can be easily constructed in a generic way. This approach is described
   in Chapter~\ref{sec:composed_bases} and it is the preferred way to construct
   a Taylor--Hood basis.
\end{itemize}

For all bases listed above, the shape functions provided by
\cpp{tree.finiteElement()} are implemented in terms of coordinates of the reference
element $T_\text{ref}$. That is, if a grid element $e$ is obtained by the transformation
$\Phi_e: T_\text{ref} \to e$, then the implemented localized shape function
representing the restriction of the basis function $\lambda$ to the
element $e$ is given by $\hat{\lambda}|_e = \lambda\circ\Phi_e$.
Finite elements that form non-affine families~\cite{ciarlet:1978}
may require additional transformations. This is the case for the following global
bases implementations.

\begin{itemize}
 \item \cpp{RaviartThomasBasis}: The standard Raviart--Thomas basis~\cite{boffi_brezzi_fortin:2013}
  for problems in $H(\text{div})$.  Available for different orders and element types.

 \item \cpp{BrezziDouglasMariniBasis}: The standard Brezzi--Douglas--Marini basis, which is an
  alternative basis for $H(\text{div})$-conforming problems~\cite{boffi_brezzi_fortin:2013}.
\end{itemize}

Both bases require the Piola transformation to properly pull back the basis functions
onto the reference element.  This transformation is \emph{not} performed by the
\dunemodule{dune\-functions} implementation, and is expected to happen in user code.
For a detailed discussion of the template parameters and constructor arguments
of the basis implementations listed above we refer to the online class documentation.

\subsection{Combining bases into trees}
\label{sec:composed_bases}

The basis implementations of the previous section can be combined by multiplication to form new bases.
This produces the tree structures described in Section~\ref{sec:finite_element_trees}.
The multiplication code resides in the \cpp{BasisFactory} namespace, which is a nested namespace
within \cpp{Dune::Functions::}. Therefore, the examples in this section need a
\begin{lstlisting}[style=Example]
using namespace Dune::Functions::BasisFactory;
\end{lstlisting}
to compile.

The methods to combine bases into trees do not operate on the basis classes of the previous section
directly.  Rather, they combine so-called \emph{pre-bases}, of which there is one for each basis.
The reason for this is
that it is technically challenging to combine the actual user-visible basis types in a
tree hierarchy that itself again implements the interface of a hierarchical function space basis.
Therefore, the multiplication operators are applied to pre-basis objects, and return pre-basis
objects of the resulting tree.
The pre-basis of the final basis tree can then be turned into an actual basis.

Since all pre-bases in the product pre-basis have to know some common information
like, e.g., the grid view, doing this hierarchic construction
manually is verbose and error prone. As a more user friendly and safer solution
a global basis can be constructed by a call to
\begin{lstlisting}[style=Interface]
template <class GridView, class PreBasis>
auto makeBasis(const GridView& gridView, PreBasis&& preBasis);
\end{lstlisting}
The pre-basis argument encodes the hierarchic product.
The actual basis is constructed automatically by the
\cpp{makeBasis} function from the pre-bases in a consistent way.
This also determines a suitable multi-index type automatically,
which otherwise would have to be done by the user.
\footnote{The interface description is in fact slightly simplified:
The user-provided arguments of \cpp{makeBasis} are not pre-bases themselves
but pre-basis-factory objects that can construct the corresponding pre-bases.
This mechanism allows to delay passing the shared information
(e.g. the grid view) to the construction of the real pre-bases which is triggered
by \cpp{makeBasis}. However, to simplify the presentation
we will ignore the technical difference of a pre-basis
and its pre-basis-facory in the following.}

In the simple-most case, the basis tree consists of a single leaf.
This leaf is then, e.g., one of the basis implementations of the previous section.
As a convention, for each global basis
\cpp{FooBarBasis} there is a function \cpp{BasisFactory::fooBar()}
(defined in the same header file as \cpp{FooBarBasis}),
creating a suitable pre-basis object which
stores all basis-specific information.
That means that in particular you can write
\begin{lstlisting}[style=Example]
auto raviartThomasBasis = makeBasis(gridView, raviartThomas<k>());
\end{lstlisting}
to obtain a Raviart--Thomas basis for the given grid view.
This call to \cpp{makeBasis} is equivalent to constructing
the basis directly:
\begin{lstlisting}[style=Example]
RaviartThomasBasis<GridView,k> raviartThomasBasis(gridView);
\end{lstlisting}
Note that the \cpp{raviartThomas} function, just like the corresponding functions for other bases, does not need
the grid view as parameter.

If \cpp{FooBarBasis} has template and/or constructor parameters, then by convention they
are given in the same order as the template and method
parameters of the \cpp{BasisFactory::fooBar()} function.
As the only difference, the former has the grid view type and object prepended.

The pre-basis combining several bases in a product is called \cpp{CompositePreBasis},
defined in the header \file{dune/functions/functionspacebases/compositebasis.hh}.
It implements a \emph{composite} tree node as introduced in Definition~\ref{def:power_node}.
Analogously to the above description, a pre-basis for a tree with a composite root
can be constructed using the global function
\begin{lstlisting}[style=Interface]
template <class... ChildPreBasis>
auto composite(ChildPreBasis&&... childPreBasis);
\end{lstlisting}
contained in the namespace \cpp{BasisFactory}.
The method has an unspecified number of arguments, of unspecified type.
The arguments are expected to be pre-basis objects themselves.
They can either be plain pre-bases constructed by, e.g.,
\cpp{lagrange<1>()} or \cpp{raviartThomas<k>()},
or composite- or power pre-bases constructed by the \cpp{composite}
or \cpp{power} function (see below), respectively.

\begin{figure}
    \begin{center}
        \begin{tikzpicture}[
                level/.style={
                    sibling distance = (3-#1)*0.3cm + 1cm,
                    level distance = 1.5cm
                }
            ]
            \node [treenode] {$\Lambda_{RT} \sqcup \Lambda_{P_0}$}
                child{ node [treenode] {$\Lambda_{RT}$} }
                child{ node [treenode] {$\Lambda_{P_0}$} };
        \end{tikzpicture}
        \hspace{0.15\textwidth}
        \begin{tikzpicture}[
                level/.style={
                    sibling distance = (3-#1)*0.3cm + 1cm,
                    level distance = 1.5cm
                }
            ]
            \node [treenode] {$\Lambda_{P_1} \sqcup \Lambda_{P_1} \sqcup \Lambda_{P_1}$}
                child{ node [treenode] {$\Lambda_{P_1}$} }
                child{ node [treenode] {$\Lambda_{P_1}$} }
                child{ node [treenode] {$\Lambda_{P_1}$} };
        \end{tikzpicture}
    \end{center}
    \caption{Example composite bases}
    \label{fig:example_composite_bases}
\end{figure}
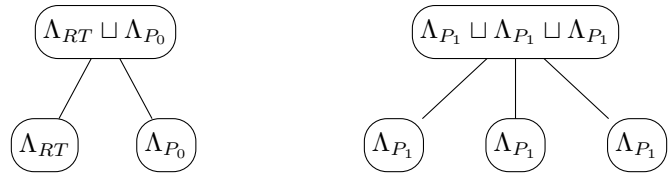

As an example, to combine a Raviart--Thomas basis with a zero-order Lagrange basis
(let's say for solving
the mixed formulation of the Poisson equation \cite{braess:2013}), the appropriate call is
\begin{lstlisting}[style=Example]
auto mixedBasis = makeBasis(
  gridView,
  composite(
    raviartThomas<0>(),
    lagrange<0>()
  ));
\end{lstlisting}
Combining three copies of a first-order Lagrange basis for a displacement field in elasticity theory is
done by
\begin{lstlisting}[style=Example]
auto displacementBasis = makeBasis(
  gridView,
  composite(
    lagrange<1>(),
    lagrange<1>(),
    lagrange<1>()
  ));
\end{lstlisting}
The examples produce the trees shown in Figure~\ref{fig:example_composite_bases}.

The second example is not as elegant as it could be.  First of all, it is inconvenient and unnecessarily
wordy to list the same scalar Lagrange basis three times.  Secondly, the required number may depend on
a parameter.
Finally, the implementation can benefit from the explicit knowledge that
all children are equal.
For these reasons, \dunemodule{dune-functions}
offers a second way to combine bases: The \cpp{PowerPreBasis}
to be constructed by the factory method \cpp{BasisFactory::power()}.
The interface is again a single method
\begin{lstlisting}[style=Interface]
template<std::size_t k, class ChildPreBasis>
auto power(ChildPreBasis&& childPreBasis)
\end{lstlisting}
provided in the file \file{dune/functions/functionspacebases/powerbasis.hh}.
It combines \cpp{k} copies of a subtree of type \cpp{ChildPreBasis} in a new tree.  Therefore, the
displacement vector field basis from above is more easily written as
\begin{lstlisting}[style=Example]
auto displacementBasis = makeBasis(
  gridView,
  power<3>(
    lagrange<1>()
  ));
\end{lstlisting}
Since \cpp{composite} and \cpp{power} create pre-bases themselves,
all these techniques can be combined. To obtain the \cpp{p}-th order Taylor--Hood basis,
write
\begin{lstlisting}[style=Example]
auto taylorHoodBasis = makeBasis(
  gridView,
  composite(
    power<dim>(
      lagrange<p+1>()),
    lagrange<p>()
  ));
\end{lstlisting}
The call to \cpp{power} produces the \cpp{dim}-component Lagrange basis of order \cpp{p+1} for the velocity,
and the call to \cpp{composite} combines this with a \cpp{p}-th order Lagrange basis for the pressure.
Note that this is the preferred way to construct a Taylor--Hood basis in contrast to
\begin{lstlisting}[style=Example]
auto taylorHoodBasis1 = makeBasis(gridView, taylorHood());
\end{lstlisting}
and
\begin{lstlisting}[style=Example]
auto taylorHoodBasis2 = TaylorHoodBasis<GridView>(gridView);
\end{lstlisting}
These variants mainly exist as an implementation example.

The previous discussion has left out the question of how the degrees of freedom in the combined tree
are numbered.  In Section~\ref{sec:index_trees} it was explained how the indices
of the degrees of freedom form a separate tree by their multi-index structure, and how this tree
is constructed from the basis tree by a set of strategies.
These ideas are reflected in the design of the \dunemodule{dune-functions} programmer interface.
First of all, each of the bases of
Section~\ref{subsec:available_bases} implements a numbering of its degrees of freedom,
and generally these numberings cannot be changed.
To select a degree of freedom numbering for a non-trivial basis,
each call to \cpp{composite} or \cpp{power} can be augmented by an additional
flag indicating an \cpp{IndexMergingStrategy}. The four implemented strategies are
\begin{itemize}
  \item
    \cpp{BlockedLexicographic}
  \item
    \cpp{BlockedInterleaved}
  \item
    \cpp{FlatLexicographic}
  \item
    \cpp{FlatInterleaved}
\end{itemize}
and have been described in Section~\ref{sec:index_strategies}.
For each strategy \cpp{FooBar} there is a function \cpp{BasisFactory::fooBar()} creating
the flag in the header \file{functionspacebases/\allowbreak basistags.hh}.
For example, a Taylor--Hood basis with the indexing listed in the
second column (labeled BL(BI)) of Table~\ref{tab:th_indexing_variants} can be created using
\begin{lstlisting}[style=Example]
auto taylorHoodBasis = makeBasis(
  gridView,
  composite(
    power<dim>(
      lagrange<p+1>(),
      blockedInterleaved()),
    lagrange<p>(),
    blockedLexicographic()
  ));
\end{lstlisting}
This will lead to multi-indices of length three and two
for velocity and pressure degrees of freedom, respectively.
The same ordering of basis functions with a uniform indexing scheme
with multi-index length two (Column~4 labeled BL(FI) in Table~\ref{tab:th_indexing_variants}) is obtained by
\begin{lstlisting}[style=Example]
auto taylorHoodBasis = makeBasis(
  gridView,
  composite(
    power<dim>(
      lagrange<p+1>(),
      flatInterleaved()),
    lagrange<p>(),
    blockedLexicographic()
  ));
\end{lstlisting}
Finally, a flat indexing scheme still preserving the same ordering
(Column~8 labeled FL(FI) in Table~\ref{tab:th_indexing_variants})
is obtained by
\begin{lstlisting}[style=Example]
auto taylorHoodBasis = makeBasis(
  gridView,
  composite(
    power<dim>(
      lagrange<p+1>(),
      flatInterleaved()),
    lagrange<p>(),
    flatLexicographic()
  ));
\end{lstlisting}
If no strategy is given, \cpp{composite} will use the \cpp{BlockedLexicographic} strategy,
where\-as \cpp{power} will use \cpp{BlockedInterleaved}.

\section{Treating subtrees as separate bases}
\label{sec:subtrees}

The previous section has shown how trees of bases can be combined to form
bigger trees.  It is also possible to extract subtrees from other trees
and treat these subtrees as basis trees in their own right.
The programmer interface for such subtree bases is called \cpp{SubspaceBasis}.
It mostly coincides with the interface of a global basis, but additionally
to the \cpp{GlobalBasis} interface the \cpp{SubspaceBasis} provides
information about how the subtree is embedded into the global basis.
More specifically, the method
\begin{lstlisting}[style=Interface]
const @@<implementation defined>@@& rootBasis() const
\end{lstlisting}
provides access to the root basis, and the method
\begin{lstlisting}[style=Interface]
using PrefixPath = TypeTree::HybridTreePath@@<implementation defined>@@;
const PrefixPath& prefixPath() const
\end{lstlisting}
returns the
path of the subtree associated to the \cpp{SubspaceBasis}
within the full tree.
For convenience a global basis behaves like a trivial \cpp{SubspaceBasis},
i.e., it has the method \cpp{rootBasis} returning the basis itself,
and \cpp{prefixPath} returning an empty tree-path.
Note that a \cpp{SubspaceBasis} differs from a full
global basis because the global multi-indices are the
same as the ones of the root basis, and thus they are in general
neither consecutive nor zero-based. Instead, those multi-indices
allow to access containers storing coefficients for the
full root basis.

\cpp{SubspaceBasis} objects are created using a global
factory function from the root basis and the path
to the desired subtree. The path can either be passed
as a single \cpp{HybridTreePath} object (see Section~\ref{sec:multi_indices}), or
as a sequence of individual indices.

\begin{lstlisting}[style=Interface]
template<class RootBasis, class... PathIndices>
auto subspaceBasis(const RootBasis& rootBasis,
                   const TypeTree::HybridTreePath<PathIndices...>& prefixPath);

template<class RootBasis, class... PathIndices>
auto subspaceBasis(const RootBasis& rootBasis, const PathIndices&... indices);
\end{lstlisting}
For example, suppose that \cpp{taylorHoodBasis} is any one of the implementations
of the Taylor--Hood basis defined in Section~\ref{sec:composed_bases}.
Then
\begin{lstlisting}[style=Example]
auto velocityBasis = subspaceBasis(taylorHoodBasis, _0);
\end{lstlisting}
will extract the subtree of velocity degrees of freedom, and
\begin{lstlisting}[style=Example]
auto pressureBasis = subspaceBasis(taylorHoodBasis, _1);
\end{lstlisting}
will extract the (trivial) subtree of pressure degrees of freedom.
The possibly non-consecutive multi-indices of a \cpp{SubspaceBasis} are
best illustrated by extracting a single velocity component
\begin{lstlisting}[style=Example]
auto velocityZBasis = subspaceBasis(taylorHoodBasis, _0, 2);
\end{lstlisting}
For this example the following table shows the multi-indices
of the \cpp{SubspaceBasis} extracted from the full basis,
with columns representing the different index merging strategies also
used in Table~\ref{tab:th_indexing_variants}:

\medskip

\noindent
\makebox[\textwidth][c]{
\footnotesize
\begin{tabular}{c|c|c|c|c|c|c|c|c}
    & BL(BL)
    & BL(BI)
    & BL(FL)
    & BL(FI)
    & FL(BL)
    & FL(BI)
    & FL(FL)
    & FL(FI)
    \\
  \hline
%  \hline
  $v_{x_2,0}$
    & $(0,2,0)$
    & $(0,0,2)$
    & $(0,2n_2+0)$
    & $(0,0+2)$
    & $(2,0)$
    & $(0,2)$
    & $(2n_2+0)$
    & $(0+2)$
    \\
  $v_{x_2,1}$
    & $(0,2,1)$
    & $(0,1,2)$
    & $(0,2n_2+1)$
    & $(0,3+2)$
    & $(2,1)$
    & $(1,2)$
    & $(2n_2+1)$
    & $(3+2)$
    \\
  $v_{x_2,2}$
    & $(0,2,2)$
    & $(0,2,2)$
    & $(0,2n_2+2)$
    & $(0,6+2)$
    & $(2,2)$
    & $(2,2)$
    & $(2n_2+2)$
    & $(6+2)$
    \\
  $v_{x_2,3}$
    & $(0,2,3)$
    & $(0,3,2)$
    & $(0,2n_2+3)$
    & $(0,9+2)$
    & $(2,3)$
    & $(3,2)$
    & $(2n_2+3)$
    & $(9+2)$
    \\
  $\vdots$ & $\vdots$ & $\vdots$ & $\vdots$ &  $\vdots$ & $\vdots$ & $\vdots$ & $\vdots$ & $\vdots$
  \\
  \hline
\end{tabular}
}

\medskip

\cpp{SubspaceBasis} objects can be combined with
coefficient vectors to represent vector- and scalar-valued discrete functions.
The interface for this construction is discussed in the next section.

\section{Combining global bases and coefficient vectors}

Function space bases and coefficient vectors are combined to yield discrete functions,
by the linear combination shown exemplarily in~\eqref{eq:linear_combination}.
Such discrete functions can then, e.g., be written to a file, or handed to some
post-processing agent. Conversely, discrete and non-discrete functions can be
projected onto the span of a basis, which yields a corresponding coefficient
vector.  In \dunemodule{dune-functions}, this process is called \emph{interpolation},
although it is not always an interpolation in the strict sense of the word.

\subsection{Vector backends}
\label{subsec:backends}

In both cases, individual basis functions need to be associated with corresponding
entries of a container data type that holds vector coefficients. While trivial
in theory, in practice there is a gap here because the multi-index types used
by \dunemodule{dune-functions} to label basis functions (Section~\ref{sec:multi_indicies_for_index_trees})
cannot be used to access entries of standard random-access containers.

The gap is bridged by a concept called \emph{vector backends}.
These are shim classes that abstract away implementation details of particular
container classes, and make them addressable by multi-indices.
The \dunemodule{dune-functions} module currently offers such a backend
for containers from \dunemodule{dune-istl} and the C++ standard library,
but others can be added easily. This makes it possible to combine \dunemodule{dune-functions}
function space bases with basically any linear algebra implementation.

There are two parts to the vector backend concept:
When interpreting a vector of given coefficients with respect
to a basis,  access is only required in a non-mutable way.
In \dunemodule{dune-functions}
this functionality is encoded in the \cpp{ConstVectorBackend} concept which
solely requires direct access by \cpp{operator[]} using the multi-indices
provided by the function space basis:
\begin{lstlisting}[style=Interface]
  auto operator[](Basis::MultiIndex) const;
\end{lstlisting}

For interpolation of given functions a corresponding mutable access is needed as well.
Furthermore, it must be possible to resize the vector to match
the index tree generated by the basis. These two additional methods make up the
\cpp{VectorBackend} concept:
\begin{lstlisting}[style=Interface]
  auto operator[](Basis::MultiIndex);
  void resize(const Basis&);
\end{lstlisting}
Note that the argument of the \cpp{resize}
member function is not a number, but the basis itself. This
is necessary because resizing nested containers requires information about the whole
index tree.

For the vector types implemented in the \dunemodule{dune-istl} and \dunemodule{dune-common}
modules, such a backend can be obtained using
\begin{lstlisting}[style=Interface]
  template<class SomeDuneISTLVector>
  auto istlVectorBackend(SomeDuneISTLVector& x);

  template<class SomeDuneISTLVector>
  auto istlVectorBackend(const SomeDuneISTLVector& x);
\end{lstlisting}
Depending on the \cpp{const}-ness of the argument, the resulting
object implements the \cpp{VectorBackend} or only the
\cpp{ConstVectorBackend} interface.  Even though these methods have the \cpp{istl} prefix
in their names, they actually also work well for containers from the C++ standard library
like \cpp{std::array} and \cpp{std::vector}.

Since a non-trivial product function space basis corresponds to functions with
a non-scalar range, we additionally have to map the components
of the spanned product function space to components of a function range type.
If, for example, the functions from the power function space generated by the basis
\begin{lstlisting}[style=Example]
auto basis = makeBasis(gridView, power<dim>(lagrange<1>()));
\end{lstlisting}
should be interpreted as vector fields, one would map the \cpp{dim}
leaf nodes of this basis to the \cpp{dim} entries of a \cpp{FieldVector<double,dim>}.
Whenever combining bases and range types \dunemodule{dune-function}
uses a default mapping generalizing this idea to more complex nested bases:
Assume that \cpp{y} is an object of the function range type.
Then a leaf node with tree path \cpp{i0}, \dots, \cpp{in} is associated
to the entry \cpp{y[i0]}\dots\cpp{[in]}.
In the exceptional case that the range type does not provide an \cpp{operator[]}
it is directly used for all leaf nodes in the ansatz space. This last rule allows
to interpolate a scalar function into all components of a basis at once.
For additional flexibility, users can also provide custom mappings to be used instead of this one.
However, we will not discuss the corresponding interface and rely
on the implicitly used default implementation in the following.

\subsection{Interpreting coefficient vectors as finite element functions}

To combine a basis and a coefficient vector to a discrete function that
can be evaluated point-wise in $\Omega$, \cpp{dune-functions} provides
the function
\begin{lstlisting}[style=Interface]
template<class Range, class B, class C>
auto makeDiscreteGlobalBasisFunction(const B& basis, const C& coefficients);
\end{lstlisting}
For given basis \cpp{basis} and coefficient vector \cpp{coefficients}
this returns an object representing the corresponding
finite element function. This object implements the \cpp{GridViewFunction}
concept for the grid view the basis is defined on, described in~\cite{engwer_graeser_muething_sander:2015}
with the range type \cpp{Range}.
The \cpp{basis} can either be a global basis or
a \cpp{SubspaceBasis}.  In the latter case the coefficient
vector has to correspond to the full basis nevertheless, but only
the coefficients associated with the subspace basis functions
will be used.

For the type \cpp{C} used to represent the coefficient vector there are two choices.
Either it implements the \cpp{ConstVectorBackend} concept, as for example all objects
returned by the \cpp{istlVectorBackend} method do.  If \cpp{C} does not implement
this concept, then the code assumes that it is a \dunemodule{dune-istl}-style
container and tries to wrap it with an \cpp{ISTLVectorBackend}.
That way, \cpp{makeDiscreteGlobalBasisFunction} can be called with \dunemodule{dune-istl}
or STL containers directly, but all others have to be wrapped in an appropriate backend
explicitly.

Notice that the range type can in general not be
determined automatically from the basis and coefficient type
because there are multiple possible types to implement this.
For example a scalar function could return \cpp{double}
or \cpp{FieldVector<double,1>}.
Hence the range type \cpp{Range} has to be given explicitly
by the user. The mapping from the different leaf nodes of the basis
to the entries of \cpp{Range} follows the procedure described
in Section~\ref{subsec:backends}.

To give an example how \cpp{makeDiscreteGlobalBasisFunction} is used,
we construct yet another instance of the Taylor--Hood basis
\begin{lstlisting}[style=Example]
auto taylorHoodBasis = makeBasis(
  gridView,
  composite(
    power<dim>(
      lagrange<p+1>()),
    lagrange<p>()
  ));
\end{lstlisting}
By default, the merging strategies are \cpp{BlockedLexicographic} for the composite node,
and \cpp{BlockedInterleaved} for the power node.  The resulting indices are the ones
from Column~2 of Table~\ref{tab:th_indexing_variants}.  An appropriate vector container
type for this is
\begin{lstlisting}[style=Example]
using Vectortype = TupleVector<
        BlockVector<FieldVector<double,dim> >,
        BlockVector<FieldVector<double,1> > >;
\end{lstlisting}
Let \cpp{x} be an object of this type.  To obtain the corresponding velocity field
as a discrete function, write
\begin{lstlisting}[style=Example]
// Create SubspaceBasis for the velocity field
auto velocityBasis = subspaceBasis(taylorHoodBasis, _0);

// Fix a range type for the velocity field
using VelocityRange = FieldVector<double,dim>;

// Create a function for the velocity field only
// but using the vector x for the full taylorHoodBasis.
auto velocityFunction
        = makeDiscreteGlobalBasisFunction<VelocityRange>(velocityBasis, x);
\end{lstlisting}

Notice that the \cpp{dim} leaf nodes of the function space
tree spanned by \cpp{velocityBasis} are automatically mapped to the
\cpp{dim} components of the \cpp{VelocityRange} type.
The resulting function created in the last line implements the full \cpp{GridViewFunction}
interface described in~\cite{engwer_graeser_muething_sander:2015}.  For example
it can be directly
passed to the \cpp{VTKWriter} class of the
\dunemodule{dune-grid} module to write the velocity field
as a VTK vector field.
See the end of Section~\ref{sec:stokes_example_main}
for how this is done.

\subsection{Interpolation}

In various parts of a finite element or finite volume simulation code, given functions need to be interpolated
into spaces spanned by a global basis.  For example, initial iterates may be given in closed form, but need to be
transferred to a finite element representation to be usable.  Similarly, Dirichlet values given in closed form
may need to be interpolated on the set of Dirichlet degrees of freedom.  Depending on the finite element space,
interpolation may take different forms.  Nodal interpolation is the natural choice for Lagrange elements, but
for other spaces $L^2$-projections or Hermite-type interpolation may be more appropriate.

The \dunemodule{dune-functions} module provides a set of methods for interpolation in the file
\file{dune/functions/functionspacebases/interpolation.hh}.  These methods are canonical in the sense that
they use the \cpp{LocalInterpolation} functionality on each element for the interpolation.  This is appropriate
for a lot, but not all finite element spaces.  For example, no reasonable local interpolation can be defined
for B-spline  bases, and therefore the standard interpolation functionality cannot be used with the
\cpp{BSplineBasis} class.
This approach also fails for non-affine finite elements, because \cpp{LocalInterpolation} is
not applying non-standard transformations to the reference element.

The interpolation functionality is implemented in two global functions.
The first deals with the simple case of a given function and basis,
where the function is to be projected onto the span of the basis, yielding
a coefficient vector describing the result.

\begin{lstlisting}[style=Interface]
template <class Basis, class C, class F>
void interpolate(const Basis& basis, C&& coefficients, const F& f);
\end{lstlisting}
Note that this will only work if the range type of \cpp{f}
and the global basis \cpp{basis} are compatible.
\dunemodule{dune-functions} implements a compatibility layer
that allows to use different vector (or matrix) types
from the dune core modules and scalar types like, e.g. \cpp{double}
for the range of \cpp{f} as long as the number of scalar entries
of this range type is the same as the dimension of the range space of
the function space spanned by the basis.
This also implies the assumption that the coefficients for
individual basis functions are scalar.
The type of the coefficient vector \cpp{coefficients}
either has to implement the \cpp{VectorBackend}
concept or to be wrappable by the
\cpp{istlVectorBackend}.
For example, consider the function
\begin{equation*}
 f_1 : \R^2 \to \R
 \qquad
 f_1 =  \exp(-\norm{x}^2)
\end{equation*}
implemented as
\begin{lstlisting}[linerange=definition_f1_begin-definition_f1_end,
                 numbers=left]
// -*- tab-width: 4; indent-tabs-mode: nil; c-basic-offset: 2 -*-
// vi: set et ts=4 sw=2 sts=2:
#include <config.h>

#include <vector>

#include <dune/common/function.hh>
#include <dune/common/bitsetvector.hh>
#include <dune/common/tuplevector.hh>

#include <dune/geometry/quadraturerules.hh>

#include <dune/grid/yaspgrid.hh>
#include <dune/grid/io/file/vtk/subsamplingvtkwriter.hh>

#include <dune/istl/bvector.hh>
#include <dune/istl/multitypeblockvector.hh>

#include <dune/functions/functionspacebases/interpolate.hh>
#include <dune/functions/functionspacebases/powerbasis.hh>
#include <dune/functions/functionspacebases/compositebasis.hh>
#include <dune/functions/functionspacebases/lagrangebasis.hh>
#include <dune/functions/functionspacebases/subspacebasis.hh>
#include <dune/functions/functionspacebases/boundarydofs.hh>
#include <dune/functions/gridfunctions/discreteglobalbasisfunction.hh>
#include <dune/functions/gridfunctions/gridviewfunction.hh>

using namespace Dune;


int main (int argc, char *argv[])
{
  //Maybe initialize Mpi
  MPIHelper::instance(argc, argv);

  // make grid
  const int dim = 2;
  FieldVector<double,dim> L(1.0);
  std::array<int,dim> N = {{1,1}};
  std::bitset<dim> periodic(false);
  YaspGrid<2> grid(L,N);
  using GridView = YaspGrid<2>::LeafGridView;
  GridView gridView = grid.leafGridView();

  // Create closed-form function to interpolate
  // { definition_f1_begin }
  auto f1 = [](const FieldVector<double,2>& x)
  {
    return exp(-1.0*x.two_norm2());
  };
  // { definition_f1_end }

  // { definition_p2basis_begin }
  Functions::LagrangeBasis<GridView,2> p2basis(gridView);
  // { definition_p2basis_end }

  // { definition_x1_begin }
  std::vector<double> x1;
  // { definition_x1_end }

  // { interpolation1_begin }
  interpolate(p2basis, x1, f1);
  // { interpolation1_end }

  // Interpolate a vector-valued function using a vector-valued basis
  // { taylorhood_basis_begin }
  using namespace Functions::BasisBuilder;

  auto taylorHoodBasis = makeBasis(
    gridView,
    composite(
      power<dim>(
        lagrange<2>(),
        flatLexicographic()),
      lagrange<1>(),
      flatLexicographic()
    ));
  // { taylorhood_basis_end }

  // { taylorhood_vector_begin }
  BlockVector<FieldVector<double,1>> x2;
  // { taylorhood_vector_end }


  // { taylorhood_pressure_begin }
  using namespace Indices;
  interpolate(subspaceBasis(taylorHoodBasis, _1), x2, f1);
  // { taylorhood_pressure_end }

  // { taylorhood_velocity_begin }
  auto f2 = [](const FieldVector<double,2>& x) {
    return x;
  };
  interpolate(subspaceBasis(taylorHoodBasis, _0), x2, f2);
  // { taylorhood_velocity_end }

  // { setup_mask_begin }
  BlockVector<FieldVector<char,1>> isBoundary;
  auto isBoundaryBackend = Functions::istlVectorBackend(isBoundary);
  isBoundaryBackend.resize(taylorHoodBasis);
  isBoundary = false;
  forEachBoundaryDOF(subspaceBasis(taylorHoodBasis, _0),
    [&] (auto&& index) {
      isBoundaryBackend[index] = true;
    });
  // { setup_mask_end }

  // { masked_interpolation_begin }
  interpolate(subspaceBasis(taylorHoodBasis, _1), x2, f2, isBoundary);
  // { masked_interpolation_end }


  return 0;
}
\end{lstlisting}
Additionally, consider a scalar second-order Lagrange space
\begin{lstlisting}[linerange=definition_p2basis_begin-definition_p2basis_end,
                 numbers=left]
// -*- tab-width: 4; indent-tabs-mode: nil; c-basic-offset: 2 -*-
// vi: set et ts=4 sw=2 sts=2:
#include <config.h>

#include <vector>

#include <dune/common/function.hh>
#include <dune/common/bitsetvector.hh>
#include <dune/common/tuplevector.hh>

#include <dune/geometry/quadraturerules.hh>

#include <dune/grid/yaspgrid.hh>
#include <dune/grid/io/file/vtk/subsamplingvtkwriter.hh>

#include <dune/istl/bvector.hh>
#include <dune/istl/multitypeblockvector.hh>

#include <dune/functions/functionspacebases/interpolate.hh>
#include <dune/functions/functionspacebases/powerbasis.hh>
#include <dune/functions/functionspacebases/compositebasis.hh>
#include <dune/functions/functionspacebases/lagrangebasis.hh>
#include <dune/functions/functionspacebases/subspacebasis.hh>
#include <dune/functions/functionspacebases/boundarydofs.hh>
#include <dune/functions/gridfunctions/discreteglobalbasisfunction.hh>
#include <dune/functions/gridfunctions/gridviewfunction.hh>

using namespace Dune;


int main (int argc, char *argv[])
{
  //Maybe initialize Mpi
  MPIHelper::instance(argc, argv);

  // make grid
  const int dim = 2;
  FieldVector<double,dim> L(1.0);
  std::array<int,dim> N = {{1,1}};
  std::bitset<dim> periodic(false);
  YaspGrid<2> grid(L,N);
  using GridView = YaspGrid<2>::LeafGridView;
  GridView gridView = grid.leafGridView();

  // Create closed-form function to interpolate
  // { definition_f1_begin }
  auto f1 = [](const FieldVector<double,2>& x)
  {
    return exp(-1.0*x.two_norm2());
  };
  // { definition_f1_end }

  // { definition_p2basis_begin }
  Functions::LagrangeBasis<GridView,2> p2basis(gridView);
  // { definition_p2basis_end }

  // { definition_x1_begin }
  std::vector<double> x1;
  // { definition_x1_end }

  // { interpolation1_begin }
  interpolate(p2basis, x1, f1);
  // { interpolation1_end }

  // Interpolate a vector-valued function using a vector-valued basis
  // { taylorhood_basis_begin }
  using namespace Functions::BasisBuilder;

  auto taylorHoodBasis = makeBasis(
    gridView,
    composite(
      power<dim>(
        lagrange<2>(),
        flatLexicographic()),
      lagrange<1>(),
      flatLexicographic()
    ));
  // { taylorhood_basis_end }

  // { taylorhood_vector_begin }
  BlockVector<FieldVector<double,1>> x2;
  // { taylorhood_vector_end }


  // { taylorhood_pressure_begin }
  using namespace Indices;
  interpolate(subspaceBasis(taylorHoodBasis, _1), x2, f1);
  // { taylorhood_pressure_end }

  // { taylorhood_velocity_begin }
  auto f2 = [](const FieldVector<double,2>& x) {
    return x;
  };
  interpolate(subspaceBasis(taylorHoodBasis, _0), x2, f2);
  // { taylorhood_velocity_end }

  // { setup_mask_begin }
  BlockVector<FieldVector<char,1>> isBoundary;
  auto isBoundaryBackend = Functions::istlVectorBackend(isBoundary);
  isBoundaryBackend.resize(taylorHoodBasis);
  isBoundary = false;
  forEachBoundaryDOF(subspaceBasis(taylorHoodBasis, _0),
    [&] (auto&& index) {
      isBoundaryBackend[index] = true;
    });
  // { setup_mask_end }

  // { masked_interpolation_begin }
  interpolate(subspaceBasis(taylorHoodBasis, _1), x2, f2, isBoundary);
  // { masked_interpolation_end }


  return 0;
}
\end{lstlisting}
and an empty coefficient vector \cpp{x1}, not necessarily of correct size:
\begin{lstlisting}[linerange=definition_x1_begin-definition_x1_end,
                 numbers=left]
// -*- tab-width: 4; indent-tabs-mode: nil; c-basic-offset: 2 -*-
// vi: set et ts=4 sw=2 sts=2:
#include <config.h>

#include <vector>

#include <dune/common/function.hh>
#include <dune/common/bitsetvector.hh>
#include <dune/common/tuplevector.hh>

#include <dune/geometry/quadraturerules.hh>

#include <dune/grid/yaspgrid.hh>
#include <dune/grid/io/file/vtk/subsamplingvtkwriter.hh>

#include <dune/istl/bvector.hh>
#include <dune/istl/multitypeblockvector.hh>

#include <dune/functions/functionspacebases/interpolate.hh>
#include <dune/functions/functionspacebases/powerbasis.hh>
#include <dune/functions/functionspacebases/compositebasis.hh>
#include <dune/functions/functionspacebases/lagrangebasis.hh>
#include <dune/functions/functionspacebases/subspacebasis.hh>
#include <dune/functions/functionspacebases/boundarydofs.hh>
#include <dune/functions/gridfunctions/discreteglobalbasisfunction.hh>
#include <dune/functions/gridfunctions/gridviewfunction.hh>

using namespace Dune;


int main (int argc, char *argv[])
{
  //Maybe initialize Mpi
  MPIHelper::instance(argc, argv);

  // make grid
  const int dim = 2;
  FieldVector<double,dim> L(1.0);
  std::array<int,dim> N = {{1,1}};
  std::bitset<dim> periodic(false);
  YaspGrid<2> grid(L,N);
  using GridView = YaspGrid<2>::LeafGridView;
  GridView gridView = grid.leafGridView();

  // Create closed-form function to interpolate
  // { definition_f1_begin }
  auto f1 = [](const FieldVector<double,2>& x)
  {
    return exp(-1.0*x.two_norm2());
  };
  // { definition_f1_end }

  // { definition_p2basis_begin }
  Functions::LagrangeBasis<GridView,2> p2basis(gridView);
  // { definition_p2basis_end }

  // { definition_x1_begin }
  std::vector<double> x1;
  // { definition_x1_end }

  // { interpolation1_begin }
  interpolate(p2basis, x1, f1);
  // { interpolation1_end }

  // Interpolate a vector-valued function using a vector-valued basis
  // { taylorhood_basis_begin }
  using namespace Functions::BasisBuilder;

  auto taylorHoodBasis = makeBasis(
    gridView,
    composite(
      power<dim>(
        lagrange<2>(),
        flatLexicographic()),
      lagrange<1>(),
      flatLexicographic()
    ));
  // { taylorhood_basis_end }

  // { taylorhood_vector_begin }
  BlockVector<FieldVector<double,1>> x2;
  // { taylorhood_vector_end }


  // { taylorhood_pressure_begin }
  using namespace Indices;
  interpolate(subspaceBasis(taylorHoodBasis, _1), x2, f1);
  // { taylorhood_pressure_end }

  // { taylorhood_velocity_begin }
  auto f2 = [](const FieldVector<double,2>& x) {
    return x;
  };
  interpolate(subspaceBasis(taylorHoodBasis, _0), x2, f2);
  // { taylorhood_velocity_end }

  // { setup_mask_begin }
  BlockVector<FieldVector<char,1>> isBoundary;
  auto isBoundaryBackend = Functions::istlVectorBackend(isBoundary);
  isBoundaryBackend.resize(taylorHoodBasis);
  isBoundary = false;
  forEachBoundaryDOF(subspaceBasis(taylorHoodBasis, _0),
    [&] (auto&& index) {
      isBoundaryBackend[index] = true;
    });
  // { setup_mask_end }

  // { masked_interpolation_begin }
  interpolate(subspaceBasis(taylorHoodBasis, _1), x2, f2, isBoundary);
  // { masked_interpolation_end }


  return 0;
}
\end{lstlisting}
Then, the single line
\begin{lstlisting}[linerange=interpolation1_begin-interpolation1_end,
                 numbers=left]
// -*- tab-width: 4; indent-tabs-mode: nil; c-basic-offset: 2 -*-
// vi: set et ts=4 sw=2 sts=2:
#include <config.h>

#include <vector>

#include <dune/common/function.hh>
#include <dune/common/bitsetvector.hh>
#include <dune/common/tuplevector.hh>

#include <dune/geometry/quadraturerules.hh>

#include <dune/grid/yaspgrid.hh>
#include <dune/grid/io/file/vtk/subsamplingvtkwriter.hh>

#include <dune/istl/bvector.hh>
#include <dune/istl/multitypeblockvector.hh>

#include <dune/functions/functionspacebases/interpolate.hh>
#include <dune/functions/functionspacebases/powerbasis.hh>
#include <dune/functions/functionspacebases/compositebasis.hh>
#include <dune/functions/functionspacebases/lagrangebasis.hh>
#include <dune/functions/functionspacebases/subspacebasis.hh>
#include <dune/functions/functionspacebases/boundarydofs.hh>
#include <dune/functions/gridfunctions/discreteglobalbasisfunction.hh>
#include <dune/functions/gridfunctions/gridviewfunction.hh>

using namespace Dune;


int main (int argc, char *argv[])
{
  //Maybe initialize Mpi
  MPIHelper::instance(argc, argv);

  // make grid
  const int dim = 2;
  FieldVector<double,dim> L(1.0);
  std::array<int,dim> N = {{1,1}};
  std::bitset<dim> periodic(false);
  YaspGrid<2> grid(L,N);
  using GridView = YaspGrid<2>::LeafGridView;
  GridView gridView = grid.leafGridView();

  // Create closed-form function to interpolate
  // { definition_f1_begin }
  auto f1 = [](const FieldVector<double,2>& x)
  {
    return exp(-1.0*x.two_norm2());
  };
  // { definition_f1_end }

  // { definition_p2basis_begin }
  Functions::LagrangeBasis<GridView,2> p2basis(gridView);
  // { definition_p2basis_end }

  // { definition_x1_begin }
  std::vector<double> x1;
  // { definition_x1_end }

  // { interpolation1_begin }
  interpolate(p2basis, x1, f1);
  // { interpolation1_end }

  // Interpolate a vector-valued function using a vector-valued basis
  // { taylorhood_basis_begin }
  using namespace Functions::BasisBuilder;

  auto taylorHoodBasis = makeBasis(
    gridView,
    composite(
      power<dim>(
        lagrange<2>(),
        flatLexicographic()),
      lagrange<1>(),
      flatLexicographic()
    ));
  // { taylorhood_basis_end }

  // { taylorhood_vector_begin }
  BlockVector<FieldVector<double,1>> x2;
  // { taylorhood_vector_end }


  // { taylorhood_pressure_begin }
  using namespace Indices;
  interpolate(subspaceBasis(taylorHoodBasis, _1), x2, f1);
  // { taylorhood_pressure_end }

  // { taylorhood_velocity_begin }
  auto f2 = [](const FieldVector<double,2>& x) {
    return x;
  };
  interpolate(subspaceBasis(taylorHoodBasis, _0), x2, f2);
  // { taylorhood_velocity_end }

  // { setup_mask_begin }
  BlockVector<FieldVector<char,1>> isBoundary;
  auto isBoundaryBackend = Functions::istlVectorBackend(isBoundary);
  isBoundaryBackend.resize(taylorHoodBasis);
  isBoundary = false;
  forEachBoundaryDOF(subspaceBasis(taylorHoodBasis, _0),
    [&] (auto&& index) {
      isBoundaryBackend[index] = true;
    });
  // { setup_mask_end }

  // { masked_interpolation_begin }
  interpolate(subspaceBasis(taylorHoodBasis, _1), x2, f2, isBoundary);
  // { masked_interpolation_end }


  return 0;
}
\end{lstlisting}
will fill \cpp{x1} with the nodal values of the function \cpp{f1}.

This interpolation works equally well for non-trivial basis trees
and subtrees obtained by the \cpp{subspaceBasis} function,
provided that the coefficient vector matches the basis
and that the function range can be mapped to the product space
associated to the basis.  Suppose there is a Taylor--Hood basis for a
two-dimensional grid that uses flat multi-indices to label its
degrees of freedom:
\begin{lstlisting}[linerange=taylorhood_basis_begin-taylorhood_basis_end,
                 numbers=left]
// -*- tab-width: 4; indent-tabs-mode: nil; c-basic-offset: 2 -*-
// vi: set et ts=4 sw=2 sts=2:
#include <config.h>

#include <vector>

#include <dune/common/function.hh>
#include <dune/common/bitsetvector.hh>
#include <dune/common/tuplevector.hh>

#include <dune/geometry/quadraturerules.hh>

#include <dune/grid/yaspgrid.hh>
#include <dune/grid/io/file/vtk/subsamplingvtkwriter.hh>

#include <dune/istl/bvector.hh>
#include <dune/istl/multitypeblockvector.hh>

#include <dune/functions/functionspacebases/interpolate.hh>
#include <dune/functions/functionspacebases/powerbasis.hh>
#include <dune/functions/functionspacebases/compositebasis.hh>
#include <dune/functions/functionspacebases/lagrangebasis.hh>
#include <dune/functions/functionspacebases/subspacebasis.hh>
#include <dune/functions/functionspacebases/boundarydofs.hh>
#include <dune/functions/gridfunctions/discreteglobalbasisfunction.hh>
#include <dune/functions/gridfunctions/gridviewfunction.hh>

using namespace Dune;


int main (int argc, char *argv[])
{
  //Maybe initialize Mpi
  MPIHelper::instance(argc, argv);

  // make grid
  const int dim = 2;
  FieldVector<double,dim> L(1.0);
  std::array<int,dim> N = {{1,1}};
  std::bitset<dim> periodic(false);
  YaspGrid<2> grid(L,N);
  using GridView = YaspGrid<2>::LeafGridView;
  GridView gridView = grid.leafGridView();

  // Create closed-form function to interpolate
  // { definition_f1_begin }
  auto f1 = [](const FieldVector<double,2>& x)
  {
    return exp(-1.0*x.two_norm2());
  };
  // { definition_f1_end }

  // { definition_p2basis_begin }
  Functions::LagrangeBasis<GridView,2> p2basis(gridView);
  // { definition_p2basis_end }

  // { definition_x1_begin }
  std::vector<double> x1;
  // { definition_x1_end }

  // { interpolation1_begin }
  interpolate(p2basis, x1, f1);
  // { interpolation1_end }

  // Interpolate a vector-valued function using a vector-valued basis
  // { taylorhood_basis_begin }
  using namespace Functions::BasisBuilder;

  auto taylorHoodBasis = makeBasis(
    gridView,
    composite(
      power<dim>(
        lagrange<2>(),
        flatLexicographic()),
      lagrange<1>(),
      flatLexicographic()
    ));
  // { taylorhood_basis_end }

  // { taylorhood_vector_begin }
  BlockVector<FieldVector<double,1>> x2;
  // { taylorhood_vector_end }


  // { taylorhood_pressure_begin }
  using namespace Indices;
  interpolate(subspaceBasis(taylorHoodBasis, _1), x2, f1);
  // { taylorhood_pressure_end }

  // { taylorhood_velocity_begin }
  auto f2 = [](const FieldVector<double,2>& x) {
    return x;
  };
  interpolate(subspaceBasis(taylorHoodBasis, _0), x2, f2);
  // { taylorhood_velocity_end }

  // { setup_mask_begin }
  BlockVector<FieldVector<char,1>> isBoundary;
  auto isBoundaryBackend = Functions::istlVectorBackend(isBoundary);
  isBoundaryBackend.resize(taylorHoodBasis);
  isBoundary = false;
  forEachBoundaryDOF(subspaceBasis(taylorHoodBasis, _0),
    [&] (auto&& index) {
      isBoundaryBackend[index] = true;
    });
  // { setup_mask_end }

  // { masked_interpolation_begin }
  interpolate(subspaceBasis(taylorHoodBasis, _1), x2, f2, isBoundary);
  // { masked_interpolation_end }


  return 0;
}
\end{lstlisting}
A suitable coefficient vector for such a basis is for example
\begin{lstlisting}[linerange=taylorhood_vector_begin-taylorhood_vector_end,
                 numbers=left]
// -*- tab-width: 4; indent-tabs-mode: nil; c-basic-offset: 2 -*-
// vi: set et ts=4 sw=2 sts=2:
#include <config.h>

#include <vector>

#include <dune/common/function.hh>
#include <dune/common/bitsetvector.hh>
#include <dune/common/tuplevector.hh>

#include <dune/geometry/quadraturerules.hh>

#include <dune/grid/yaspgrid.hh>
#include <dune/grid/io/file/vtk/subsamplingvtkwriter.hh>

#include <dune/istl/bvector.hh>
#include <dune/istl/multitypeblockvector.hh>

#include <dune/functions/functionspacebases/interpolate.hh>
#include <dune/functions/functionspacebases/powerbasis.hh>
#include <dune/functions/functionspacebases/compositebasis.hh>
#include <dune/functions/functionspacebases/lagrangebasis.hh>
#include <dune/functions/functionspacebases/subspacebasis.hh>
#include <dune/functions/functionspacebases/boundarydofs.hh>
#include <dune/functions/gridfunctions/discreteglobalbasisfunction.hh>
#include <dune/functions/gridfunctions/gridviewfunction.hh>

using namespace Dune;


int main (int argc, char *argv[])
{
  //Maybe initialize Mpi
  MPIHelper::instance(argc, argv);

  // make grid
  const int dim = 2;
  FieldVector<double,dim> L(1.0);
  std::array<int,dim> N = {{1,1}};
  std::bitset<dim> periodic(false);
  YaspGrid<2> grid(L,N);
  using GridView = YaspGrid<2>::LeafGridView;
  GridView gridView = grid.leafGridView();

  // Create closed-form function to interpolate
  // { definition_f1_begin }
  auto f1 = [](const FieldVector<double,2>& x)
  {
    return exp(-1.0*x.two_norm2());
  };
  // { definition_f1_end }

  // { definition_p2basis_begin }
  Functions::LagrangeBasis<GridView,2> p2basis(gridView);
  // { definition_p2basis_end }

  // { definition_x1_begin }
  std::vector<double> x1;
  // { definition_x1_end }

  // { interpolation1_begin }
  interpolate(p2basis, x1, f1);
  // { interpolation1_end }

  // Interpolate a vector-valued function using a vector-valued basis
  // { taylorhood_basis_begin }
  using namespace Functions::BasisBuilder;

  auto taylorHoodBasis = makeBasis(
    gridView,
    composite(
      power<dim>(
        lagrange<2>(),
        flatLexicographic()),
      lagrange<1>(),
      flatLexicographic()
    ));
  // { taylorhood_basis_end }

  // { taylorhood_vector_begin }
  BlockVector<FieldVector<double,1>> x2;
  // { taylorhood_vector_end }


  // { taylorhood_pressure_begin }
  using namespace Indices;
  interpolate(subspaceBasis(taylorHoodBasis, _1), x2, f1);
  // { taylorhood_pressure_end }

  // { taylorhood_velocity_begin }
  auto f2 = [](const FieldVector<double,2>& x) {
    return x;
  };
  interpolate(subspaceBasis(taylorHoodBasis, _0), x2, f2);
  // { taylorhood_velocity_end }

  // { setup_mask_begin }
  BlockVector<FieldVector<char,1>> isBoundary;
  auto isBoundaryBackend = Functions::istlVectorBackend(isBoundary);
  isBoundaryBackend.resize(taylorHoodBasis);
  isBoundary = false;
  forEachBoundaryDOF(subspaceBasis(taylorHoodBasis, _0),
    [&] (auto&& index) {
      isBoundaryBackend[index] = true;
    });
  // { setup_mask_end }

  // { masked_interpolation_begin }
  interpolate(subspaceBasis(taylorHoodBasis, _1), x2, f2, isBoundary);
  // { masked_interpolation_end }


  return 0;
}
\end{lstlisting}
In this situation, interpolation of \cpp{f1} into the pressure components
of the Taylor--Hood basis can be achieved by
\begin{lstlisting}[linerange=taylorhood_pressure_begin-taylorhood_pressure_end,
                 numbers=left]
// -*- tab-width: 4; indent-tabs-mode: nil; c-basic-offset: 2 -*-
// vi: set et ts=4 sw=2 sts=2:
#include <config.h>

#include <vector>

#include <dune/common/function.hh>
#include <dune/common/bitsetvector.hh>
#include <dune/common/tuplevector.hh>

#include <dune/geometry/quadraturerules.hh>

#include <dune/grid/yaspgrid.hh>
#include <dune/grid/io/file/vtk/subsamplingvtkwriter.hh>

#include <dune/istl/bvector.hh>
#include <dune/istl/multitypeblockvector.hh>

#include <dune/functions/functionspacebases/interpolate.hh>
#include <dune/functions/functionspacebases/powerbasis.hh>
#include <dune/functions/functionspacebases/compositebasis.hh>
#include <dune/functions/functionspacebases/lagrangebasis.hh>
#include <dune/functions/functionspacebases/subspacebasis.hh>
#include <dune/functions/functionspacebases/boundarydofs.hh>
#include <dune/functions/gridfunctions/discreteglobalbasisfunction.hh>
#include <dune/functions/gridfunctions/gridviewfunction.hh>

using namespace Dune;


int main (int argc, char *argv[])
{
  //Maybe initialize Mpi
  MPIHelper::instance(argc, argv);

  // make grid
  const int dim = 2;
  FieldVector<double,dim> L(1.0);
  std::array<int,dim> N = {{1,1}};
  std::bitset<dim> periodic(false);
  YaspGrid<2> grid(L,N);
  using GridView = YaspGrid<2>::LeafGridView;
  GridView gridView = grid.leafGridView();

  // Create closed-form function to interpolate
  // { definition_f1_begin }
  auto f1 = [](const FieldVector<double,2>& x)
  {
    return exp(-1.0*x.two_norm2());
  };
  // { definition_f1_end }

  // { definition_p2basis_begin }
  Functions::LagrangeBasis<GridView,2> p2basis(gridView);
  // { definition_p2basis_end }

  // { definition_x1_begin }
  std::vector<double> x1;
  // { definition_x1_end }

  // { interpolation1_begin }
  interpolate(p2basis, x1, f1);
  // { interpolation1_end }

  // Interpolate a vector-valued function using a vector-valued basis
  // { taylorhood_basis_begin }
  using namespace Functions::BasisBuilder;

  auto taylorHoodBasis = makeBasis(
    gridView,
    composite(
      power<dim>(
        lagrange<2>(),
        flatLexicographic()),
      lagrange<1>(),
      flatLexicographic()
    ));
  // { taylorhood_basis_end }

  // { taylorhood_vector_begin }
  BlockVector<FieldVector<double,1>> x2;
  // { taylorhood_vector_end }


  // { taylorhood_pressure_begin }
  using namespace Indices;
  interpolate(subspaceBasis(taylorHoodBasis, _1), x2, f1);
  // { taylorhood_pressure_end }

  // { taylorhood_velocity_begin }
  auto f2 = [](const FieldVector<double,2>& x) {
    return x;
  };
  interpolate(subspaceBasis(taylorHoodBasis, _0), x2, f2);
  // { taylorhood_velocity_end }

  // { setup_mask_begin }
  BlockVector<FieldVector<char,1>> isBoundary;
  auto isBoundaryBackend = Functions::istlVectorBackend(isBoundary);
  isBoundaryBackend.resize(taylorHoodBasis);
  isBoundary = false;
  forEachBoundaryDOF(subspaceBasis(taylorHoodBasis, _0),
    [&] (auto&& index) {
      isBoundaryBackend[index] = true;
    });
  // { setup_mask_end }

  // { masked_interpolation_begin }
  interpolate(subspaceBasis(taylorHoodBasis, _1), x2, f2, isBoundary);
  // { masked_interpolation_end }


  return 0;
}
\end{lstlisting}
Similarly we can interpolate a given vector field \cpp{f2}
into the non-trivial subtree representing the velocity using
\begin{lstlisting}[linerange=taylorhood_velocity_begin-taylorhood_velocity_end,
                 numbers=left]
// -*- tab-width: 4; indent-tabs-mode: nil; c-basic-offset: 2 -*-
// vi: set et ts=4 sw=2 sts=2:
#include <config.h>

#include <vector>

#include <dune/common/function.hh>
#include <dune/common/bitsetvector.hh>
#include <dune/common/tuplevector.hh>

#include <dune/geometry/quadraturerules.hh>

#include <dune/grid/yaspgrid.hh>
#include <dune/grid/io/file/vtk/subsamplingvtkwriter.hh>

#include <dune/istl/bvector.hh>
#include <dune/istl/multitypeblockvector.hh>

#include <dune/functions/functionspacebases/interpolate.hh>
#include <dune/functions/functionspacebases/powerbasis.hh>
#include <dune/functions/functionspacebases/compositebasis.hh>
#include <dune/functions/functionspacebases/lagrangebasis.hh>
#include <dune/functions/functionspacebases/subspacebasis.hh>
#include <dune/functions/functionspacebases/boundarydofs.hh>
#include <dune/functions/gridfunctions/discreteglobalbasisfunction.hh>
#include <dune/functions/gridfunctions/gridviewfunction.hh>

using namespace Dune;


int main (int argc, char *argv[])
{
  //Maybe initialize Mpi
  MPIHelper::instance(argc, argv);

  // make grid
  const int dim = 2;
  FieldVector<double,dim> L(1.0);
  std::array<int,dim> N = {{1,1}};
  std::bitset<dim> periodic(false);
  YaspGrid<2> grid(L,N);
  using GridView = YaspGrid<2>::LeafGridView;
  GridView gridView = grid.leafGridView();

  // Create closed-form function to interpolate
  // { definition_f1_begin }
  auto f1 = [](const FieldVector<double,2>& x)
  {
    return exp(-1.0*x.two_norm2());
  };
  // { definition_f1_end }

  // { definition_p2basis_begin }
  Functions::LagrangeBasis<GridView,2> p2basis(gridView);
  // { definition_p2basis_end }

  // { definition_x1_begin }
  std::vector<double> x1;
  // { definition_x1_end }

  // { interpolation1_begin }
  interpolate(p2basis, x1, f1);
  // { interpolation1_end }

  // Interpolate a vector-valued function using a vector-valued basis
  // { taylorhood_basis_begin }
  using namespace Functions::BasisBuilder;

  auto taylorHoodBasis = makeBasis(
    gridView,
    composite(
      power<dim>(
        lagrange<2>(),
        flatLexicographic()),
      lagrange<1>(),
      flatLexicographic()
    ));
  // { taylorhood_basis_end }

  // { taylorhood_vector_begin }
  BlockVector<FieldVector<double,1>> x2;
  // { taylorhood_vector_end }


  // { taylorhood_pressure_begin }
  using namespace Indices;
  interpolate(subspaceBasis(taylorHoodBasis, _1), x2, f1);
  // { taylorhood_pressure_end }

  // { taylorhood_velocity_begin }
  auto f2 = [](const FieldVector<double,2>& x) {
    return x;
  };
  interpolate(subspaceBasis(taylorHoodBasis, _0), x2, f2);
  // { taylorhood_velocity_end }

  // { setup_mask_begin }
  BlockVector<FieldVector<char,1>> isBoundary;
  auto isBoundaryBackend = Functions::istlVectorBackend(isBoundary);
  isBoundaryBackend.resize(taylorHoodBasis);
  isBoundary = false;
  forEachBoundaryDOF(subspaceBasis(taylorHoodBasis, _0),
    [&] (auto&& index) {
      isBoundaryBackend[index] = true;
    });
  // { setup_mask_end }

  // { masked_interpolation_begin }
  interpolate(subspaceBasis(taylorHoodBasis, _1), x2, f2, isBoundary);
  // { masked_interpolation_end }


  return 0;
}
\end{lstlisting}
It is even possible to interpolate into the full \cpp{taylorHoodBasis}
if the range type of the provided function has the same nesting structure
as the basis.

In some situations it is also desirable to interpolate only on a part of the domain.  Algebraically, the interpolation
is then performed as before, but only a subset of all coefficients are written.  The most frequent use-case is the interpolation
of Dirichlet data onto the algebraic degrees of freedom on the Dirichlet boundary.  All others
degrees of freedom must not be touched, as they contain, e.g., a suitable initial iterate obtained by some other
means.

To support this kind of interpolation, a variant of the
\cpp{interpolate} method allows to explicitly mark a subset of
coefficient vector entries to be written.
\begin{lstlisting}[style=Interface]
template <class B, class C, class F, class BV>
void interpolate(const B& basis,
        C&& coefficients,
        const F& f,
        const BV& bitVector);
\end{lstlisting}
Conceptually, the additional \cpp{bitVector}
argument must be a container of booleans having
the same nesting structure as \cpp{coefficients}.
Its entries are treated as boolean
values indicating if the corresponding entry of \cpp{coefficients}
should be written.
For example, for flat global indices \cpp{std::vector<bool>} and
\cpp{std::vector<char>} work nicely.
The class \cpp{BitSetVector<N>} (from the \dunemodule{dune-common} module) can be used
as a space-optimized alternative to \cpp{std::vector<std::bitset<N>>}.
For example, to interpolate the \cpp{f2} function defined above
only into the boundary velocity degrees of freedom,
first set up a suitable bit-vector:
\begin{lstlisting}[linerange=setup_mask_begin-setup_mask_end,
                 numbers=left]
// -*- tab-width: 4; indent-tabs-mode: nil; c-basic-offset: 2 -*-
// vi: set et ts=4 sw=2 sts=2:
#include <config.h>

#include <vector>

#include <dune/common/function.hh>
#include <dune/common/bitsetvector.hh>
#include <dune/common/tuplevector.hh>

#include <dune/geometry/quadraturerules.hh>

#include <dune/grid/yaspgrid.hh>
#include <dune/grid/io/file/vtk/subsamplingvtkwriter.hh>

#include <dune/istl/bvector.hh>
#include <dune/istl/multitypeblockvector.hh>

#include <dune/functions/functionspacebases/interpolate.hh>
#include <dune/functions/functionspacebases/powerbasis.hh>
#include <dune/functions/functionspacebases/compositebasis.hh>
#include <dune/functions/functionspacebases/lagrangebasis.hh>
#include <dune/functions/functionspacebases/subspacebasis.hh>
#include <dune/functions/functionspacebases/boundarydofs.hh>
#include <dune/functions/gridfunctions/discreteglobalbasisfunction.hh>
#include <dune/functions/gridfunctions/gridviewfunction.hh>

using namespace Dune;


int main (int argc, char *argv[])
{
  //Maybe initialize Mpi
  MPIHelper::instance(argc, argv);

  // make grid
  const int dim = 2;
  FieldVector<double,dim> L(1.0);
  std::array<int,dim> N = {{1,1}};
  std::bitset<dim> periodic(false);
  YaspGrid<2> grid(L,N);
  using GridView = YaspGrid<2>::LeafGridView;
  GridView gridView = grid.leafGridView();

  // Create closed-form function to interpolate
  // { definition_f1_begin }
  auto f1 = [](const FieldVector<double,2>& x)
  {
    return exp(-1.0*x.two_norm2());
  };
  // { definition_f1_end }

  // { definition_p2basis_begin }
  Functions::LagrangeBasis<GridView,2> p2basis(gridView);
  // { definition_p2basis_end }

  // { definition_x1_begin }
  std::vector<double> x1;
  // { definition_x1_end }

  // { interpolation1_begin }
  interpolate(p2basis, x1, f1);
  // { interpolation1_end }

  // Interpolate a vector-valued function using a vector-valued basis
  // { taylorhood_basis_begin }
  using namespace Functions::BasisBuilder;

  auto taylorHoodBasis = makeBasis(
    gridView,
    composite(
      power<dim>(
        lagrange<2>(),
        flatLexicographic()),
      lagrange<1>(),
      flatLexicographic()
    ));
  // { taylorhood_basis_end }

  // { taylorhood_vector_begin }
  BlockVector<FieldVector<double,1>> x2;
  // { taylorhood_vector_end }


  // { taylorhood_pressure_begin }
  using namespace Indices;
  interpolate(subspaceBasis(taylorHoodBasis, _1), x2, f1);
  // { taylorhood_pressure_end }

  // { taylorhood_velocity_begin }
  auto f2 = [](const FieldVector<double,2>& x) {
    return x;
  };
  interpolate(subspaceBasis(taylorHoodBasis, _0), x2, f2);
  // { taylorhood_velocity_end }

  // { setup_mask_begin }
  BlockVector<FieldVector<char,1>> isBoundary;
  auto isBoundaryBackend = Functions::istlVectorBackend(isBoundary);
  isBoundaryBackend.resize(taylorHoodBasis);
  isBoundary = false;
  forEachBoundaryDOF(subspaceBasis(taylorHoodBasis, _0),
    [&] (auto&& index) {
      isBoundaryBackend[index] = true;
    });
  // { setup_mask_end }

  // { masked_interpolation_begin }
  interpolate(subspaceBasis(taylorHoodBasis, _1), x2, f2, isBoundary);
  // { masked_interpolation_end }


  return 0;
}
\end{lstlisting}
The actual interpolation is then a single line:
\begin{lstlisting}[linerange=masked_interpolation_begin-masked_interpolation_end,
                 numbers=left]
// -*- tab-width: 4; indent-tabs-mode: nil; c-basic-offset: 2 -*-
// vi: set et ts=4 sw=2 sts=2:
#include <config.h>

#include <vector>

#include <dune/common/function.hh>
#include <dune/common/bitsetvector.hh>
#include <dune/common/tuplevector.hh>

#include <dune/geometry/quadraturerules.hh>

#include <dune/grid/yaspgrid.hh>
#include <dune/grid/io/file/vtk/subsamplingvtkwriter.hh>

#include <dune/istl/bvector.hh>
#include <dune/istl/multitypeblockvector.hh>

#include <dune/functions/functionspacebases/interpolate.hh>
#include <dune/functions/functionspacebases/powerbasis.hh>
#include <dune/functions/functionspacebases/compositebasis.hh>
#include <dune/functions/functionspacebases/lagrangebasis.hh>
#include <dune/functions/functionspacebases/subspacebasis.hh>
#include <dune/functions/functionspacebases/boundarydofs.hh>
#include <dune/functions/gridfunctions/discreteglobalbasisfunction.hh>
#include <dune/functions/gridfunctions/gridviewfunction.hh>

using namespace Dune;


int main (int argc, char *argv[])
{
  //Maybe initialize Mpi
  MPIHelper::instance(argc, argv);

  // make grid
  const int dim = 2;
  FieldVector<double,dim> L(1.0);
  std::array<int,dim> N = {{1,1}};
  std::bitset<dim> periodic(false);
  YaspGrid<2> grid(L,N);
  using GridView = YaspGrid<2>::LeafGridView;
  GridView gridView = grid.leafGridView();

  // Create closed-form function to interpolate
  // { definition_f1_begin }
  auto f1 = [](const FieldVector<double,2>& x)
  {
    return exp(-1.0*x.two_norm2());
  };
  // { definition_f1_end }

  // { definition_p2basis_begin }
  Functions::LagrangeBasis<GridView,2> p2basis(gridView);
  // { definition_p2basis_end }

  // { definition_x1_begin }
  std::vector<double> x1;
  // { definition_x1_end }

  // { interpolation1_begin }
  interpolate(p2basis, x1, f1);
  // { interpolation1_end }

  // Interpolate a vector-valued function using a vector-valued basis
  // { taylorhood_basis_begin }
  using namespace Functions::BasisBuilder;

  auto taylorHoodBasis = makeBasis(
    gridView,
    composite(
      power<dim>(
        lagrange<2>(),
        flatLexicographic()),
      lagrange<1>(),
      flatLexicographic()
    ));
  // { taylorhood_basis_end }

  // { taylorhood_vector_begin }
  BlockVector<FieldVector<double,1>> x2;
  // { taylorhood_vector_end }


  // { taylorhood_pressure_begin }
  using namespace Indices;
  interpolate(subspaceBasis(taylorHoodBasis, _1), x2, f1);
  // { taylorhood_pressure_end }

  // { taylorhood_velocity_begin }
  auto f2 = [](const FieldVector<double,2>& x) {
    return x;
  };
  interpolate(subspaceBasis(taylorHoodBasis, _0), x2, f2);
  // { taylorhood_velocity_end }

  // { setup_mask_begin }
  BlockVector<FieldVector<char,1>> isBoundary;
  auto isBoundaryBackend = Functions::istlVectorBackend(isBoundary);
  isBoundaryBackend.resize(taylorHoodBasis);
  isBoundary = false;
  forEachBoundaryDOF(subspaceBasis(taylorHoodBasis, _0),
    [&] (auto&& index) {
      isBoundaryBackend[index] = true;
    });
  // { setup_mask_end }

  // { masked_interpolation_begin }
  interpolate(subspaceBasis(taylorHoodBasis, _1), x2, f2, isBoundary);
  // { masked_interpolation_end }


  return 0;
}
\end{lstlisting}
See Chapter~\ref{sec:stokes_example_main} for a more involved example.
The \cpp{forEachBoundaryDOF} method that loops over all boundary degrees of freedom
is defined in the file~\file{dune/functions/\allowbreak functionspacebases/boundarydofs.hh}.

\section{Example: Solving the Stokes equation with \dunemodule{dune-functions}}

We close the paper by showing a complete example program that demonstrates a lot
of the techniques presented so far.  The example program will solve the
stationary Stokes problem using Taylor--Hood finite elements.
The example is contained in a single file, which comes as part of the \dunemodule{dune-functions}
source tree, in \file{dune-functions/examples/stokes-taylorhood.cc}.  If you read this document in electronic form,
the file can also be accessed by clicking on the icon in the margin.%
\marginpar{\attachfile[author=The dune-functions team,
                       color = 1 0 0,
                       mimetype=text/plain,
                       description=Complete source code of the Stokes/Taylor-Hood example]
                       {stokes-taylorhood.cc}}

\subsection{The Stokes equation}

\begin{figure}
 \begin{center}
  \begin{overpic}[height=0.25\textheight]{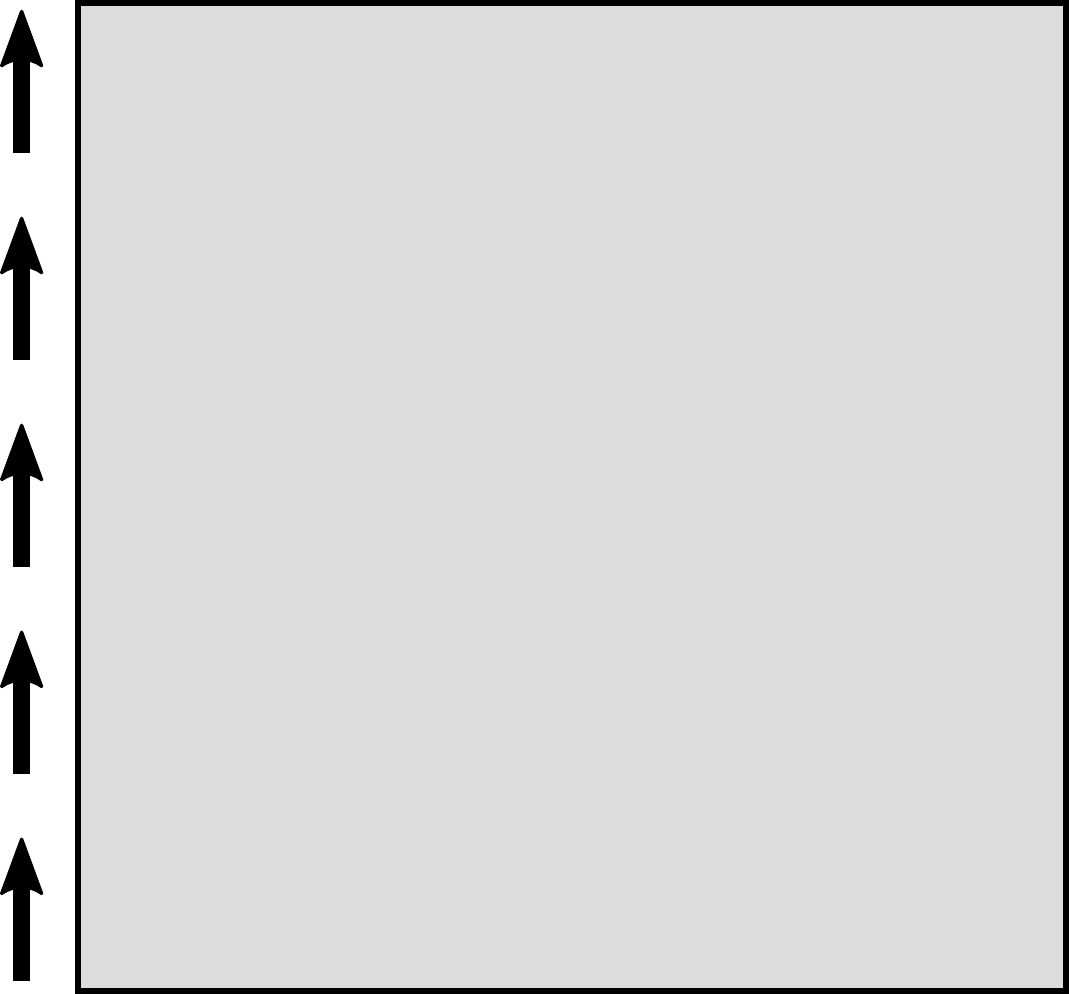}
   \put(50,45){$\Omega$}
  \end{overpic}
  \qquad
  \includegraphics[height=0.25\textheight]{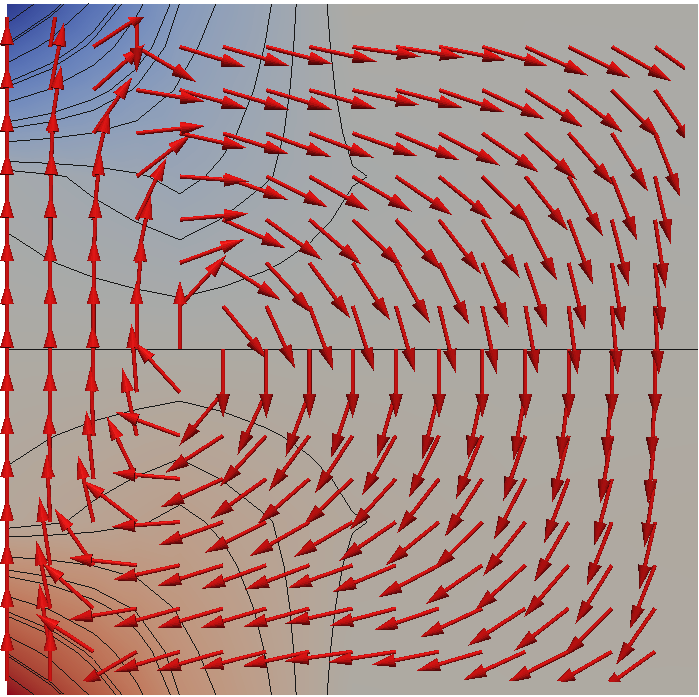}
 \end{center}
 \caption{Driven cavity. Left: domain and boundary conditions;
     right: simulation result.  The arrows show the {\em normalized} velocity.}
 \label{fig:driven_cavity}
\end{figure}

The Stokes equation models a viscous incompressible
fluid in a $d$-dimensional domain~$\Omega$.
There are two unknowns in this problem: a stationary
fluid velocity field $\mathbf{u} : \Omega \to \R^d$, and the fluid pressure $p : \Omega \to \R$.
Together, they have to solve the boundary value problem
\begin{alignat*}{2}
 -\Delta \mathbf{u} - \nabla p & = 0  & \qquad & \text{in $\Omega$}, \\
 \div \mathbf{u} & = 0                &        & \text{in $\Omega$}, \\
                    \mathbf{u} & = \mathbf{g}  &        & \text{on $\partial \Omega$},
\end{alignat*}
where we have omitted the physical parameters.  The boundary value problem only determines the
pressure $p$ up to a constant function.  The pressure is therefore usually normalized such
that $\int_\Omega p\,dx = 0$.

Due to the constraint $\div \mathbf{u} = 0$, the corresponding weak form of the equation is a saddle-point problem.
Introduce the spaces
\begin{align*}
  \mathbf{H}^1_{\mathbf{g}}(\Omega)
      & \colonequals
      \big\{ \mathbf{v} \in H^1(\Omega,\R^d) \; :\; \operatorname{tr}{\mathbf{v}} = \mathbf{g} \big\}, \\
 L_{2,0}(\Omega) & \colonequals  \Big\{ q \in L_2(\Omega) \; :\; \int_\Omega q\,dx = 0 \Big\},
\end{align*}
and the bilinear forms
\begin{equation*}
 a(\mathbf{u},\mathbf{v}) \colonequals \int_\Omega \nabla \mathbf{u} \nabla \mathbf{v} \,dx,
 \qquad \text{and} \qquad
 b(\mathbf{v},q) \colonequals \int_\Omega \div \mathbf{v} \cdot q \,dx.
\end{equation*}
Then the weak form of the Stokes equation is: Find $(\mathbf{u},p) \in \mathbf{H}_{\mathbf{g}}^1(\Omega) \times L_{2,0}(\Omega)$ such that
\begin{alignat*}{2}
 a(\mathbf{u},\mathbf{v}) + b(\mathbf{v},p) & = 0 & \qquad & \text{for all $\mathbf{v} \in \mathbf{H}_0^1(\Omega)$} \\
 b(\mathbf{u},q)\qquad\qquad & = 0       &        & \text{for all $q \in L_{2,0}(\Omega)$}.
\end{alignat*}
If $\mathbf{g}$ is sufficiently smooth, this variational problem has a unique solution.
The Taylor--Hood element is the standard way to discretize this saddle point problem~\cite{braess:2013},
and will be used in the following implementation.

\subsection{The driven-cavity benchmark}

For our example we choose to simulate a two-dimensional driven-cavity.  This is a standard benchmark
for the Stokes problem in the literature \cite{schreiber1983driven}.
Let $\Omega$ be the unit square $[0,1]^2$, and set the Dirichlet
boundary conditions for the velocity $\mathbf{u}$ to
\begin{equation*}
 \mathbf{u}(x)
 =\mathbf{g}(x)
 =
 \begin{cases}
  (0,1)^T & \text{if $x \in \{0\} \times [0,1]$} \\
  (0,0)^T & \text{elsewhere on $\partial \Omega$}.
 \end{cases}
\end{equation*}
The interpretation of this is a fluid container that is closed on all but one side.  While the fluid remains
motionless on the closed sides, an external agent drives a constant upward motion on the left vertical side.
The domain and boundary conditions are depicted in Figure~\ref{fig:driven_cavity}, left.
The corresponding solution is shown on the right side of the same figure.  The velocity forms a vortex,
while the pressure forms extrema in the two left corners.

\subsection{Implementation}

The implementation consists of an assembler for the Stokes problem and a \cpp{main}
method. Both will be discussed in the following.

\subsubsection{The \texorpdfstring{\cpp{main}}{main} method}
\label{sec:stokes_example_main}

The \cpp{main} method sets up the algebraic Stokes problem, calls a linear solver,
and writes the result to a VTK file. It begins by setting up MPI and the grid.
We choose to discretize the domain using a structured $4 \times 4$ quadrilateral
grid, which we get by using the \cpp{YaspGrid} grid implementation from the
\dunemodule{dune-grid} module.  Note that there is the line
\lstinputlisting[linerange={using_namespace_dune_begin-using_namespace_dune_end},
                 numbers=left]{\PATH/stokes-taylorhood.cc}
at the top of the file, so this namespace is imported completely.  Additionally, everything in the \dunemodule{dune-functions}
module is contained in the namespace \cpp{Functions}.  This namespace is not imported; instead, the prefix \cpp{Functions::} is always
given explicitly.

\lstinputlisting[linerange={main_begin-mpi_setup_end},
                 numbers=left]{\PATH/stokes-taylorhood.cc}
\lstinputlisting[linerange={grid_setup_begin-grid_setup_end},
                 numbers=left]{\PATH/stokes-taylorhood.cc}
The \cpp{gridView} object is the non-hierarchical finite element grid that we will use for
the computation.
On this grid view, we then set up the function space basis for the Taylor--Hood element.  This is as simple as
\lstinputlisting[linerange={function_space_basis_begin-function_space_basis_end},
                 numbers=left]{\PATH/stokes-taylorhood.cc}
This way of constructing a Taylor--Hood basis from instances of Lagrange bases
has been discussed in Section~\ref{sec:composed_bases}.  The indexing strategies
used here are \cpp{BlockedInterleaved} for the velocity subtree and \cpp{BlockedLexicographic}
(the default) for the root.  This results in the indexing scheme spelled out in
Column~2 of Table~\ref{tab:th_indexing_variants}.

Before being able to assemble the stiffness matrix of the Stokes system we need to pick suitable data structures
for the linear algebra.
These data structures should have a blocking structure that matches the multi-indices
used by the Taylor--Hood basis we just constructed.  More concretely,
the appropriate vector container will be a pair of vectors, where the first one
has entries in $\mathbb{R}^d$ for the velocity and the second one has entries in $\mathbb{R}$
for the pressure degrees of freedom.  Analogously, the matrix must consist of
$2 \times 2$ large sparse matrices where
the $(0,0)$-block has entries in $\R^{d\times d}$,
the $(0,1)$-block has entries in $\R^{d\times 1}$,
the $(1,0)$-block has entries in $\R^{1\times d}$,
and the $(1,1)$-block has entries in $\R$
(as on the right side of Figure~\ref{fig:matrix_occupation_patterns}).
The following code sets up such vector and matrix types for this.
It uses the nesting machinery from \dunemodule{dune-istl},
but data types from other linear algebra libraries could be used as well.
\lstinputlisting[linerange={linear_algebra_setup_begin-linear_algebra_setup_end},
                 numbers=left]{\PATH/stokes-taylorhood.cc}
Note that \cpp{VectorType} and \cpp{MatrixType}
are no classical containers, because the entries have non-uniform types.
Rather, they are constructed similarly to \cpp{std::tuple} from the C++ standard library.
However, it must be emphasized that the use of such advanced data structures is by no
means mandatory. As detailed in Sections~\ref{sec:index_strategies}
and~\ref{sec:composed_bases} it is trivial to make the Taylor--Hood basis
use flat global indices, which work directly with standard container types like \cpp{std::vector}.

Now that we have chosen the C++ types for the matrix and vector data structures we can actually assemble the system.
Assembling the right-hand-side vector \cpp{rhs} is easy, because, apart from the Dirichlet boundary data (which we
will insert later), all its entries are zero.  An all-zero vector of the correct type and size is set up by the
following lines
\lstinputlisting[linerange={rhs_assembly_begin-rhs_assembly_end},
                 numbers=left]{\PATH/stokes-taylorhood.cc}
The object returned by \cpp{istlVectorBackend} connects the \dunemodule{dune-functions}
basis with \dunemodule{dune-istl} linear algebra containers.
In particular, it offers convenient resizing of an entire hierarchy of nested vectors
from given function space basis trees.
Line~\ref{li:stokes_taylorhood_set_rhs_to_zero} fills the entire vector with zeros
in one go, but observe that this is actually a \dunemodule{dune-istl} feature.

To obtain the stiffness matrix we first create an empty matrix object of the correct type.  The actual assembly
is factored out into a separate method.
\lstinputlisting[linerange={matrix_assembly_begin-matrix_assembly_end},
                 numbers=left]{\PATH/stokes-taylorhood.cc}
As the matrix assembly is a central part of this example we explain it in detail below, after having covered the \cpp{main} method.

Suppose now that we have the correct stiffness matrix assembled in the object \cpp{stiffnessMatrix}.  We still need
to modify the linear system to include the Dirichlet boundary information.
In a first step we need to determine all degrees of freedom with Dirichlet boundary conditions.
To store this information we use a vector of flags which has the same structure
as \cpp{VectorType} and is again initialized using the \cpp{ISTLVectorBackend}.
\lstinputlisting[linerange={initialize_boundary_dofs_vector_begin-initialize_boundary_dofs_vector_end},
                 numbers=left]{\PATH/stokes-taylorhood.cc}
We now want to mark all the velocity degrees of freedom on the Dirichlet boundary.
In the driven-cavity example, the entire boundary is Dirichlet boundary.
For convenience, \dunemodule{dune-functions} provides the method
\begin{lstlisting}[style=Interface]
template<class Basis, class F>
void forEachBoundaryDOF(const Basis& basis, F&& f);
\end{lstlisting}
(in the file \file{dune/functions/functionspacebases/boundarydofs.hh}),
which implements a loop over all degrees of freedom associated to entities located
on the domain boundary. The algorithm will invoke the callback function \cpp{f}
for each such degree of freedom, passing its global index as the callback argument.
To mark the boundary degrees of freedom for the velocity subtree write:
\lstinputlisting[linerange={determine_boundary_dofs_begin-determine_boundary_dofs_end},
                 numbers=left]{\PATH/stokes-taylorhood.cc}
The \cpp{forEachBoundaryDOF} algorithm only considers velocity
degrees of freedom because we called it with the corresponding subspace basis.
Nevertheless, the global indices handed out by \cpp{forEachBoundaryDOF} correspond
to the full tree, and can therefore by used to access the \cpp{isBoundary}
container via the \cpp{ISTLVectorBackend} (Section~\ref{sec:subtrees}).

Now that we have determined the set of Dirichlet degrees of freedom,
we define a method implementing the actual Dirichlet values function $\mathbf{g}$, and interpolate
that into the right-hand-side vector \cpp{rhs}.
\lstinputlisting[linerange={interpolate_dirichlet_values_begin-interpolate_dirichlet_values_end},
                 numbers=left]{\PATH/stokes-taylorhood.cc}
Observe how the \dunemodule{dune-functions} interface allows to interpolate C++11 lambdas, which makes the code
very short and readable. Again the operation is constrained to the velocity degrees of freedom
by passing the corresponding subspace basis only.
The \cpp{isBoundary} vector given as the last argument restricts the interpolation
to only the boundary degrees of freedom which we marked before.

The stiffness matrix is modified in a more manual fashion.  For each Dirichlet degree of freedom we need to fill the corresponding matrix row
with zeros, and write a~1 on the diagonal.
The following algorithm does this by looping over all grid elements, and for each element
looping over all Dirichlet degrees of freedom.  This is less efficient than simply
looping over all matrix rows, but it allows to avoid implementing iterators for the
nested sparse matrix data types.
\lstinputlisting[linerange={set_dirichlet_matrix_begin-set_dirichlet_matrix_end},
                 numbers=left]{\PATH/stokes-taylorhood.cc}
Access to the matrix entries needs a matrix analogue to the vector backends---a translation
layer that converts multi-indices to the correct sequence of instructions required to
access the matrix data structure.  Such a backend is more challenging to write, as it
requires handling row and column indices at the same time.  At the time of writing \dunemodule{dune-functions}
does not provide matrix backends.  Instead, the example code uses a small helper method
\cpp{matrixEntry} that is defined in the example file itself.
It is much simpler than a generic matrix backend, because it is written
directly for the matrix data type of the Stokes problem.
\lstinputlisting[linerange={matrixentry_begin-matrixentry_end},
                 numbers=left]{\PATH/stokes-taylorhood.cc}
Notice that the outer indices for the \cpp{MultiTypeBlockMatrix} have to be
encoded statically, because different matrix entries have different types.
The additional \cpp{[0][0]} in Line~\ref{li:matrixentry_pressure_pressure} is necessary
because the entries of the lower-right matrix diagonal block are of type \cpp{FieldMatrix<double,1,1>}
(see Line~\ref{li:matrix_type_pressure_pressure}), and the \cpp{[0][0]} is needed to get the \cpp{double} from that.
Similarly the entries of the $(0,1)$- and $(1,0)$-blocks of the matrix are
interpreted as single-column and single-row matrices, respectively, such that
corresponding \cpp{[0]} indices have to be inserted.

Finally, we can solve the linear system.
Dedicated Stokes solvers frequently operate on some sort of Schur complement,
and hence they need direct access to the submatrices~\cite{john:2016}.
This can be elegantly done using the nested matrix type used in this example.
However, efficiently solving the Stokes system is an art,
which we do not want to get into here.
Instead, we use a GMRes solver, without any preconditioner at all.  This is known to converge,
albeit slowly.
The advantage is that it can be written down in very few lines.
The following code shows a typical way of using \dunemodule{dune-istl} to solve
a linear system of equations, and is not particular to \dunemodule{dune-functions}
at all.
\lstinputlisting[linerange={stokes_solve_begin-stokes_solve_end},
                 numbers=left]{\PATH/stokes-taylorhood.cc}
Observe how the \cpp{RestartedGMResSolver} object is completely oblivious to the fact that the matrix
has a nesting structure.

Once the iterative solver has terminated, the result is written to a VTK file.
However, as the \cpp{VTKWriter} class from the \dunemodule{dune-grid} module
expects discrete functions rather than coefficient vectors, we construct
velocity and pressure discrete functions by combining the appropriate
coefficient vectors and basis subtrees:
\lstinputlisting[linerange={make_result_functions_begin-make_result_functions_end},
                 numbers=left]{\PATH/stokes-taylorhood.cc}
Then, we write the resulting velocity as a vector field,
and the resulting pressure as a scalar field.  We subsample the grid twice, because the \cpp{VTKWriter}
class natively only displays piecewise linear functions.
\lstinputlisting[linerange={vtk_output_begin-vtk_output_end},
                 numbers=left]{\PATH/stokes-taylorhood.cc}
When run, this program produces a file called \file{stokes-taylorhood-result.vtu}.  The file can be opened in
\program{ParaView}, and the outcome looks like the image on the right in Figure~\ref{fig:driven_cavity}.

\subsubsection{The global assembler}

Now that we have covered the \cpp{main} method, we can turn to the assembler for the Stokes stiffness matrix.
We begin with the global assembler,
which is implemented in the method \cpp{assembleStokesMatrix} called in Line~\ref{li:stokes_taylorhood_call_to_assemblestokesmatrix}
of the \cpp{main} method.
The global assembler sets up the matrix pattern, loops over all elements, and accumulates the element stiffness
matrices in the global matrix. The signature of the method is
\lstinputlisting[linerange={global_assembler_signature_begin-global_assembler_signature_end},
                 numbers=left]{\PATH/stokes-taylorhood.cc}
The only arguments it gets are the finite element basis and the matrix to fill.  Observe that the Taylor--Hood basis is not
hard-wired here, so we could call the method with a different basis.
However, not surprisingly the local assembler for the Stokes problem makes relatively tight assumptions on the basis tree
structure, so there is relatively little practical freedom here.  Ideally, a global assembler should be fully
generic, and all knowledge about the current spaces and differential operators should be confined to the local
assembler.  Real discretization frameworks like \dunemodule{dune-pdelab} do achieve this separation,
but for our example here we are less strict, to avoid technicalities.

The first few lines of the \cpp{assembleStokesMatrix} method set up the matrix occupation pattern,
and initialize all matrix entries with zero.
\lstinputlisting[linerange={setup_matrix_pattern_begin-setup_matrix_pattern_end},
                 numbers=left]{\PATH/stokes-taylorhood.cc}
The method \cpp{setOccupationPattern} that constructs the matrix pattern is included
in the example file itself.  It is easy to understand for everyone who understands
the rest of the assembly code, and we therefore omit a detailed description.

Next comes the actual element loop.  We first request a \cpp{localView} object
from the finite element basis:
\lstinputlisting[linerange={get_localview_begin-get_localview_end},
                 numbers=left]{\PATH/stokes-taylorhood.cc}
After that starts the loop over the grid elements.  For each element, we bind the \cpp{localView} object
to the element.
From now on all enquiries to the local view will implicitly refer to this element.
\lstinputlisting[linerange={element_loop_and_bind_begin-element_loop_and_bind_end},
                 numbers=left]{\PATH/stokes-taylorhood.cc}
We then create the element stiffness matrix, and call the separate method \cpp{getLocalMatrix} to fill it.
For simplicity the code uses a dense matrix type even though it is known a priori
that the stationary Stokes matrix does not contain entries in the pressure diagonal block.
As local shape function indices are flat, a matrix data type without nesting is used
for the element stiffness matrix:
\lstinputlisting[linerange={setup_element_stiffness_begin-setup_element_stiffness_end},
                 numbers=left]{\PATH/stokes-taylorhood.cc}
The \cpp{getLocalMatrix} method is discussed in detail below.
It gets only the \cpp{localView} object in addition to the \cpp{elementMatrix}.  The former object contains
all necessary information.
After the call to \cpp{getLocalMatrix} the \cpp{elementMatrix} object contains the
element stiffness matrix for the current element.
The code loops over the entries of the element stiffness matrix and adds them onto the global matrix.
\lstinputlisting[linerange={accumulate_global_matrix_begin-accumulate_global_matrix_end},
                 numbers=left]{\PATH/stokes-taylorhood.cc}
The type returned in Lines~\ref{li:stokes_taylorhood_get_global_row_index} and~\ref{li:stokes_taylorhood_get_global_column_index}
for the global row and column indices is a multi-index.  It has length~3 for velocity degrees of freedom and
length~2 for pressure degrees of freedom.
Line~\ref{li:stokes_taylorhood_scatter_matrix_indices} uses the helper function
\cpp{matrixEntry} again to access the nested global stiffness matrix using those multi-indices.

The preceding loops write into all four of the matrix blocks, even though
it is known that for the Stokes
system the lower right block contains only zeros.  A more optimized version of the code would leave out the lower right
submatrix altogether.

\subsubsection{The local assembler}

It remains to investigate the method that assembles the element stiffness matrices.  Its signature is
\lstinputlisting[linerange={local_assembler_signature_begin-local_assembler_signature_end},
                 numbers=left]{\PATH/stokes-taylorhood.cc}
It only receives the local view of the Taylor--Hood basis, expected to be bound to an element,
and the empty element matrix.  There is no explicit requirement that the \cpp{LocalView} object
be a local view of a Taylor--Hood basis, but the assumption is made implicitly in various
parts of the local assembler.
The first few lines of the \cpp{getLocalMatrix} method gather some information about the element the method is to work on.
In particular, from the \cpp{localView} object it extracts the element itself, and the element's dimension and
geometry
\lstinputlisting[linerange={local_assembler_get_element_information_begin-local_assembler_get_element_information_end},
                 numbers=left]{\PATH/stokes-taylorhood.cc}
Next, the element stiffness matrix is initialized.  The \cpp{localView} object knows the total number of
degrees of freedom of the element it is bound to, and since the matrix has only scalar entries this is the correct
number of matrix rows and columns:
\lstinputlisting[linerange={initialize_element_matrix_begin-initialize_element_matrix_end},
                 numbers=left]{\PATH/stokes-taylorhood.cc}
Finally, we ask for the set of velocity and pressure shape functions:
\lstinputlisting[linerange={get_local_fe_begin-get_local_fe_end},
                 numbers=left]{\PATH/stokes-taylorhood.cc}
The two objects returned in
Lines~\ref{li:stokes_taylorhood_get_velocity_lfe}--\ref{li:stokes_taylorhood_get_pressure_lfe}
are \cpp{LocalFiniteElement}s in the \dunemodule{dune\-localfunctions} sense of the word.
These lines also
show the tree structure of the Taylor--Hood basis in action:
The expression
\begin{lstlisting}[style=Example]
localView.tree().child(_0,0)
\end{lstlisting}
returns the first child of the first child of the root, i.e., the basis for the first component of the velocity field,
and
\begin{lstlisting}[style=Example]
localView.tree().child(_1)
\end{lstlisting}
is the basis for the pressure space.
As the root of the tree combines two different bases, the static identifiers \cpp{_0} and \cpp{_1}
from the \cpp{Dune::TypeTree::Indices} namespace are needed to specify its children.  The inner node for the velocities
combines $d$ times the same basis, and hence the normal integer \cpp{0} can be used to address its first child.
This particular implementation of the local Stokes assembler is actually ``cheating'', because it exploits the knowledge
that the same basis is used for all velocity components.  Therefore, only the first leaf of the velocity
subtree is acquired in Line~\ref{li:stokes_taylorhood_get_velocity_lfe}, and then used for all components.
Using separate local finite elements is wasteful because the same shape function values and gradients
would be computed multiple times.

Next, the code constructs a suitable quadrature rule and loops over the quadrature points.  The formula for the quadrature
order combines information about the element type, the shape functions, and the differential operator.
It computes the lowest order that will integrate the weak form of the Stokes equation exactly
on a cube grid.
\lstinputlisting[linerange={begin_quad_loop_begin-begin_quad_loop_end},
                 numbers=left]{\PATH/stokes-taylorhood.cc}
The quadrature loop starts like similar local assembler codes seen elsewhere.
First, we get the inverse transposed Jacobian
of the map from the reference element to the grid element, and the Jacobian determinant for the integral
transformation formula:
\lstinputlisting[linerange={quad_loop_preamble_begin-quad_loop_preamble_end},
                 numbers=left]{\PATH/stokes-taylorhood.cc}
With these preparations done, we can assemble the first part of the stiffness matrix,  corresponding to the
velocity--velocity coupling.  For two $d$-valued velocity basis functions $\bm{\varphi}_i^k = \mathbf{e}_k \varphi_i$
and $\bm{\varphi}_j^l = \mathbf{e}_l \varphi_j$ we need to compute
\begin{equation*}
 a_e(\bm{\varphi}_i^k, \bm{\varphi}_j^l)
 \colonequals
 \int_e \nabla \bm{\varphi}_i^k \nabla \bm{\varphi}_j^l \,dx
 =
 \delta_{kl} \int_e \nabla \varphi_i \nabla \varphi_j \,dx
\end{equation*}
on the current element $e$,
where $\varphi_i$ and $\varphi_j$ are the corresponding scalar basis functions.
The code first computes the derivatives of the velocity
shape functions at the current quadrature point,
and then uses the matrix in \cpp{jacobianInverseTransposed} to transform the shape functions gradients to
gradients of the actual basis functions defined on the grid element.
\lstinputlisting[linerange={velocity_gradients_begin-velocity_gradients_end},
                 numbers=left]{\PATH/stokes-taylorhood.cc}
With the velocity basis function gradients at hand we can assemble the velocity contribution
to the stiffness matrix:
\lstinputlisting[linerange={velocity_velocity_coupling_begin-velocity_velocity_coupling_end},
                 numbers=left]{\PATH/stokes-taylorhood.cc}
Noteworthy here are the Lines~\ref{li:stokes_taylorhood_compute_vv_element_matrix_row}--\ref{li:stokes_taylorhood_compute_vv_element_matrix_column} which,
for two given shape functions from the finite element basis tree, obtain the flat numbering
used to index the element stiffness matrix.  The expression \cpp{child(_0,k)} singles out the tree leaf
for the \cpp{k}-th component of the velocity basis.  The loop variables \cpp{i} and \cpp{j} run over
the shape functions in this set, and
\begin{lstlisting}[style=Example]
localView.tree().child(_0,k).localIndex(i);
\end{lstlisting}
returns the corresponding scalar index for this shape function in the set of \emph{all} shape functions
of the Taylor--Hood basis on this element.  This is the \emph{local} index
$\iota^{\text{local}}_{\Lambda|_e}(\cdot)$ of Section~\ref{sec:localization}.
Line~\ref{li:stokes_taylorhood_update_vv_element_matrix} then updates the corresponding (scalar)
element matrix entry with the correctly weighted product of the two gradients $\nabla \varphi_i$
and $\nabla \varphi_j$.

Once this part is understood, computing the velocity--pressure coupling terms is easy.
For a given velocity shape function $\bm{\varphi}_i^k$ and pressure shape function $\theta_j$ we need
to compute
\begin{equation*}
 b_e(\bm{\varphi}_i^k,\theta_j)
 \colonequals
 \int_e \operatorname{div} \bm{\varphi}_i^k \cdot \theta_j\,dx
 =
 \int_e \sum_{l=1}^d \frac{\partial (\bm{\varphi}_i^k)_l}{\partial x_l} \cdot \theta_j\,dx
 =
 \int_e \frac{\partial \varphi_i}{\partial x_k} \cdot \theta_j\,dx
 =
 \int_e (\nabla \varphi_i)_k \cdot \theta_j\,dx.
\end{equation*}
At this point in the code the value of $\nabla \varphi_i$ at the current quadrature point
has been computed already, but value of $\theta_i$ is still unknown.
The values for all $i$ are evaluated by the following two lines:
\lstinputlisting[linerange={pressure_values_begin-pressure_values_end},
                 numbers=left]{\PATH/stokes-taylorhood.cc}
Then, the actual matrix assembly of the bilinear form $b_e(\cdot,\cdot)$ is
\lstinputlisting[linerange={velocity_pressure_coupling_begin-velocity_pressure_coupling_end},
                 numbers=left]{\PATH/stokes-taylorhood.cc}
Line~\ref{li:stokes_taylorhood_compute_vp_element_matrix_row} computes the
flat local index of $\bm{\varphi}_i^k$ again,
and Line~\ref{li:stokes_taylorhood_compute_vp_element_matrix_column} computes the index for $\theta_j$ (remember that \cpp{_1} denotes
the pressure basis).  Finally, Lines~\ref{li:stokes_taylorhood_update_vp_element_matrix_a}--\ref{li:stokes_taylorhood_update_vp_element_matrix_b}
then compute the integrand value $(\nabla \varphi_i)_k \cdot \theta_j$,
and add the resulting terms to the matrix.
This concludes the implementation of the local assembler for the Stokes problem.

\bibliographystyle{plainnat}
\bibliography{dune-functions-bases}

\end{document}